\documentclass[11pt]{article}
\usepackage{style}
\usepackage{lipsum, xcolor}
\title{High-frequency Density Nowcasts of U.S. State-Level Carbon Dioxide Emissions\thanks{The authors thank Esther Ruiz and Vladimir Rodriguez for their comments and suggestions. Our gratitude extends as well to the attendees of the 44th International Symposium on Forecasting and the Seminario Aleatorio at the Autonomous Technological Institute of Mexico.}}

\author{Ignacio Garrón\\Carlos III University of Madrid \\
    \href{igarron@est-econ.uc3m.es}{\texttt{igarron@est-econ.uc3m.es}} 
\and Andrey Ramos\\Carlos III University of Madrid \\
    \href{anramosr@eco.uc3m.es}{\texttt{anramosr@eco.uc3m.es}}
    }

\date{\today}

\begin{document}

{\setstretch{1}
\maketitle

\begin{abstract}
\vspace{0.3 cm}
\noindent

Accurate tracking of anthropogenic carbon dioxide (CO2) emissions is crucial for shaping climate policies and meeting global decarbonization targets. However, energy consumption and emissions data are released annually and with substantial publication lags, hindering timely decision-making. This paper introduces a panel nowcasting framework to produce higher-frequency predictions of the state-level growth rate of per-capita energy consumption and CO2 emissions in the United States (U.S.). Our approach employs a panel mixed-data sampling (MIDAS) model to predict per-capita energy consumption growth, considering quarterly personal income, monthly electricity consumption, and a weekly economic conditions index as predictors. A bridge equation linking per-capita CO2 emissions growth with the nowcasts of energy consumption is estimated using panel quantile regression methods. A pseudo out-of-sample study (2009–2018), simulating the real-time data release calendar, confirms the improved accuracy of our nowcasts with respect to a historical benchmark. Our results suggest that by leveraging the availability of higher-frequency indicators, we not only enhance predictive accuracy for per-capita energy consumption growth but also provide more reliable estimates of the distribution of CO2 emissions growth.\\ \\

\hspace{-0.5 cm}\textit{\textbf{Keywords:} Nowcasting, Quantile regressions, CO2 emissions, Energy consumption, MIDAS model}  
\end{abstract}
}
\newpage

\section{Introduction}

Anthropogenic emissions of carbon dioxide (CO2) and other greenhouse gases (GHGs) are the primary drivers of climate change since the pre-industrial era \citep{ipcc, jones2023national}. The combustion of fossil carbon sources across various sectors—energy, industry, transportation, waste management, and others—combined with land use changes and forestry practices, has led to increased atmospheric CO2 concentrations, significantly disturbing the Earth's surface energy balance \citep{SmithRF, Yueetal, gcb2023}. According to the most recent assessment report from the Intergovernmental Panel on Climate Change (IPCC) \citep{ipcc}, the rise in atmospheric levels of CO2 and other GHGs attributable to human activities has resulted in a net global average temperature increase of 1.1°C during the industrial period. 

The growing concern about the economic and ecological impacts of anthropogenic climate change has rapidly increased the need for policies aimed at reducing CO2 emissions. Currently, national emissions of CO2 and other GHGs are extensively regulated by the United Nations Framework Convention on Climate Change (UNFCCC). Parties to the convention are required to set annual CO2 emission targets in the form of nationally determined contributions under the Paris Agreement \citep{jones2023national}. In this context, accurate tracking anthropogenic CO2 emissions at national and sub-national levels is essential for the formulation of effective climate policies and for fulfilling long-term international commitments to mitigate climate change impacts. However, in practice, calculation of annual CO2 emissions requires information on energy consumption that is published approximately 18 months after the end of the reference period. These significant delay poses challenges to a timely and informed decision-making process and underscore the necessity of employing techniques that utilize economic indicators with a more timely publication schedule.

In this paper, we introduce a panel nowcasting methodology to simultaneously obtain high-frequency state-level nowcasts of annual energy consumption and CO2 emissions growth rate in the United States (U.S.). The set of economic predictors includes the quarterly real personal income from the Bureau of Economic Analysis (BEA), monthly electricity sales from the Energy Information Administration (EIA), and the weekly economic conditions index developed by \cite{Baumeister}. These indicators are characterized by a higher sampling frequency and a significantly shorter publication delay with respect to the energy consumption data. Building on the recent contribution of \cite{Fosten_Nandi}, our approach is implemented in two stages. In the first stage, a panel mixed-data sampling (MIDAS) model is estimated, employing a restricted Almon lag polynomial approximation of the weekly high-frequency component \cite[see,][]{Mogliani2021, Ferrara2022, Chulia2024}, and an unrestricted MIDAS for the monthly and quarterly higher-frequency indicators, as illustrated in \cite{Fosten_Nandi}. This strategy is based on \cite{Foroni2015} which indicate that while distributed lag functions, such as the Almon lag functions, are effective for high-frequency indicators, the unrestricted MIDAS performs better for small differences in sampling frequencies. Distinguishing our work from \cite{Fosten_Nandi}, the use of weekly economic indicator allow us to capture a broader dimension of economic activity potentially enhancing the accuracy of energy consumption predictions. In addition to electricity consumption, this indicator captures labor market conditions, real economic activity, mobility, and expectations. The inclusion of data at weekly frequency is what lends the "high frequency" character to our analysis and enables to produce a most timely monitoring of environmental variables.

In the second stage, a bridge equation linking CO2 emissions growth and the timely predictions of energy consumption obtained from the panel MIDAS model is estimated using the quantile regression for longitudinal data approach of \cite{Koenker2004}. The bridge equation is directly motivated by the procedure implemented by the EIA to compute CO2 emissions based on energy consumption statistics. The obtained density nowcasts provide important information regarding the distortion of the entire expected CO2 growth distribution with respect to economic condition changes. Based on the estimates of the conditional quantile function over a discrete number of quantile levels, we estimate the full continuous conditional distribution of CO2 emissions growth. This approximation allows us to provide not only the expected path of CO2 growth, but also the uncertainty surrounding the central trajectory, which distinguishes our work from that of \cite{Fosten_Nandi}, who employed conditional mean regressions. Following \cite{Adrian2019}, we chose to fit a flexible generalized skewed Student's distribution allowing for fat tails and asymmetry. This distribution has also been used with a mixed frequency model in \cite{Ferrara2022} in the context of output growth.

The predictive accuracy of each alternative model is assessed through a pseudo out-of-sample nowcasting study that simulates the real-time release schedule of the data. Several alternatives to combine the higher frequency economic indicators in the panel MIDAS equation are considered. To align closely with the methodology of \cite{Fosten_Nandi}, we perform nowcasting exercises that include the quarterly personal income and the monthly electricity sales separately. We also examine models that start with weekly economic indicators and sequentially expand the information set to include monthly and quarterly variables. Additionally, an exercise that directly nowcasts CO2 emissions growth using annual energy consumption data and all higher-frequency economic indicators, avoiding the bridge step, is conducted. This model mimics the model of \cite{Ferrara2022}, in that it introduces variables with different frequencies into a quantile regression model. The nowcasting ability of each alternative model is assessed against the historical mean and quantiles of the nowcasted variables.

A summary of the results of the  pseudo-out-of-sample exercise is as follows. Combining predictors with different sampling frequencies is beneficial for nowcasting energy consumption growth. Overall, models that include both the WECI and the monthly electricity sales simultaneously exhibit the best predictive performance across all alternatives. Substantial variation in the performance of the nowcasting exercise between states is obtained, with some states showing improvements of about 60\% relative to the historical unconditional mean benchmark. The gains in predictive accuracy observed in energy consumption growth are translated into the nowcasting of CO2 emissions when using a bridge equation. The best-performing models are again those including the WECI and the monthly electricity sales.  A model that directly produces density nowcasts of CO2 growth without relying on a bridge equation shows slightly superior performance, particularly at the lower quantiles.

\begin{figure}[H]
\centering
\begin{subfigure}{.50\textwidth}
  \centering
  \includegraphics[width=.95\linewidth]{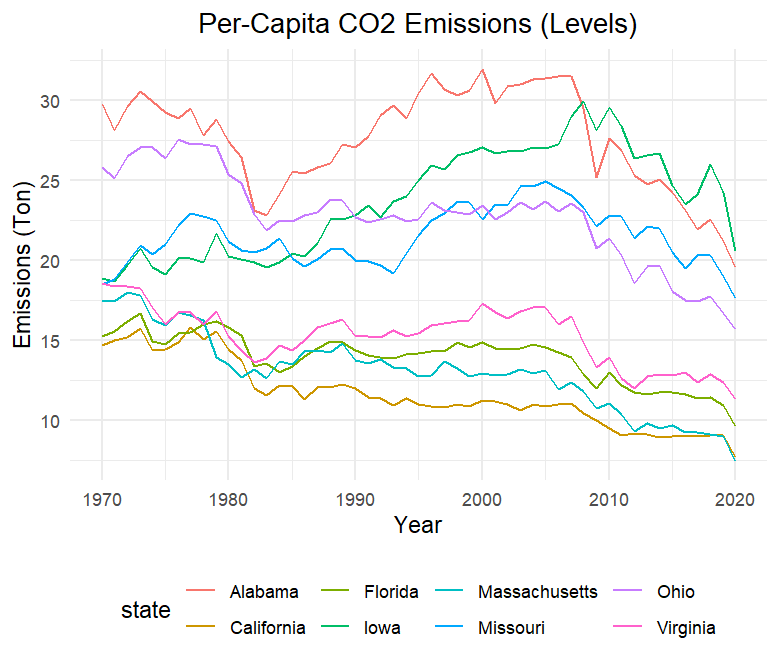}
\end{subfigure}%
\begin{subfigure}{.50\textwidth}
  \centering
  \includegraphics[width=.95\linewidth]{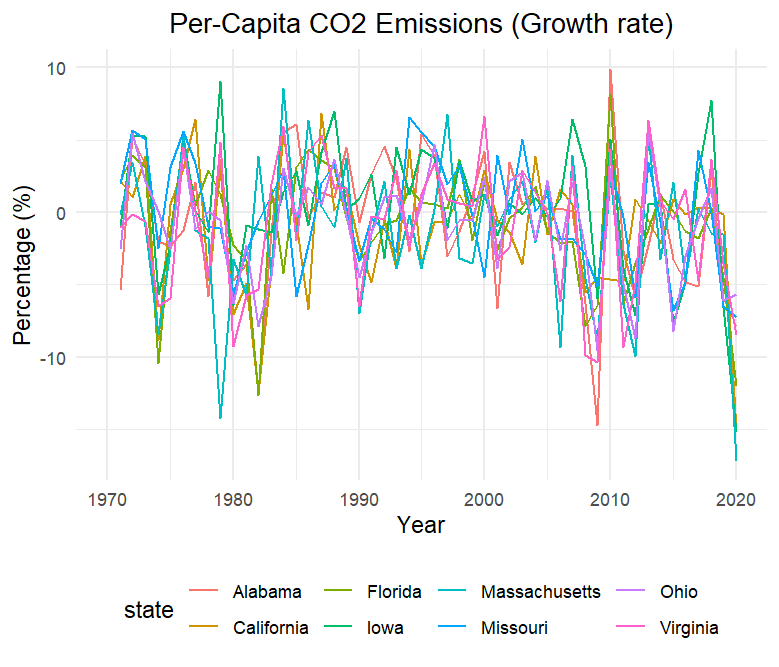}
\end{subfigure}
\caption{Per-capita CO2 emissions for selected states}
\label{CO2_Emissions}
\end{figure}

Our focus on sub-national variables provides a more detailed perspective on environmental degradation, which cannot be captured by aggregated national analyses like those in \cite{Bennedsen_CO2} or \cite{Jensen}. Figure \ref{CO2_Emissions} illustrates the annual per-capita CO2 emissions levels and growth rates for eight selected states from 1970 to the present. The data reveal considerable heterogeneity in emissions trajectories. For instance, states like California and Ohio show a consistent decrease in CO2 emissions levels, whereas in states like Iowa and Missouri, emissions continued to rise until the mid-2000s. These diverse trajectories reflect the varied policies and technologies implemented across different states, underscoring the significance of analyzing individual units to inform collective environmental goals.

\textbf{Related Literature}. This paper is related to the literature on the modeling and forecasting of the relationship between economic activity and CO2 emissions. The seminal contribution by \cite{GrossmanKrueguer} established the concept of an inverse U-shaped relationship between income and various air pollutants, a relationship now widely referred to as the Environmental Kuznets Curve (EKC). The key idea of the EKC postulates that as income increases, emissions initially rise and then eventually decline. Several mechanisms drive the complex interplay of economic and environmental factors: the scale effect, which suggests that economic growth tends to increase emissions through heightened consumption of natural resources and increased waste production; the composition effect, that implies changes in the economic output mix leading to varying emissions levels; and the technique effect, according to which advances in technology and shifts in the input mix potentially reduce emissions. \cite{Jayachandran} provides a recent review of the micro-empirical literature studying how economic development affects the environment.

From a methodological standpoint, the relationship between economic activity and CO2 emissions is typically modeled through two approaches: theoretical Integrated Assessment Models (IAMs) and empirical reduced-form econometric models. Our study aligns with the latter. A prevalent method in this domain involves specifying a panel data model that incorporates country and time fixed effects, along with a Almon lag polynomial specification of the income-pollution relationship. However, \cite{Bennedsen_ECK} highlight several econometric challenges associated with this approach, including functional misspecification, cross-sectional heterogeneity, and structural changes. Semi-parametric panel data models represent one valid alternative to face these challenges by combining parametric fixed effects with a nonparametric regression component, often employing splines or kernels \citep{Azomahou, Auffhammer, Magazzino}, as well as neural networks \citep{Bennedsen_ECK}.

A complementary interest to modeling is the development of statistical models designed to address practical challenges such as forecasting and nowcasting. For example, \cite{Bennedsen_CO2} introduce a structural augmented dynamic factor model to analyze the relationship between U.S. CO2 emissions and a large macroeconomic dataset. This model is utilized to explain, forecast, and nowcast industrial production indices and, consequently, CO2 emissions through a structural equation. The relevance of nowcasting in this context is underscored by the significant delays in the publication of emissions data. Similarly, \cite{Jensen} explores the use of machine learning methods applied to a high-dimensional panel of macroeconomic variables sampled at mixed frequencies. This approach aims to nowcast the yearly growth rate of U.S. CO2 emissions for the period 2000-2019. Both studies focus on the aggregate level of CO2 emissions. In contrast, \cite{Fosten_Nandi} adopt a different approach by proposing panel nowcasting methods to provide timely predictions of CO2 emissions and energy consumption growth across all U.S. states. Their methodology employs a panel MIDAS model that uses quarterly and monthly economic indicators as predictors for energy consumption, coupled with a bridge equation that projects CO2 emissions based on these energy consumption forecasts.

Our research contributes to the existing literature on CO2 emissions nowcasting in two ways. First, we expand the methodology established by \cite{Fosten_Nandi} by introducing quantile high-frequency density nowcasts through a panel quantile regression in the bridge equation for CO2 emissions. This procedure enables the examination of not only the central trajectory of CO2 emissions, which has been previously investigated in the literature, but also the uncertainty surrounding that trajectory. Indeed, there has been a notable increase in recent years in the focus of policymakers on uncertainty. This is evidenced by the growing body of literature on methodologies for assessing the likelihood of distress scenarios using quantile regressions, which builds on the seminal work of \cite{Adrian2019}. Second, we advance the timeliness of our predictions by incorporating high-frequency economic indicators. Unlike \cite{Fosten_Nandi}, who limited their analysis to quarterly personal income and monthly electricity sales data, we include the state-level weekly economic indicator from \cite{Baumeister} to predict energy consumption in the panel MIDAS model. This inclusion allows us to capture a wider set of dimensions of economic activity with important capacity to predict energy consumption.

The remainder of the paper is organized as follows. Section 2 describes the data sources and variables used in our analysis. Section 3 outlines the different steps of our nowcasting methodology and the models proposed. The results of the empirical analysis are presented in Section 4. Finally, Section 5 concludes.

\section{Data}

\subsection{State-level CO2 emissions and energy consumption data}

Our primary variable of interest is state-level energy-related CO2 emissions in the U.S. Data for this variable, sourced from the U.S. EIA, are available annually starting from 1970. Total state CO2 emissions aggregates emissions from direct fuel use across all sectors, including residential, commercial, industrial, and transportation, as well as from primary fuels consumed for electricity generation. The panel consists of \(N = 51\) units, which include the 50 states and the District of Columbia. The publication delay for CO2 emissions data is approximately two years and three months after the end of the reference year, a notably longer lag compared to other state-level economic data. Our analysis focuses on nowcasting the growth rate of per-capita CO2 emissions.

Annual energy consumption (EC) data at the state-level is obtained from the State Energy Data System (SEDS) also produced by the EIA. This dataset, available from 1960 onwards, is the main input to compute the state-level CO2 emissions. In particular, the SEDS collects detailed data on the consumption of coal, natural gas, and petroleum across the different economic sectors. To estimate CO2 emissions, the EIA applies specific energy content and carbon emission factors to each type of consumed fuel. These factors convert the quantity of fuel used into energy produced and corresponding CO2 emissions. The calculations are periodically adjusted to reflect changes in fuel composition and new scientific findings. Regarding timeliness of the data, the publication lag of energy consumption is approximately 18 months, considerably shorter than that for CO2 emissions. As with CO2 emissions, we focus on the growth rate of per-capita energy consumption.

\subsection{Higher-frequency economic indicators}

State-level economic indicators are available at higher frequency and are published in a more timely fashion than CO2 emissions or energy consumption. Concretely, quarterly real and per-capita personal income (PI) is obtained from the BEA since 1950 and features a publication lag of approximately three months. Monthly electricity consumption (ELEC), computed as total electricity sales to end-users across all U.S. states, is published by the EIA since 1990 with a publication lag of about two months following the end of the reference month. For the analysis, we consider the year-on-year log difference of both variables.

Both variables, PI and ELEC, are employed as predictors in the analysis of \cite{Fosten_Nandi}. We extend the scope of the analysis by including the weekly economic conditions index (WECI) developed by \cite{Baumeister}. Starting in 1987, the WECI is derived from a mixed-frequency dynamic factor model that integrates a wide set of weekly, monthly, and quarterly economic variables. It encompasses a comprehensive range of economic dimensions, including labor market indicators, household spending, real economic activity, mobility, financial indicators, and expectations. The publication lag is approximately one month, as indicated on the author's website. The high-frequency nature of our nowcasting approach stems from the use of this variable.

\section{Nowcasting Methodology}

Our nowcasting approach is implemented in two stages. In the first stage, we propose a panel MIDAS model to predict energy consumption growth, utilizing quarterly, monthly, and weekly economic predictors. A restricted Almon lag polynomial approximation of the weekly high-frequency component is employed, as outlined in \cite{Mogliani2021}, \cite{Ferrara2022}, and \cite{Chulia2024}. This is combined with an unrestricted MIDAS model for the monthly and quarterly indicators, as in \cite{Fosten_Nandi}. In the second stage, we employ a bridge equation to generate predictions for CO2 emissions growth, using the first-stage nowcasts of energy consumption growth as the predictor variable. Distinctively from \cite{Fosten_Nandi}, we move beyond the conditional-mean predictions framework and produce density forecasts adopting the panel quantile regressions approach of \cite{Koenker2004}. Out-of-sample density nowcasts are obtained for the period 2009 to 2018, and their performance both on aggregate and individual levels are evaluated using different metrics. The subsequent subsections provide a detail each of these steps. A note on notation: we use bold letters to refer to vectors and matrices.

\subsection{Nowcasting energy consumption growth using a panel MIDAS model}

As explained earlier, to compute state-level CO2 emissions, the EIA uses the energy content and carbon emission factors for each consumed fuel type, converting fuel use into energy produced and CO2 emitted. These calculations are periodically updated by changes in fuel composition and scientific advances in terms of energy efficiency. Energy consumption data, published with an 18-month lag, is timelier than CO2 emissions data. Thus, we first nowcast energy consumption growth.

Consider the following general model,

  \begin{equation}
    \underbrace{c_{i,t}}_{Annual} = g_v(c_{i,t-d_{v}},\underbrace{{\mathbf{x}_{i,t-\frac{q_{v}}{4}}^{(q)}}}_{Quarterly},\underbrace{{\mathbf{z}_{i,t-\frac{m_{v}}{12}}^{(m)}}}_{Monthly},\underbrace{{\mathbf{w}_{i,t-\frac{w_{v}}{52}}^{(w)}})}_{Weekly} + u_{i,t},
\end{equation}

where $c_{i,t}$ is the annual growth rate of per-capita energy consumption at state $i$, $i = 1,2,...N$, and year $t$, $t = 1,2,...,T$. We define $\mathbf{x}_{i,t-\frac{q_{v}}{4}}^{(q)}$ as a $(4\times1)$ vector of quarterly PI growth lags, $\mathbf{z}_{i,t-\frac{m_v}{12}}^{(m)}$ as a $(12\times1)$ vector of monthly ELEC growth lags, and $\mathbf{w}_{i,t-\frac{w_v}{52}}^{(w)}$ as a $(52\times1)$ high-frequency vector of WECI growth lags, which are observed $q=4$, $m=12$, $w=52$ times between year $t-1$ and $t$. We denote $v$ as the calendar date of prediction. The available lag of the dependent variable is represented by $d_{v}$, while for the quarterly, monthly, and weekly predictors are denoted by $q_{v}$, $m_{v}$, and $w_{v}$, respectively. The target function $g_v(.)$ maps our covariates for a given calendar $v$, and $u_{i,t}$ is a zero mean random error.

The general nowcasting equation is,

\begin{equation}
    \widehat{c_{i,t}} = \widehat{g_v}(c_{i,t-d_{v}},{\mathbf{x}_{i,t-\frac{q_{v}}{4}}^{(q)}},{\mathbf{z}_{i,t-\frac{m_{v}}{12}}^{(m)}},{\mathbf{w}_{i,t-\frac{w_{v}}{52}}^{(w)}}),
\end{equation}

where $\widehat{g_v}(.)$ is the predicted target function for the given calendar date of prediction $v$. As is common in the nowcasting literature, the information contained in $v$ is set to reflect the release schedule of the variables in real time. This allows for the generation of multiple nowcasts and backcasts for each period under consideration. This approach enables an assessment of how the performance of our model changes as new information is incorporated into the nowcasting model; see, for instance, \cite{Giannone2008,Banbura2013,Fosten_Nandi}.

It should be noted that the number of parameters in this specification is considerable large, particularly in the case of the weekly variable. This may give rise to a proliferation of parameters, which could render standard regression procedures invalid. In the following section, we present our MIDAS framework, which we employed to estimate the aforementioned regression.

\subsection{MIDAS setup}

Let us consider a linear panel MIDAS model for $g_v(.)$:
\begin{equation}
    c_{i,t} = \alpha_{i} +\phi_{v}c_{i,t-d_{v}}+
    { \mathcal{B}(L^{q_v/q},\pmb{\beta}_{q})x_{i,t}^{(q)}}+
    { \mathcal{B}(L^{m_v/m},\pmb{\beta}_{m})z_{i,t}^{(m)}} +
    { \mathcal{B}(L^{w_v/w},\pmb{\theta}_{w})w_{i,t}^{(w)}}+ u_{i,t},
    \label{eq_AR_W_M_Q_MIDAS}
\end{equation}

and,
\begin{align*}
{\mathcal{B}(L^{q_v/q},\pmb{\beta}_{q})x_{i,t}^{(q)}}&=\beta_{1,q}{x}_{i,t-\frac{q_v}{4}}^{(q)}+\beta_{2,q}{x}_{i,t-\frac{q_v-1}{4}}^{(q)}+
\beta_{3,q}{x}_{i,t-\frac{q_v-2}{4}}^{(q)}+\beta_{4,q} {x}_{i,t-\frac{q_v-3}{4}}^{(q)}, \\
{\mathcal{B}(L^{m_v/m},\pmb{\beta}_{m})z_{i,t}^{(m)}}&=\beta_{1,m}z_{i,t-\frac{m_v}{12}}^{(m)}+\beta_{2,m}z_{i,t-\frac{m_v-1}{12}}^{(m)}+\beta_{3,m}z_{i,t-\frac{m_v-2}{12}}^{(m)}+...+\beta_{1,m}z_{i,t-\frac{m_v-11}{12}}^{(m)}, \\
\mathcal{B}(L^{w_v/w},\pmb{\theta}_{w})w_{i,t}^{(w)}&=
   {\sum_{c=0}^{C_{w}-1}} {\sum_{l = 0}^{p}} \tilde{B}(c;\pmb{\theta}_{w})L^{{c}/w}w_{i,t-w_v}^{(w)},
\end{align*}

where the lag coefficients in $\mathcal{B}(L^{q_v/q},\pmb{\beta}_{q}),\mathcal{B}(L^{m_v/m},\pmb{\beta}_{m})$, and $\mathcal{B}(L^{w_v/w},\pmb{\theta}_{w})$ are 
parameterized as a function of a low dimensional vector of parameters. Here, $\alpha_{i}$ represents the individual fixed effects, $\phi$ is the autoregressive parameter, and $u_{i,t}$ is a zero-mean random error term. 

As noted in \autoref{eq_AR_W_M_Q_MIDAS}, no restrictions are placed on the quarterly and monthly variables, while an Almon lag polynomial is applied to the weekly indicator. Specifically, $\tilde{B}(c;\pmb{\theta}_{w})$ is a weighting function, normalized to sum up to 1, which depends on a vector of parameters $\theta_{w}$ and lag-order $c$. For the Almon-lags choice, $\tilde{B}(c;\pmb{\theta}_{w}) = \sum_{l = 0}^{p} \theta_{l,w} c^{l}$, where $\pmb{\theta}_{w} := (\theta_{0,w}, \theta_{1,w},...,\theta_{p,w})^{\prime}$. Also, it is desirable to further consider restrictions on the value and slope of the lag polynomial $\tilde{B}(c;\pmb{\theta}_{w})$. In particular, by imposing $\tilde{B}(C_w-1;\pmb{\theta}_{w})=0$ and $\nabla_c \tilde{B}(C_w-1;\pmb{\theta}_{w}|_{c=C_w-1})=0$, we consider a lag structure with good economic properties, as it slowly decays towards zero \cite[see][]{Mogliani2021}.

Under the so-called "direct-method", \autoref{eq_AR_W_M_Q1} can be re-parameterized as:
\begin{equation}
    c_{i,t} = \alpha_{i} +\phi c_{i,t-d_{v}}+
    \underbrace{\mathbf{x}_{i,t-\frac{q_{v}}{4}}^{(q)\prime} \pmb{{\beta}}_{q}}_{unrestricted}+
    \underbrace{\mathbf{z}_{i,t-\frac{k_{v}}{12}}^{(m)\prime} \pmb{{\beta}}_{m}}_{unrestricted} +
    \underbrace{\mathbf{\tilde{w}}_{i,t-\frac{w_{v}}{52}}^{(w)\prime} \pmb{\widehat{\theta}}_{w}}_{{Almon}} + u_{i,t}.
    \label{eq_AR_W_M_Q1}
\end{equation}

where $\pmb{\beta}_{q}$ is a $(4\times1)$ vector, $\pmb{\beta}_{m}$ is a $(12\times1)$ vector, and $\pmb{\Tilde{\theta}}_{v}^{(w)}$ is a vector featuring $(p+1-r\times1)$ parameters. The vector $\mathbf{\Tilde{w}}_{i,t}:=\mathbf{Q}_{w}\mathbf{w}_{i,t}$ with size $(p + 1-r\times 1)$ are linear combinations of the WECI lags, and $\mathbf{Q}_{w}$ is a $(p+1-r\times C_w)$ polynomial weighting matrix defined accordingly. In our application, which we consider a third-degree Almon lag polynomial ($p=3$) with two end-point restrictions $r=2$, so that the number of parameters of the high-frequency indicator is reduced substantially to $p+1-r=2$; see \cite{Ferrara2022} and \cite{Chulia2024} who consider the same parametrization. As the model is linear in parameters, \autoref{eq_AR_W_M_Q1} can be estimated by panel least squares \citep[see, ][]{Fosten_Nandi,Foroni2015}.   In terms of consistency, MIDAS models require that lag order of $c_{i,t}$ and the lag polynomials are sufficiently large to make $u_{i,t}$ white noise \citep[see][]{Ghysels2006,Foroni2015}.

Thus, the nowcasting equation becomes,
\begin{equation}
    \widehat{c}_{i,t} = \widehat{\alpha_{i}} +\widehat{\phi} c_{i,t-d_{v}}+
    \mathbf{x}_{i,t-\frac{q_{v}}{4}}^{(q)\prime} \pmb{\widehat{\beta}}_{q}+
    \mathbf{z}_{i,t-\frac{k_{v}}{12}}^{(m)\prime} \pmb{\widehat{\beta}}_{m} +
    \mathbf{\tilde{w}}_{i,t-\frac{w_{v}}{52}}^{(w)\prime} \pmb{\widehat{\theta}}_{w}.
    \label{eq_AR_W_M_Q_Q2}
\end{equation}

The combination of frequencies in \autoref{eq_AR_W_M_Q_Q2} offers the advantage of considering all the information in a single regression. Our strategy is based on the montecarlo results of \cite{Foroni2015}, which indicate that while distributed lag functions, such as the Almon lag functions, are effective for high-frequency indicators, whereas the unrestricted function performs better for small differences in sampling frequencies. Also, similar approaches has been implemented and demonstrated to enhance accuracy in the literature, as evidenced by the works of \cite{Ferrara2022} and \cite{Carriero2022}. To this end, we employ a restricted Almon lag polynomial for the weekly high-frequency indicator \cite{Mogliani2021}, while allowing for an unrestricted polynomial for the monthly and quarterly higher-frequency indicators, as illustrated in \cite{Fosten_Nandi}.

\subsection{Nowcasting CO2 emissions growth using a bridge equation}

CO2 emissions growth is the main target variable in our analysis. Timely predictions of this variable are produced via a bridge equation linking energy consumption and CO2 emissions. Let $\widehat{c}_{v,i,t}$ the predicted value of $c_{i,t}$ for state $i$ at year $t$ and nowcast time $v$. The conditional mean forecast panel bridge equation with a multi-factor error structure, adopted from \cite{Fosten_Nandi}, is given by:
\begin{equation}
    e_{i,t} = \gamma_{i} + \rho e_{i,t-g_{v}} + \delta \widehat{c}_{v,i,t}+ \pmb{\lambda}^{\prime}\pmb{f}_{t-d_v}+ \epsilon_{i,t},
\end{equation}
where $e_{i,t}$ denotes the per-capita CO2 emissions growth, $d_{v}$ is the last available lag of $e_{i,t}$, $f_{t-d_v}$ are unknown common factors with loading vector $\pmb{\lambda}$. In a similar fashion to \cite{Fosten_Nandi}, the factors estimates are obtained by taking an average of the covariates, such that $\pmb{f}_{t}=[\bar{e}_{i,t-d_{v}},\bar{c}_{i,t-d_v}]\prime$. The parameters $\gamma_i$, $\rho$, $\delta$ and $\pmb{\lambda}$ capture the fixed effects, the autoregressive coefficient, the effect of EC growth prediction from the first step, and the factors coefficients, respectively. Once the common factors are obtained, the model can be estimated using panel least squares \citep[see,][]{Fosten2023}. 

Our main contribution to this framework is to produce density nowcasts for $e_{i,t}$ using the quantile regression for longitudinal data proposed by \cite{Koenker2004}. The conditional quantile of $e_{i,t}$ is modeled as:
\begin{equation}
    Q_{e_{i,t}}(\tau|e_{i,t-g_{v}},\widehat{c}_{v,i,t},f_{t}) = \gamma_i(\tau) + \rho(\tau)e_{i,t-g_{v}} + \delta (\tau)\widehat{c}_{v,i,t} + \pmb{\lambda}_{v}^{\prime}(\tau)\pmb{f}_{t-d_v},
    \label{eq_co2_quantile1}
\end{equation}
where $\tau=0.25,0.50,0.75$ denotes the quantile level. Consequently, our framework considers the effect of our covariates on the conditional distribution of $e_{i,t}$. \cite{Koenker2004} proposes the estimation of Equation \ref{eq_co2_quantile1} employing $\ell_1$ regularization methods. As \cite{Koenker2004} points out, the introduction of individual fixed effects may result in a notable increase in the variability of estimates of other covariate effects. The application of regularization of these individual effects towards a common value, can assist in modifying this inflation effect. Nevertheless, determining the optimal degree of shrinkage presents a significant practical challenge when dealing with multiple calendar dates.\footnote{In the context of our application, we have set this parameter equal to one, and the resulting nowcasting exercise has yielded satisfactory results.}

We consider another alternative to produce high-frequency density nowcasts of the growth rate of per-capita CO2 emissions without relying on the bridge equation, in the spirit of \cite{Ferrara2022}. This alternative directly predicts the conditional quantile function of $e_{i,t}$, directly conditioning on the full set of weekly, monthly, and quarterly predictors. Adapting the equation in \cite{Ferrara2022} to a panel data framework, the estimated model is given by:
\begin{equation}
\begin{split}
    Q_{e_{i,t}}(\tau|e_{i,t-g_{v}},\mathbf{x}_{i,t-\frac{q_{v}}{4}}^{(q)},
     \mathbf{z}_{i,t-\frac{k_{v}}{12}}^{(m)},
     \pmb{\tilde{w}}_{i,t-\frac{w_{v}}{52}}^{(w)},f_{t}) = \gamma_{i}(\tau) + \rho(\tau)e_{i,t-g_{v}} +
    \mathbf{x}_{i,t-\frac{q_{v}}{4}}^{(q)\prime} \pmb{\widehat{\beta}}_{q}(\tau)+\\
    \mathbf{z}_{i,t-\frac{k_{v}}{12}}^{(m)\prime} \pmb{\widehat{\beta}}_{m}(\tau) +
    \pmb{\tilde{w}}_{i,t-\frac{w_{v}}{52}}^{(w)\prime} \pmb{\widehat{\theta}}_{w}(\tau)+
        {\lambda}(\tau)f_{t-d_v}.
    \label{eq_co2_quantile2}
\end{split}
\end{equation}

Based on these estimates, a full continuous conditional distribution is estimated as in \cite{Adrian2019} and \cite{Ferrara2022}. We choose to fit a generalized skewed Student's $t$ distribution \citep{Azzalini2003}, which depends on four parameters associated to location, scale, fatness, and shape. These parameters are obtained through a quantile matching approach aiming at minimizing the squared distance between the estimated discrete conditional quantiles and the corresponding quantiles of the skewed Student's distribution, as in \cite{Adrian2019}. This procedure is flexible to accommodate fat tails and asymmetry potentially present in the context of our application.

\subsection{Competing models}

Table \ref{tab:model_list} illustrates the various specifications considered in our empirical exercise. As simple benchmarks, we consider the historical mean and a one-lag autoregressive (AR) model, as in \cite{Fosten2023}. In addition, models AR-Q and AR-M, which include only the monthly and quarterly variables, respectively, are defined as in \cite{Fosten_Nandi}. The rest of the models include the weekly information (AR-W), jointly with the monthly (AR-W-M), and all variables sampled at different frequencies (AR-W-M-Q). Finally, the direct AR-W-M-Q considers the direct quantile nowcast approach as in \autoref{eq_co2_quantile2}.

\begin{table}[!htbp]
    \centering
    \caption{Competing models}
    \begin{adjustbox}{width=1\textwidth}
    \begin{tabular}{ll}
    \hline
        Model & Information considered for the bridge model\\
        \hline \hline
        Benchmark model & Historical mean\\
        AR & Autoregressive model with order 1\\
        AR-M & AR + monthly variable (UMIDAS)\\
        AR-Q & AR + quarterly variable (UMIDAS)\\
        AR-W & AR + weekly variable (Almon MIDAS) \\
        AR-W-M & AR + monthly (UMIDAS) + weekly variables (Almon MIDAS) \\
        AR-W-M-Q & AR + monthly and quarterly (UMIDAS) + weekly variables (Almon MIDAS) \\
        \hline
        Direct AR-W-M-Q & Defined as in \autoref{eq_co2_quantile2}. \\
         \hline \hline
    \end{tabular}
    \end{adjustbox}
    \label{tab:model_list}
\end{table}

\subsection{Out-of-sample exercise}

To maintain comparability with \cite{Fosten_Nandi}, we generate out-of-sample nowcasts for the period 2009 to 2018 across the 50 individual states. The estimation period starts from 1990 and it considers a expanding window. For each year in the evaluation period, we use a weekly calendar to make multiple nowcast and backcast updates on different dates, $v$. This approach aims to replicate the ragged edge in the data, emulating real-time releases. For each data release, we incorporate the new data lag available, adjust the lag structure of the model, re-estimate the model, and obtained energy consumption and CO2 predictions. This enables us to observe the behavior of our predictions as we incorporate additional information as it becomes available.

A distinctive feature of our proposed calendar is its consideration of a weekly calendar.  Our calendar is based on \cite{Fosten_Nandi}, and is extended to accommodate WECI information. Table \ref{tab_calendar} illustrates the full information flow for the year 2021 as an illustrative example. As the year progresses, new observations of WECI are incorporated each week, new observations of ELEC are added each month, and new observations of PI are included each quarter. Beginning from the backcasting period, WECI data is incorporated until the end of week 4 (January), at which point all data up to December of the previous year are available. Subsequently, the nowcasting period is initiated, where weekly data from the current year is incorporated at every week. Monthly data for the current year is introduced from week 8 (March), while quarterly data for the PI is added from week 21 (June). Annual data for previous years (specifically 2018 and 2019) for CO2 and EC are incorporated in week 8 (June) and week 20 (March), respectively. In total, we consider 48 weeks for the prediction period.

For the EC growth bridge model, the prediction accuracy is evaluated by comparing the average root mean squared forecast error (RMSFE) of the nowcast model with those of a benchmark consisting of the historic average using the data available at the time of the nowcast. The RMSFE is tracked across multiple nowcast dates, $v$, and is defined as:

\begin{equation}
    RMSFE_{v} = \frac{1}{N}\sum_{i = 1}^{N}\sqrt{\frac{1}{P}\sum_{t = T-P+1}^{T}\hat{c}_{v,i,t}^{2}},
\end{equation}

where $\hat{\epsilon}_{v,i,t}^{2}$ is the prediction error of a model on nowcast date $v$ for state $i$ and year $t$, $T$ is the last year in the sample and $P$ is the number of out-of-sample predictions.

For the density forecasts of per-capita CO2 emissions growth, we consider the Quantile Score (QS), which is a common metric used to evaluate a particular quantile forecast \citep[see][]{Gneiting2007,Giacomini2005}, defined as follows,

\begin{equation}
	QS_{v,\tau}=\frac{1}{N}\sum_{i = 1}^{N}\frac{1}{P}\sum_{t = T-P+1}^{T}(\hat{Q}_{e_{v,i,t}}(\tau|X)-e_{v,i,t})(1(\hat{Q}_{e_{v,i,t}}(\tau|X)<e_{v,i,t}) - \tau),  
\end{equation}

where and $1(\hat{Q}_{e_{v,i,t}}(\tau|X)<e_{v,i,t})$ is the indicator function that takes a value of 1 if the outcome is below the nowcast of the conditional quantile $\hat{Q}_{e_{v,i,t}}(\tau|X)$ and 0 otherwise. Then, when the quantile scores are aggregated, we define a discrete version of the Continuous Ranked Probability Score (CRPS) \citep{Gneiting2011}, as follows,

\begin{equation}
	CRPS_{v}= \frac{1}{J}\sum_{j=1}^{J}\omega(\tau_j)QS(v,\tau_j),
\end{equation}

where $J=3$, $\tau_1=0.25$, $\tau_2=0.50$, $\tau_3=0.75$, and the weights $\omega(\tau_j)$ are set to 1 to account for equal weights across quantiles. As \cite{Gneiting2011} point out, the discrete version of the CRPS constitutes a proper scoring rule that emerges as a particular instance of the continuous version. Furthermore, we examine the distribution of the aforementioned accuracy measures for each calendar, without averaging over the states. However, as \cite{Fosten_Nandi} observed, these results are merely indicative, as they are calculated with a small sample size.

\begin{table}[!htbp]
\caption{Weekly calendar of information availability (example for 2021)}
\centering
\begin{adjustbox}{width=0.8\textwidth}
\begin{threeparttable}
\begin{tabular}{lrcccccc}
\hline \hline
        \multicolumn{3}{c}{Calendar date} & \multicolumn{5}{c}{Latest information avalable for:} \\ \hline
        ~ & \multicolumn{2}{c}{Weekly calendar ($v$)} & CO2 & EC & PI & ELEC & WECI   \\ \hline
        Backcast & 1 & 2021:W1 & 2017 & 2018 & 2020:Q3 & 2020:M11 & 2020:W49  \\ 
        ~ & 2 & 2021:W2 & 2017 & 2018 & 2020:Q3 & 2020:M11 & 2020:W50   \\ 
        ~ & 3 & 2021:W3 & 2017 & 2018 & 2020:Q3 & 2020:M11 & 2020:W51   \\ 
        ~ & 4 & 2021:W4 & 2017 & 2018 & 2020:Q3 & 2020:M11 & 2020:W52   \\  \hline
         Nowcast & 5 & 2021:W5 & 2017 & 2018 & 2020:Q3 & 2020:M12 & 2021:W1  \\ 
        ~ & 6 & 2021:W6 & 2017 & 2018 & 2020:Q3 & 2020:M12 & 2021:W2  \\ 
        ~ & 7 & 2021:W7 & 2017 & 2018 & 2020:Q3 & 2020:M12 & 2021:W3   \\ 
        ~ & 8 & 2021:W8 & 2017 & 2018 & 2020:Q3 & 2020:M12 & 2021:W4   \\ 
        ~ & 9 & 2021:W9 & 2018 & 2018 & 2020:Q4 & 2021:M1 & 2021:W5  \\ 
        ~ & 10 & 2021:W10 & 2018 & 2018 & 2020:Q4 & 2021:M1 & 2021:W6  \\ 
        ~ & 11 & 2021:W11 & 2018 & 2018 & 2020:Q4 & 2021:M1 & 2021:W7  \\ 
        ~ & 12 & 2021:W12 & 2018 & 2018 & 2020:Q4 & 2021:M1 & 2021:W8  \\ 
        ~ & 13 & 2021:W13 & 2018 & 2018 & 2020:Q4 & 2021:M2 & 2021:W9  \\ 
        ~ & 14 & 2021:W14 & 2018 & 2018 & 2020:Q4 & 2021:M2 & 2021:W10  \\ 
        ~ & 15 & 2021:W15 & 2018 & 2018 & 2020:Q4 & 2021:M2 & 2021:W11  \\ 
        ~ & 16 & 2021:W16 & 2018 & 2018 & 2020:Q4 & 2021:M2 & 2021:W12  \\ 
        ~ & 17 & 2021:W17 & 2018 & 2018 & 2020:Q4 & 2021:M3 & 2021:W13  \\ 
        ~ & 18 & 2021:W18 & 2018 & 2018 & 2020:Q4 & 2021:M3 & 2021:W14  \\ 
        ~ & 19 & 2021:W19 & 2018 & 2018 & 2020:Q4 & 2021:M3 & 2021:W15  \\ 
        ~ & 20 & 2021:W20 & 2018 & 2018 & 2020:Q4 & 2021:M3 & 2021:W16  \\ 
        ~ & 21 & 2021:W21 & 2018 & 2019 & 2021:Q1 & 2021:M4 & 2021:W17  \\ 
        ~ & 22 & 2021:W22 & 2018 & 2019 & 2021:Q1 & 2021:M4 & 2021:W18  \\ 
        ~ & 23 & 2021:W23 & 2018 & 2019 & 2021:Q1 & 2021:M4 & 2021:W19  \\ 
        ~ & 24 & 2021:W24 & 2018 & 2019 & 2021:Q1 & 2021:M4 & 2021:W20 \\ 
        ~ & 25 & 2021:W25 & 2018 & 2019 & 2021:Q1 & 2021:M5 & 2021:W21  \\ 
        ~ & 26 & 2021:W26 & 2018 & 2019 & 2021:Q1 & 2021:M5 & 2021:W22  \\ 
        ~ & 27 & 2021:W27 & 2018 & 2019 & 2021:Q1 & 2021:M5 & 2021:W23  \\ 
        ~ & 28 & 2021:W28 & 2018 & 2019 & 2021:Q1 & 2021:M5 & 2021:W24  \\ 
        ~ & 29 & 2021:W29 & 2018 & 2019 & 2021:Q1 & 2021:M6 & 2021:W25  \\ 
        ~ & 30 & 2021:W30 & 2018 & 2019 & 2021:Q1 & 2021:M6 & 2021:W26  \\ 
        ~ & 31 & 2021:W31 & 2018 & 2019 & 2021:Q1 & 2021:M6 & 2021:W27  \\ 
        ~ & 32 & 2021:W32 & 2018 & 2019 & 2021:Q1 & 2021:M6 & 2021:W28  \\ 
        ~ & 33 & 2021:W33 & 2018 & 2019 & 2021:Q2 & 2021:M7 & 2021:W29  \\ 
        ~ & 34 & 2021:W34 & 2018 & 2019 & 2021:Q2 & 2021:M7 & 2021:W30  \\ 
        ~ & 35 & 2021:W35 & 2018 & 2019 & 2021:Q2 & 2021:M7 & 2021:W31 \\ 
        ~ & 36 & 2021:W36 & 2018 & 2019 & 2021:Q2 & 2021:M7 & 2021:W32  \\ 
        ~ & 37 & 2021:W37 & 2018 & 2019 & 2021:Q2 & 2021:M8 & 2021:W33  \\ 
        ~ & 38 & 2021:W38 & 2018 & 2019 & 2021:Q2 & 2021:M8 & 2021:W34  \\ 
        ~ & 39 & 2021:W39 & 2018 & 2019 & 2021:Q2 & 2021:M8 & 2021:W35  \\ 
        ~ & 40 & 2021:W40 & 2018 & 2019 & 2021:Q2 & 2021:M8 & 2021:W36 \\ 
        ~ & 41 & 2021:W41 & 2018 & 2019 & 2021:Q2 & 2021:M9 & 2021:W37  \\ 
        ~ & 42 & 2021:W42 & 2018 & 2019 & 2021:Q2 & 2021:M9 & 2021:W38  \\ 
        ~ & 43 & 2021:W43 & 2018 & 2019 & 2021:Q2 & 2021:M9 & 2021:W39  \\ 
        ~ & 44 & 2021:W44 & 2018 & 2019 & 2021:Q2 & 2021:M9 & 2021:W40  \\ 
        ~ & 45 & 2021:W45 & 2018 & 2019 & 2021:Q3 & 2021:M10 & 2021:W41  \\ 
        ~ & 46 & 2021:W46 & 2018 & 2019 & 2021:Q3 & 2021:M10 & 2021:W42  \\ 
        ~ & 47 & 2021:W47 & 2018 & 2019 & 2021:Q3 & 2021:M10 & 2021:W43  \\ 
        ~ & 48 & 2021:W48 & 2018 & 2019 & 2021:Q3 & 2021:M10 & 2021:W44  \\ 
        
        \hline \hline
\end{tabular}
\begin{tablenotes}
\item \textit{Notes: This calendar is based on \cite{Fosten_Nandi}, and extended to accommodate weekly information of the WECI.} 
\end{tablenotes} 
\end{threeparttable}
\end{adjustbox}
\label{tab_calendar}
\end{table}

\section{Nowcasting Results}

This section presents the results of the pseudo-out-of-sample exercise described in the previous subsection. We report the results for the growth rate of per-capita EC and CO2 emissions. In Appendix \ref{appendix_3}, we present the figures for the original growth rates of energy consumption and CO2 emissions, where similar results are documented.

\subsection{Nowcasting energy consumption growth using panel MIDAS}

\begin{figure}[h]
    \caption{RMSFE across states for EC per capita growth}
    \includegraphics[width=15 cm,keepaspectratio]{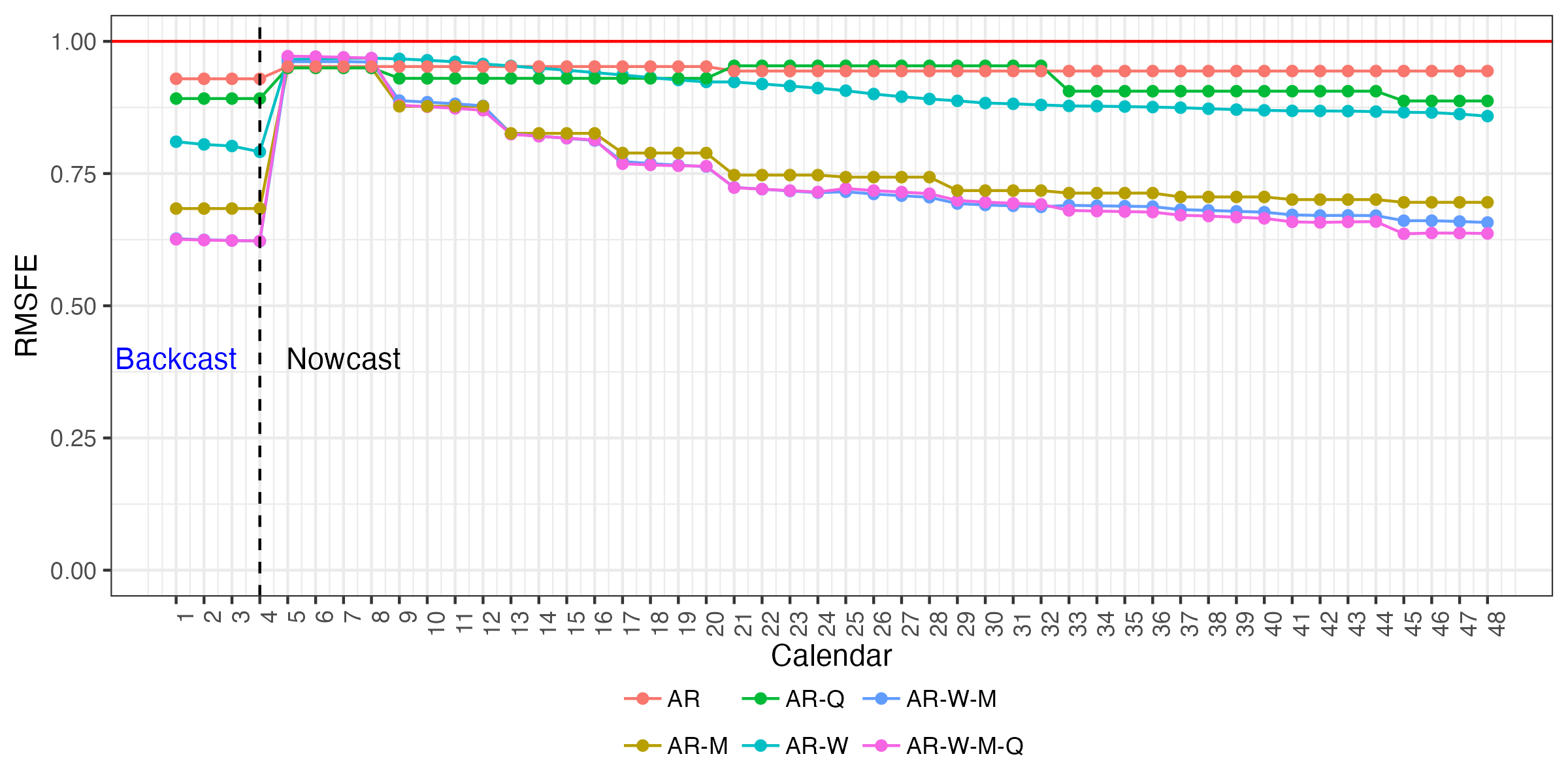}
    \textit{Notes: The AR-W-M-Q model incorporates an autoregressive (AR) component, the weekly economic condition index (W), the monthly electricity sales (M), and the quarterly PI data (Q). The benchmark model is a historic mean for electricity consumption growth. The benchmark normalizes the RMSFE figures at the first release date. Consequently, any points below 1 indicate that the RMSFE is lower than that of the benchmark.} 
    \label{fig_RMSE_monthly}
\end{figure}

We begin our discussion with the results from the panel MIDAS model used to nowcast the per-capita EC growth rate. Figure \ref{fig_RMSE_monthly} plots the average RMSFE across states at various nowcast release points and for the different proposed models. RMSFE values are normalized by that of the historic unconditional mean benchmark model, with values below 1 indicating superior predictive accuracy. Our results indicate that, on average, incorporating quarterly, monthly, and/or weekly predictors is useful to better nowcast state-level energy consumption growth. We also observe that the RMSFE typically decreases as additional information is incorporated into the models, thus supporting the nowcast monotonicity evidenced in other studies \citep[see, ][]{Giannone2008,Fosten_Nandi}. For all competing models, the relative RMSFE is below 1.

Monthly electricity sales stands out as the most significant individual predictor for energy consumption growth. When analyzing the predictive performance of each model using only one predictor at a time, we observe that the RMSFE for the AR-M model consistently falls below that of the AR-Q and AR-W models. At the beginning of the third month of the calendar, when the first data on current-year electricity sales are released, the AR-M model shows a 20\% improvement in predictive accuracy relative to the historical benchmark. Throughout the year, as additional data become available and are incorporated into the information set, the nowcast accuracy of the AR-M model further improves, reaching a peak of approximately 35\% gains with respect to the benchmark.

Combining predictors with different sampling frequencies is beneficial for nowcasting in this application. Overall, models that include both the WECI and the monthly electricity sales simultaneously exhibit the best predictive performance. At all weeks of the calendar, the AR-W-M-Q model reports the smallest RMSFE, surpassing the AR-M model. The disparity between these two models increases over the year as more quarterly and weekly data are incorporated. By year-end, the difference in performance peaks at approximately 10\%.

There is substantial variation in the performance of the nowcasting exercise across states. Table \ref{tab_dist_RMSE_AR_W_M_Q} detail the quantiles of the RMSFE distribution for the AR-W-M-Q model. Similar Tables for the other models are presented in Appendix \ref{appendix_1}. As observed, almost all quantiles of the relative RMSFE fall below 1, demonstrating improvements over the benchmark model in nearly every state. In some states at the left tail of the distribution, the improvement exceeds 60\%. Additionally, a consistent decrease in all quantiles is observed as more information becomes available throughout the year.

\begin{table}[!htbp]
\caption{Distribution of relative RMSFE across states for AR-W-M-Q model}
\label{tab_dist_RMSE_AR_W_M_Q}
\centering
\begin{adjustbox}{width=0.7\textwidth}
\begin{threeparttable}
\begin{tabular}{llccccccc}
\hline \hline
\multicolumn{2}{c}{\textbf{Calendar ($v$})} & \textbf{10\%} & \textbf{25\%} & \textbf{50\%} & \textbf{75\%} & \textbf{90\%} & \textbf{RMSE} \\ \hline
Backcast          & 2021:W1           & 0.379 & 0.456  & 0.605 & 0.759  & 0.955 & 0.626 \\
                  & 2021:W2           & 0.369 & 0.452  & 0.595 & 0.758  & 0.950 & 0.624 \\
                  & 2021:W3           & 0.370 & 0.452  & 0.588 & 0.757  & 0.945 & 0.623 \\
                  & 2021:W4           & 0.378 & 0.452  & 0.584 & 0.750  & 0.938 & 0.623 \\ \hline
Nowcast           & 2021:W5           & 0.848 & 0.904  & 0.965 & 1.042  & 1.157 & 0.972 \\
                  & 2021:W6           & 0.849 & 0.904  & 0.969 & 1.044  & 1.152 & 0.971 \\
                  & 2021:W7           & 0.845 & 0.901  & 0.971 & 1.047  & 1.145 & 0.970 \\
                  & 2021:W8           & 0.842 & 0.900  & 0.971 & 1.050  & 1.139 & 0.968 \\
                  & 2021:W9           & 0.724 & 0.790  & 0.880 & 0.962  & 1.057 & 0.879 \\
                  & 2021:W10          & 0.716 & 0.787  & 0.877 & 0.960  & 1.055 & 0.876 \\
                  & 2021:W11          & 0.708 & 0.787  & 0.873 & 0.960  & 1.053 & 0.873 \\
                  & 2021:W12          & 0.699 & 0.785  & 0.870 & 0.959  & 1.050 & 0.870 \\
                  & 2021:W13          & 0.637 & 0.696  & 0.848 & 0.930  & 1.030 & 0.824 \\
                  & 2021:W14          & 0.638 & 0.691  & 0.838 & 0.932  & 1.023 & 0.821 \\
                  & 2021:W15          & 0.638 & 0.687  & 0.827 & 0.935  & 1.018 & 0.817 \\
                  & 2021:W16          & 0.630 & 0.684  & 0.819 & 0.929  & 1.012 & 0.814 \\
                  & 2021:W17          & 0.562 & 0.691  & 0.760 & 0.892  & 0.986 & 0.769 \\
                  & 2021:W18          & 0.561 & 0.689  & 0.758 & 0.891  & 0.989 & 0.766 \\
                  & 2021:W19          & 0.560 & 0.687  & 0.757 & 0.891  & 0.990 & 0.765 \\
                  & 2021:W20          & 0.561 & 0.685  & 0.757 & 0.890  & 0.987 & 0.764 \\
                  & 2021:W21          & 0.512 & 0.620  & 0.739 & 0.859  & 0.940 & 0.723 \\
                  & 2021:W22          & 0.505 & 0.617  & 0.734 & 0.862  & 0.941 & 0.721 \\
                  & 2021:W23          & 0.505 & 0.613  & 0.729 & 0.855  & 0.941 & 0.718 \\
                  & 2021:W24          & 0.502 & 0.609  & 0.717 & 0.854  & 0.945 & 0.715 \\
                  & 2021:W25          & 0.528 & 0.624  & 0.734 & 0.838  & 0.935 & 0.722 \\
                  & 2021:W26          & 0.525 & 0.622  & 0.725 & 0.838  & 0.941 & 0.718 \\
                  & 2021:W27          & 0.519 & 0.620  & 0.718 & 0.840  & 0.939 & 0.715 \\
                  & 2021:W28          & 0.513 & 0.616  & 0.709 & 0.842  & 0.946 & 0.712 \\
                  & 2021:W29          & 0.490 & 0.585  & 0.716 & 0.816  & 0.975 & 0.699 \\
                  & 2021:W30          & 0.482 & 0.583  & 0.708 & 0.813  & 0.981 & 0.696 \\
                  & 2021:W31          & 0.475 & 0.581  & 0.704 & 0.808  & 0.982 & 0.694 \\
                  & 2021:W32          & 0.480 & 0.574  & 0.702 & 0.805  & 0.989 & 0.692 \\
                  & 2021:W33          & 0.493 & 0.549  & 0.688 & 0.828  & 0.947 & 0.680 \\
                  & 2021:W34          & 0.493 & 0.548  & 0.687 & 0.826  & 0.948 & 0.679 \\
                  & 2021:W35          & 0.490 & 0.551  & 0.681 & 0.821  & 0.948 & 0.678 \\
                  & 2021:W36          & 0.487 & 0.553  & 0.675 & 0.817  & 0.949 & 0.677 \\
                  & 2021:W37          & 0.465 & 0.543  & 0.673 & 0.798  & 0.967 & 0.671 \\
                  & 2021:W38          & 0.464 & 0.541  & 0.669 & 0.793  & 0.963 & 0.670 \\
                  & 2021:W39          & 0.467 & 0.538  & 0.663 & 0.795  & 0.975 & 0.667 \\
                  & 2021:W40          & 0.466 & 0.533  & 0.659 & 0.797  & 0.977 & 0.665 \\
                  & 2021:W41          & 0.413 & 0.522  & 0.660 & 0.827  & 0.966 & 0.659 \\
                  & 2021:W42          & 0.409 & 0.519  & 0.659 & 0.825  & 0.961 & 0.658 \\
                  & 2021:W43          & 0.408 & 0.520  & 0.659 & 0.822  & 0.958 & 0.659 \\
                  & 2021:W44          & 0.408 & 0.519  & 0.659 & 0.819  & 0.970 & 0.659 \\
                  & 2021:W45          & 0.384 & 0.491  & 0.615 & 0.804  & 1.014 & 0.636 \\
                  & 2021:W46          & 0.382 & 0.491  & 0.620 & 0.806  & 1.023 & 0.638 \\
                  & 2021:W47          & 0.382 & 0.486  & 0.620 & 0.806  & 1.026 & 0.637 \\
                  & 2021:W48          & 0.383 & 0.483  & 0.619 & 0.806  & 1.031 & 0.637 \\ \hline \hline
\end{tabular}
\begin{tablenotes}
\item \textit{Note: The dependent variable is the state-level energy consumption per capita growth. RMSFE relative to a historical mean benchmark.}
\end{tablenotes} 
\end{threeparttable}
\end{adjustbox}
\end{table}

This exercise demonstrates that the use of timely information is valuable for a better nowcasting of state-level energy consumption. Compared to the study by \cite{Fosten_Nandi}, our analysis reveals better nowcasting accuracy, possibly due to recent data revisions and the strategic combination of indicators sampled at different frequencies.

\subsection{Density nowcast of CO2 emissions growth using a bridge equation}

Using the timely predictions of per-capita energy consumption growth for the target year, we now generate density nowcasts of per-capita CO2 emissions growth through a bridge equation. The quantile levels, set at $\tau_1 = 0.25$, $\tau_2 = 0.50$, $\tau_3 = 0.75$, are chosen to capture predictions for the central part and the tails of the distribution. These density nowcasts are compared to the historical quantile benchmark for each week of the calendar year.

\begin{figure}[h]

    \caption{Quantile accuracy measures for CO2 per capita growth}
    
    \includegraphics[width=15 cm,keepaspectratio]{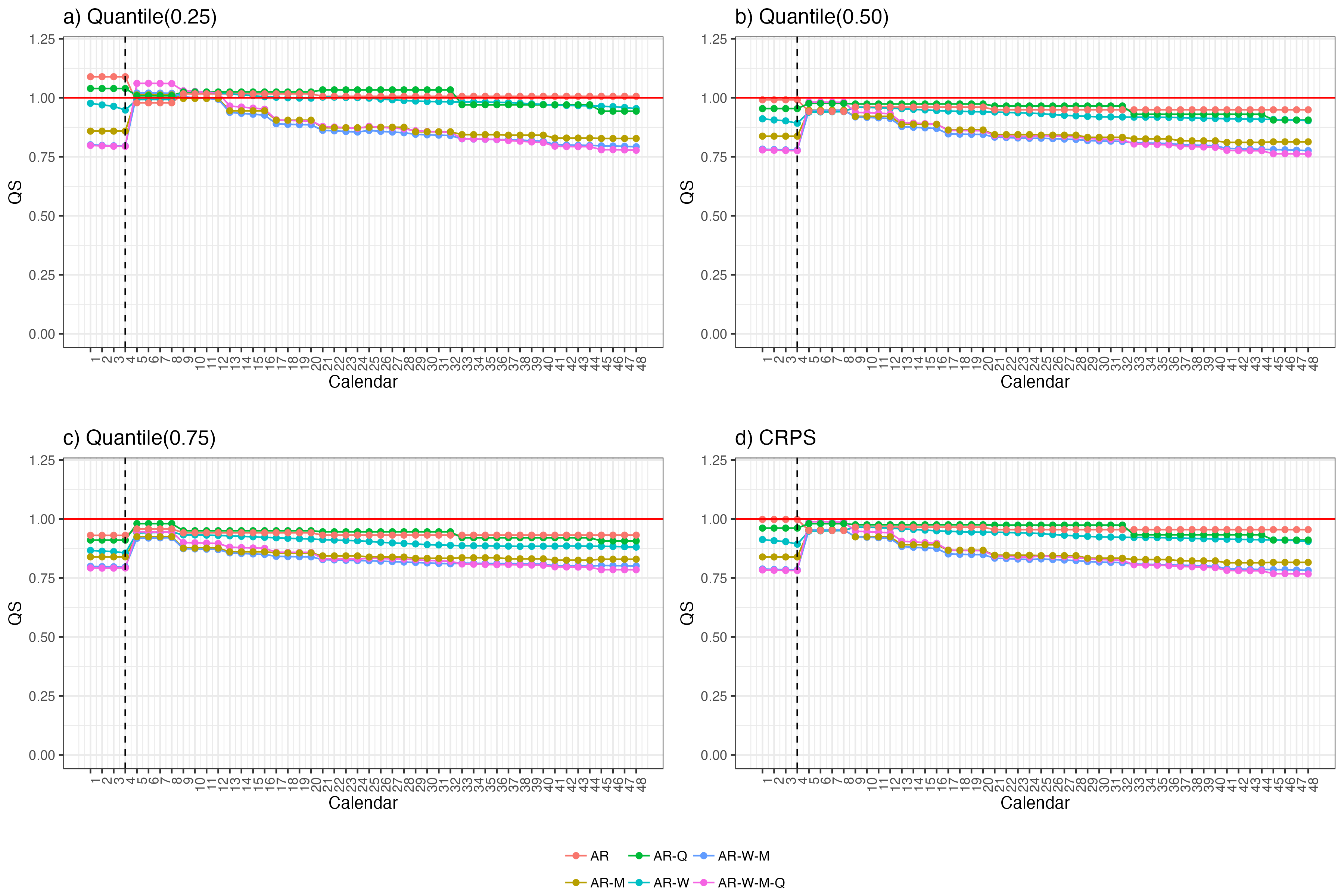}

      \textit{Notes: CO2 per capita growth nowcasts based on the AR-W-M-Q model for EC growth, which incorporates an autoregressive (AR) component, the weekly economic condition index (W), the monthly electricity sales (M), and the quarterly PI data (Q). The benchmark model is a historic mean for CO2 per capita growth. The benchmark normalizes the figures at the first release date. Consequently, any points below 1 indicate that the respective accuracy measures is lower than that of the benchmark.} 
    \label{fig_QuantEval_monthly}
\end{figure}

Figure \ref{fig_QuantEval_monthly} presents quantile scores and the CRPS measuring the predictive accuracy of each proposed model throughout the calendar year. In all models and weeks, the CRPS values are below 1, indicating that our density quantile approach provides an improvement over the benchmark. The gains in predictive accuracy observed in energy consumption growth are translated into the nowcasting of CO2 emissions. The best-performing models are AR-M, AR-W-M, and AR-W-M-Q, with the latter showing marginally superior performance due to the simultaneous inclusion of all predictors. Moreover, notice that between weeks 4 and 12, when limited information on current-year predictors is available, the CRSP are close to 1. As more data becomes available throughout the year, the precision of the nowcasts increases monotonically until reaching a maximum gain of around 25\% at the end of the year in our best model. Nowcasts show better performance at the median and the 0.75 quantile, while they are less effective at the 0.25 quantile, particularly in the early part of the calendar year. As more information is incorporated, the gains in predictive accuracy become similar across the distribution. 

Table \ref{tab_distribution_CRPS_ARWMQ} complements our analysis by presenting the distribution of relative CRPS across states for the AR-W-M-Q model. Appendix \ref{appendix_2} provides detailed tables reporting the distribution of the QS for each individual quantile level. Results indicate significant improvements in predictive accuracy over the historical quantile benchmark across nearly all states and quantile levels.

\begin{table}[ht!]
\caption{Distribution of relative CRSP across states for AR-W-M-Q model}
\label{tab_distribution_CRPS_ARWMQ}
\centering
\begin{adjustbox}{width=0.7\textwidth}
\begin{threeparttable}
\begin{tabular}{llccccccc}
\hline \hline
\multicolumn{2}{c}{\textbf{Calendar ($v$})} & \textbf{10\%} & \textbf{25\%} & \textbf{50\%} & \textbf{75\%} & \textbf{90\%} & \textbf{CRSP}\\ \hline
Backcast          & 2021:W1           & 0.628 & 0.662  & 0.754 & 0.870  & 0.966 & 0.779 \\
                  & 2021:W2           & 0.625 & 0.659  & 0.756 & 0.868  & 0.963 & 0.778 \\
                  & 2021:W3           & 0.621 & 0.660  & 0.756 & 0.868  & 0.962 & 0.777 \\
                  & 2021:W4           & 0.620 & 0.663  & 0.756 & 0.868  & 0.960 & 0.776 \\ \hline
Nowcast           & 2021:W5           & 0.785 & 0.874  & 0.988 & 1.051  & 1.189 & 0.982 \\
                  & 2021:W6           & 0.783 & 0.876  & 0.986 & 1.052  & 1.189 & 0.982 \\
                  & 2021:W7           & 0.781 & 0.876  & 0.983 & 1.053  & 1.188 & 0.982 \\
                  & 2021:W8           & 0.778 & 0.878  & 0.982 & 1.054  & 1.186 & 0.981 \\
                  & 2021:W9           & 0.727 & 0.855  & 0.939 & 1.008  & 1.143 & 0.939 \\
                  & 2021:W10          & 0.723 & 0.856  & 0.938 & 1.001  & 1.144 & 0.937 \\
                  & 2021:W11          & 0.717 & 0.856  & 0.936 & 0.996  & 1.145 & 0.935 \\
                  & 2021:W12          & 0.714 & 0.852  & 0.933 & 0.994  & 1.144 & 0.933 \\
                  & 2021:W13          & 0.705 & 0.833  & 0.893 & 0.967  & 1.071 & 0.896 \\
                  & 2021:W14          & 0.702 & 0.831  & 0.887 & 0.964  & 1.071 & 0.893 \\
                  & 2021:W15          & 0.698 & 0.831  & 0.883 & 0.963  & 1.070 & 0.890 \\
                  & 2021:W16          & 0.693 & 0.832  & 0.878 & 0.962  & 1.069 & 0.888 \\
                  & 2021:W17          & 0.684 & 0.770  & 0.840 & 0.958  & 1.043 & 0.862 \\
                  & 2021:W18          & 0.681 & 0.772  & 0.841 & 0.957  & 1.045 & 0.861 \\
                  & 2021:W19          & 0.680 & 0.776  & 0.840 & 0.955  & 1.047 & 0.861 \\
                  & 2021:W20          & 0.680 & 0.777  & 0.839 & 0.958  & 1.050 & 0.861 \\
                  & 2021:W21          & 0.676 & 0.731  & 0.834 & 0.930  & 1.015 & 0.840 \\
                  & 2021:W22          & 0.669 & 0.730  & 0.836 & 0.927  & 1.016 & 0.838 \\
                  & 2021:W23          & 0.664 & 0.731  & 0.834 & 0.924  & 1.009 & 0.837 \\
                  & 2021:W24          & 0.661 & 0.727  & 0.831 & 0.924  & 1.016 & 0.836 \\
                  & 2021:W25          & 0.667 & 0.728  & 0.832 & 0.928  & 1.018 & 0.840 \\
                  & 2021:W26          & 0.663 & 0.725  & 0.829 & 0.928  & 1.018 & 0.838 \\
                  & 2021:W27          & 0.659 & 0.717  & 0.830 & 0.925  & 1.019 & 0.836 \\
                  & 2021:W28          & 0.656 & 0.710  & 0.829 & 0.919  & 1.018 & 0.835 \\
                  & 2021:W29          & 0.652 & 0.723  & 0.790 & 0.923  & 1.020 & 0.828 \\
                  & 2021:W30          & 0.650 & 0.718  & 0.786 & 0.920  & 1.026 & 0.826 \\
                  & 2021:W31          & 0.648 & 0.716  & 0.780 & 0.918  & 1.034 & 0.824 \\
                  & 2021:W32          & 0.645 & 0.714  & 0.776 & 0.913  & 1.039 & 0.823 \\
                  & 2021:W33          & 0.606 & 0.675  & 0.773 & 0.908  & 1.023 & 0.804 \\
                  & 2021:W34          & 0.612 & 0.678  & 0.770 & 0.905  & 1.024 & 0.803 \\
                  & 2021:W35          & 0.619 & 0.677  & 0.766 & 0.901  & 1.024 & 0.802 \\
                  & 2021:W36          & 0.621 & 0.674  & 0.765 & 0.896  & 1.024 & 0.801 \\
                  & 2021:W37          & 0.613 & 0.671  & 0.776 & 0.859  & 1.011 & 0.795 \\
                  & 2021:W38          & 0.610 & 0.664  & 0.776 & 0.856  & 1.013 & 0.793 \\
                  & 2021:W39          & 0.607 & 0.661  & 0.777 & 0.856  & 1.014 & 0.792 \\
                  & 2021:W40          & 0.605 & 0.662  & 0.776 & 0.857  & 1.016 & 0.790 \\
                  & 2021:W41          & 0.597 & 0.670  & 0.769 & 0.851  & 1.005 & 0.777 \\
                  & 2021:W42          & 0.595 & 0.672  & 0.767 & 0.849  & 1.002 & 0.776 \\
                  & 2021:W43          & 0.595 & 0.668  & 0.767 & 0.846  & 1.003 & 0.776 \\
                  & 2021:W44          & 0.595 & 0.667  & 0.766 & 0.843  & 1.002 & 0.776 \\
                  & 2021:W45          & 0.567 & 0.663  & 0.751 & 0.812  & 0.972 & 0.763 \\
                  & 2021:W46          & 0.568 & 0.660  & 0.749 & 0.812  & 0.973 & 0.763 \\
                  & 2021:W47          & 0.567 & 0.657  & 0.747 & 0.812  & 0.973 & 0.763 \\
                  & 2021:W48          & 0.566 & 0.653  & 0.746 & 0.811  & 0.972 & 0.762  \\ \hline \hline
\end{tabular}
\begin{tablenotes}
\item \textit{Note: The dependent variable is the state-level per-capita CO2 emissions growth.}
\end{tablenotes} 
\end{threeparttable}
\end{adjustbox}
\end{table}

\subsection{Direct density nowcast of CO2 emissions growth}

\begin{figure}[!htbp]
    \caption{Direct vs bridge approaches}
    
    \includegraphics[width=15 cm,keepaspectratio]{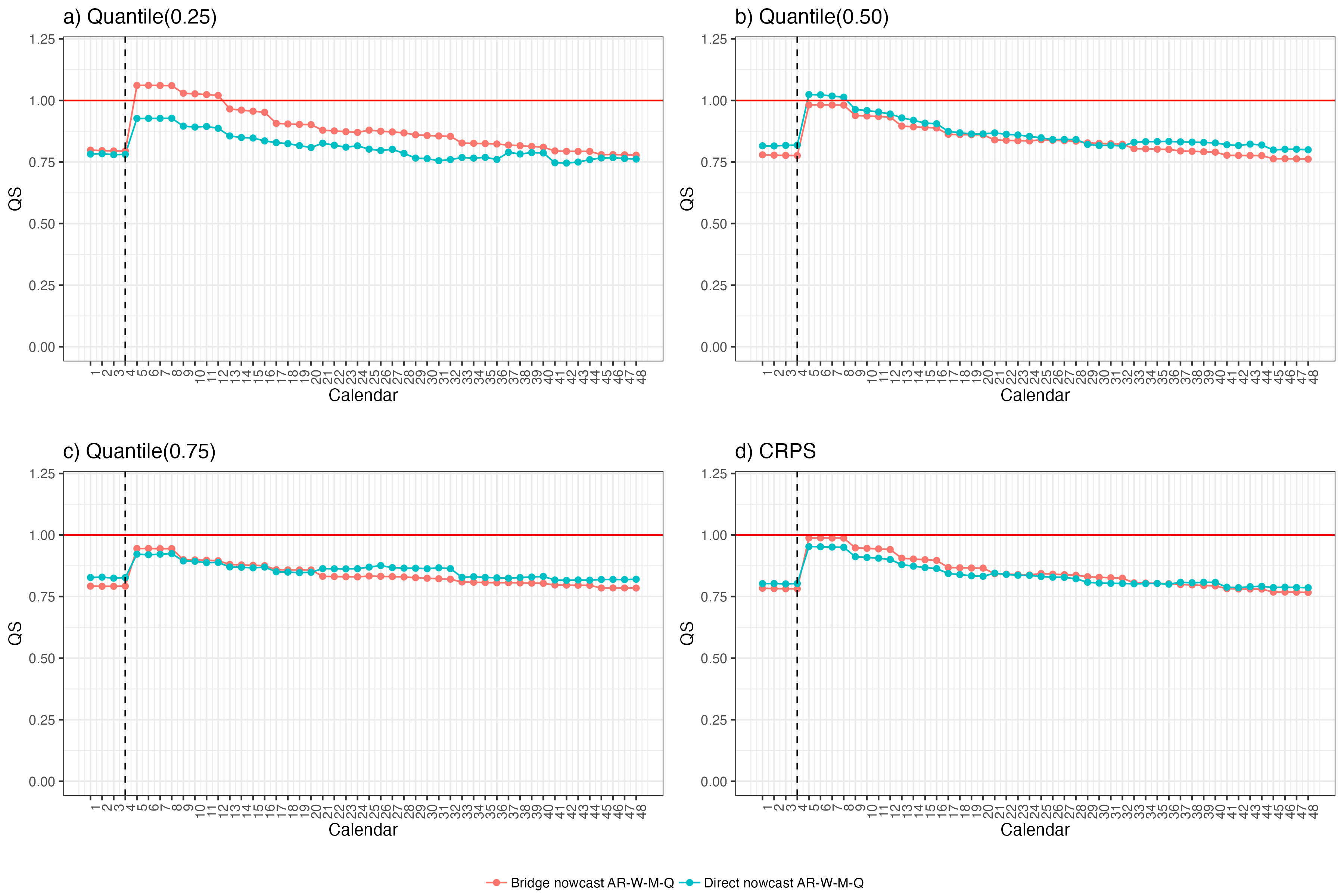}
    
\textit{Notes: CO2 per capita growth nowcasts based on the AR-W-M-Q model for EC growth, which incorporates an autoregressive (AR) component, the weekly economic condition index (W), the monthly electricity sales (M), and the quarterly PI data (Q). The direct model incorporates the W-M-Q component instead of EC per capita growth predictions, and the common correlated effect of CO2 per capita growth. The benchmark model is a historic mean for CO2 per capita growth. The benchmark normalizes the figures at the first release date. Consequently, any points below 1 indicate that the respective accuracy measures is lower than that of the benchmark.} 
    \label{fig_Bridge_vs_Direct}
\end{figure}

This subsection evaluates the nowcasts from the best-performing model that utilizes a bridge equation against those derived from a model that applies direct panel quantile regression to CO2 emissions growth data, without relying on predictions of energy consumption growth. Figure \ref{fig_Bridge_vs_Direct} presents a comparative analysis of the quantile scores and the CRPS between the AR-W-M-Q model and the direct approach. Both methods show similar performance in terms of CRPS, with the best-case scenario indicating an approximate 25\% improvement relative to the historical unconditional benchmark.  However, at the beginning of the calendar year, the direct method demonstrates slightly superior performance. The enhanced early-year performance of the direct approach is attributed to its better accuracy in nowcasting at the 0.25 quantile level. This advantage persists throughout the year, suggesting that the direct method may be more effective in capturing lower tail dynamics in the distribution of CO2 emissions growth. 

The observed difference in performance between the direct method and the AR-W-M-Q model can be attributed to the prediction errors inherent in the bridge equation employed by the latter. These errors likely influence the overall accuracy of the AR-W-M-Q model, particularly affecting its efficiency in early-year predictions and at lower quantile levels. Despite these challenges, a valid argument for continuing to use a bridge equation lies in its capacity to generate timely predictions of energy consumption growth, object of specific interest to policymakers and researchers.

\subsection{High-frequency density nowcasts for selected states}

\begin{figure}[H]
\caption{Density nowcast for per-capita CO2 emissions for California, Texas, and New York}
\centering
\begin{subfigure}{1\textwidth}
  \centering
  \includegraphics[width=.95\linewidth]{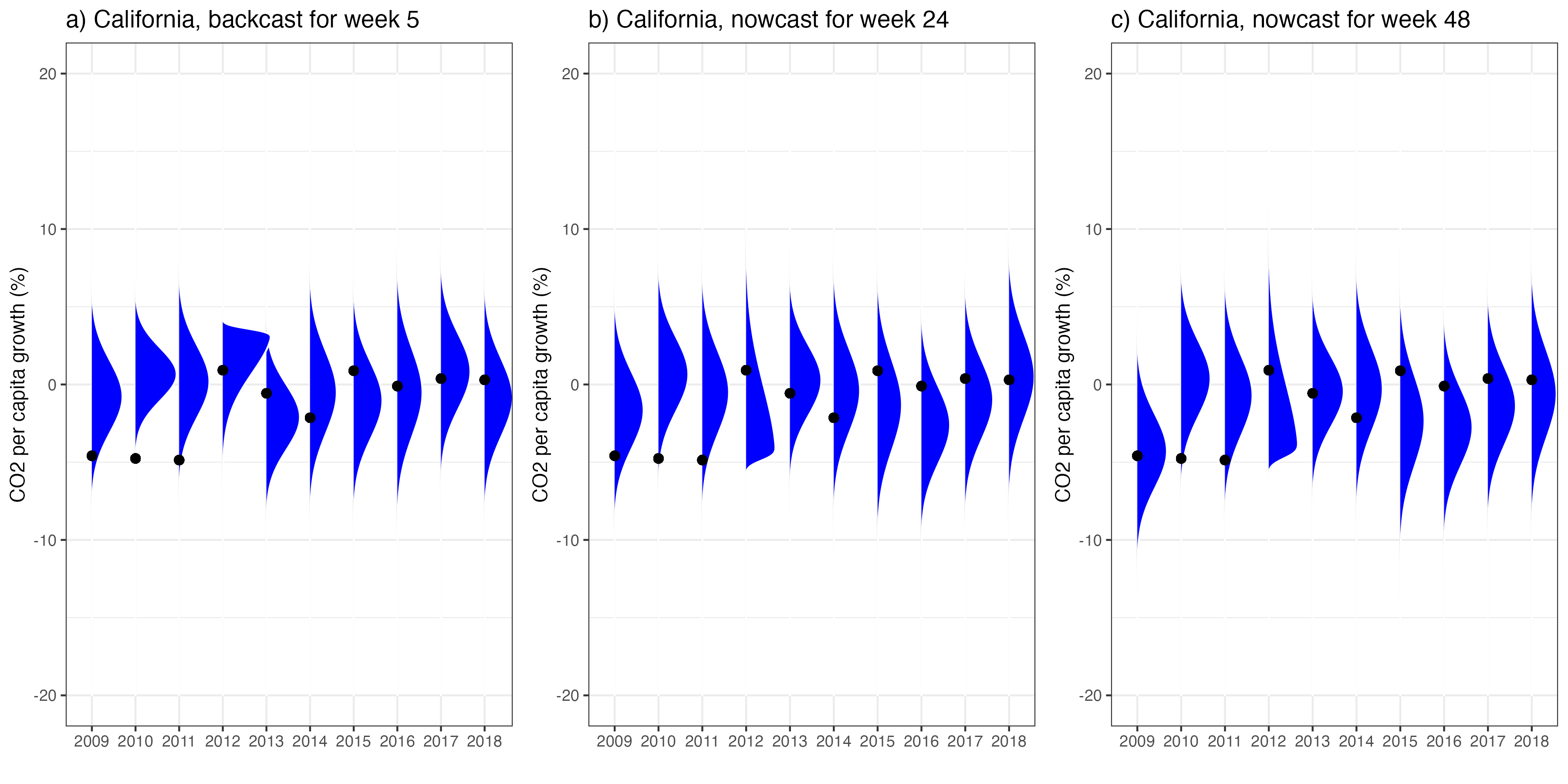}
\end{subfigure}%
  \hfill
\begin{subfigure}{1\textwidth}
  \centering
  \includegraphics[width=.95\linewidth]{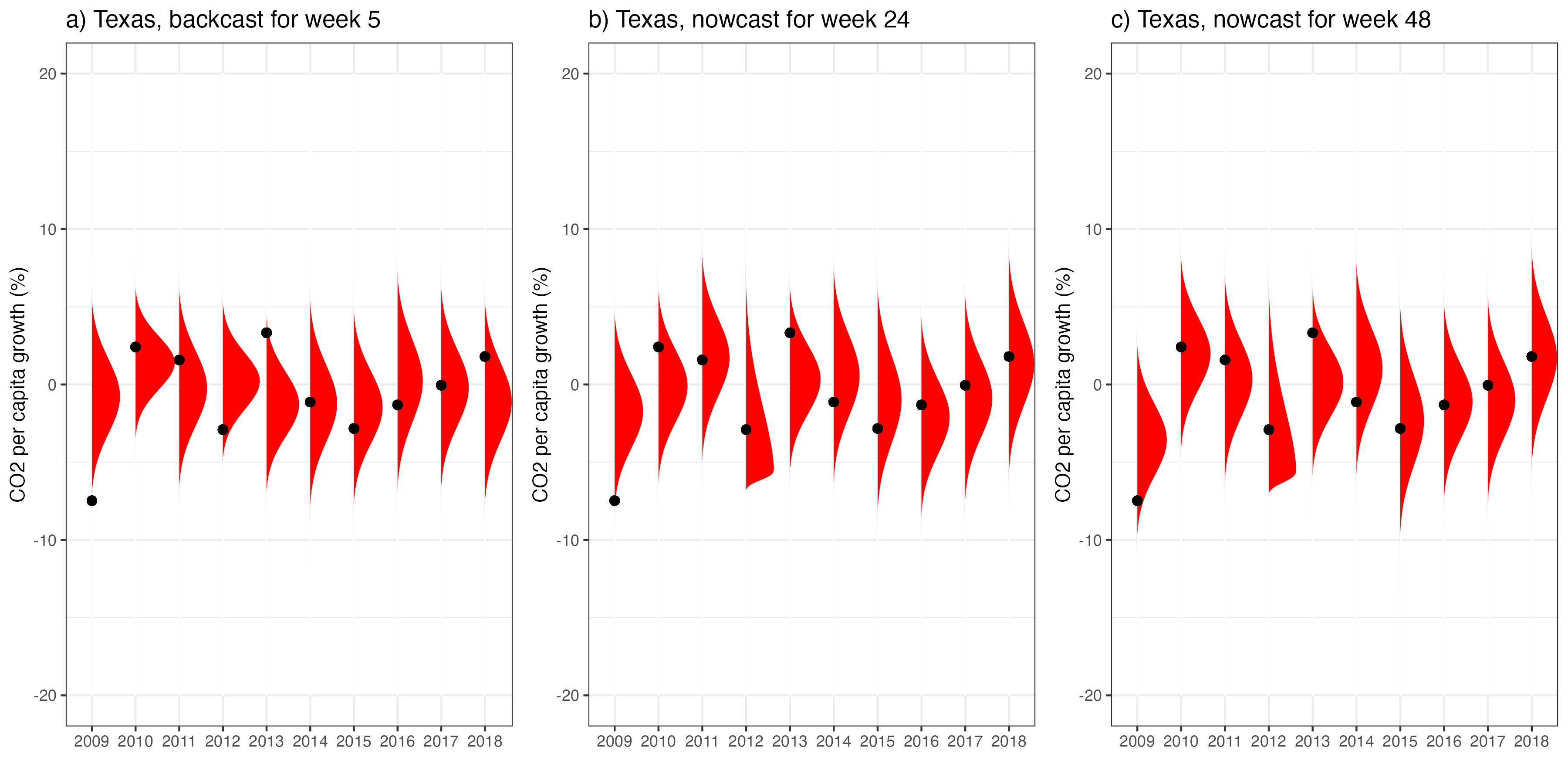}
\end{subfigure}
\begin{subfigure}{1\textwidth}
  \centering
  \includegraphics[width=.95\linewidth]{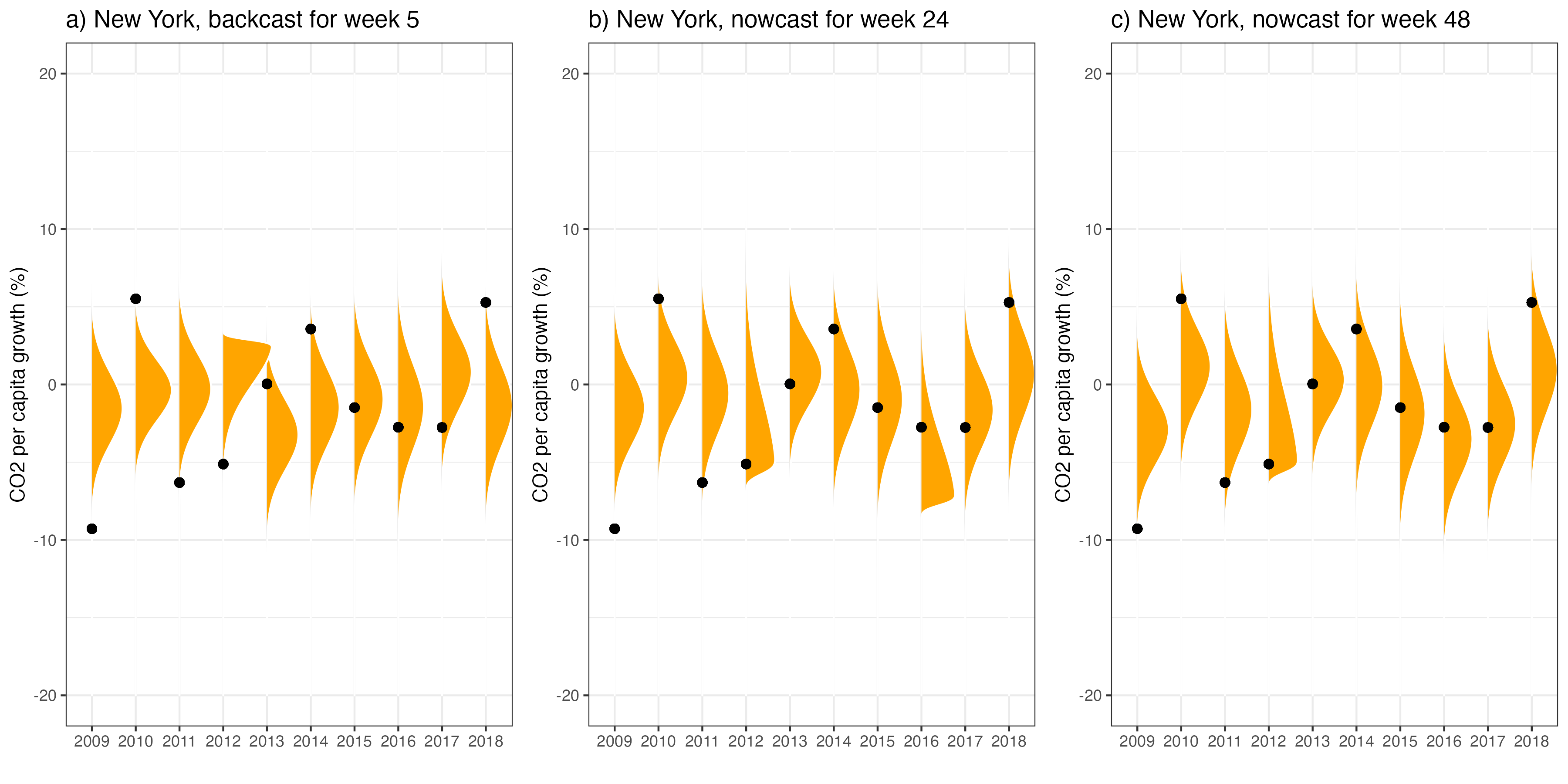}
\end{subfigure}
\label{Density_CO2_Emissions}
      \textit{Notes: Black dots are the realized value of CO2 per capita growth. The densities are obtained from the quantile predictions obtained through the AR-W-M-Q model using a bridge equation.} 
    \label{fig_nowcasts_states}
\end{figure}

This section illustrates the behavior of our timely quantile predictions of per-capita CO2 emissions growth throughout the calendar year. Figure \ref{fig_nowcasts_states} displays the predicted densities for weeks 5, 24, and 48, and compare these predictions with the realized values of the target variable. Following \cite{Adrian2019}, full continuous conditional densities are constructed by fitting a generalized skewed Student's distribution to the discrete conditional quantiles predicted at each nowcasting point. Specific weeks are chosen to examine how predictions for the current year improve as more information becomes available over the calendar. Results are presented for the AR-W-M-Q model and focus on California, Texas, and New York, states that annually report the highest CO2 emissions records.

Our analysis indicates that in most cases, the actual values of CO2 emissions growth (represented by black dots) fall within the range of the predicted densities. Improvements in predictive accuracy across the calendar are evident, particularly if we compare density predictions in week 5 versus week 24. As additional data become available, the median of the predicted density moves closer to the actual observed value, providing a visual confirmation of the decreasing CRPS highlighted earlier in this paper. The performance of our approach exhibits considerable variability across different states and years. For Texas and New York,  the model demonstrates relatively strong predictive accuracy between 2014 and 2017. In contrast, for California, the model achieves better results between 2013 and 2015.

\subsection{Nowcasting during the COVID 19 period}

\begin{figure}[H]
    \caption{COVID-19: RMSFE across states for EC per capita growth}
    \includegraphics[width=15 cm,keepaspectratio]{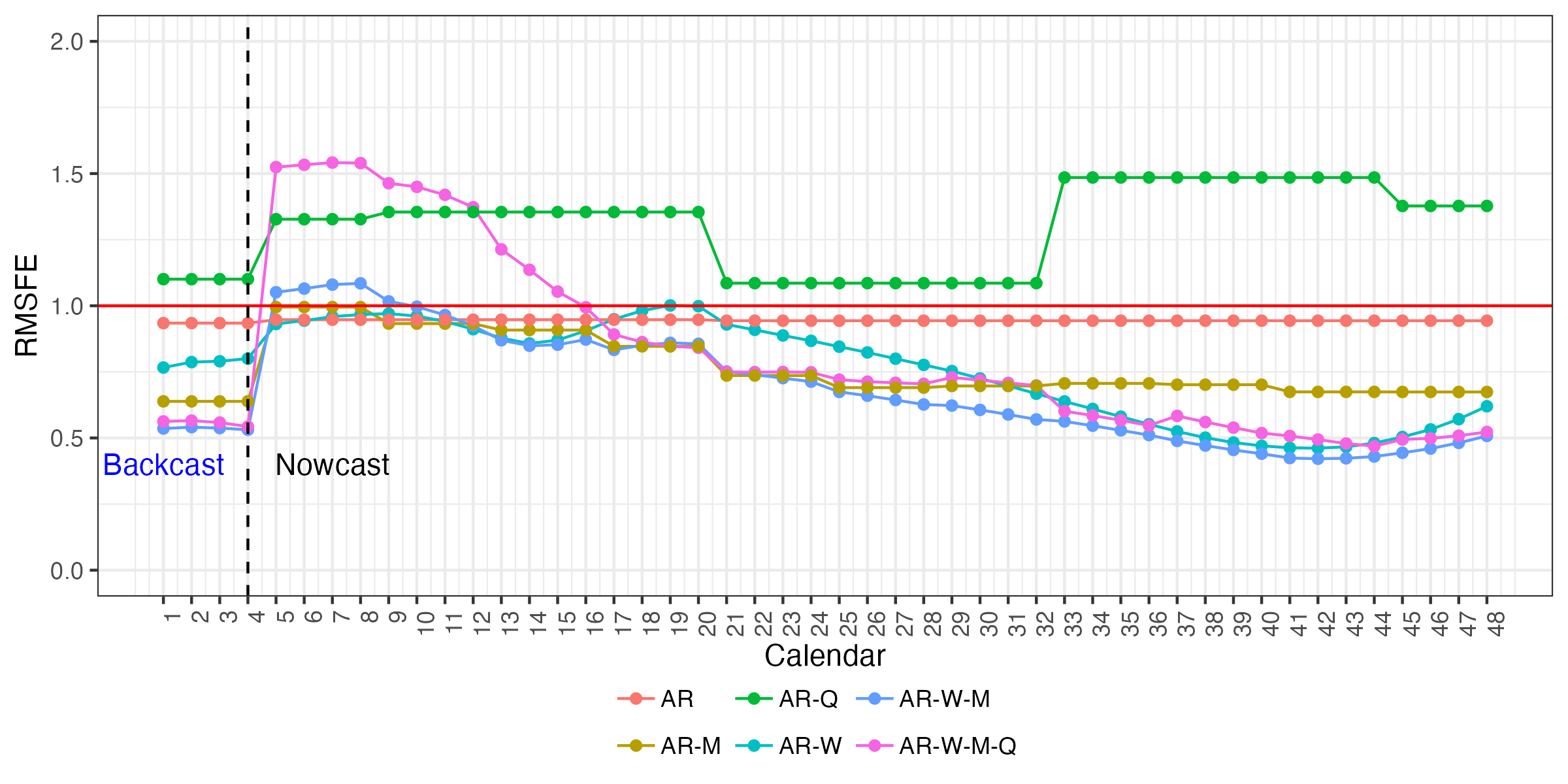}
\textit{Notes: The AR-W-M-Q model incorporates an autoregressive (AR) component, the weekly economic condition index (W), the monthly electricity sales (M), and the quarterly PI data (Q). The benchmark model is a historic mean for electricity consumption growth. The benchmark normalizes the RMSFE figures at the first release date. Consequently, any points below 1 indicate that the RMSFE is lower than that of the benchmark.} 
    \label{fig_RMSE_monthly_covid_19}
\end{figure}

\begin{figure}[H]
    \caption{COVID-19: Quantile accuracy measures for CO2 per capita growth}
    \includegraphics[width=15 cm,keepaspectratio]{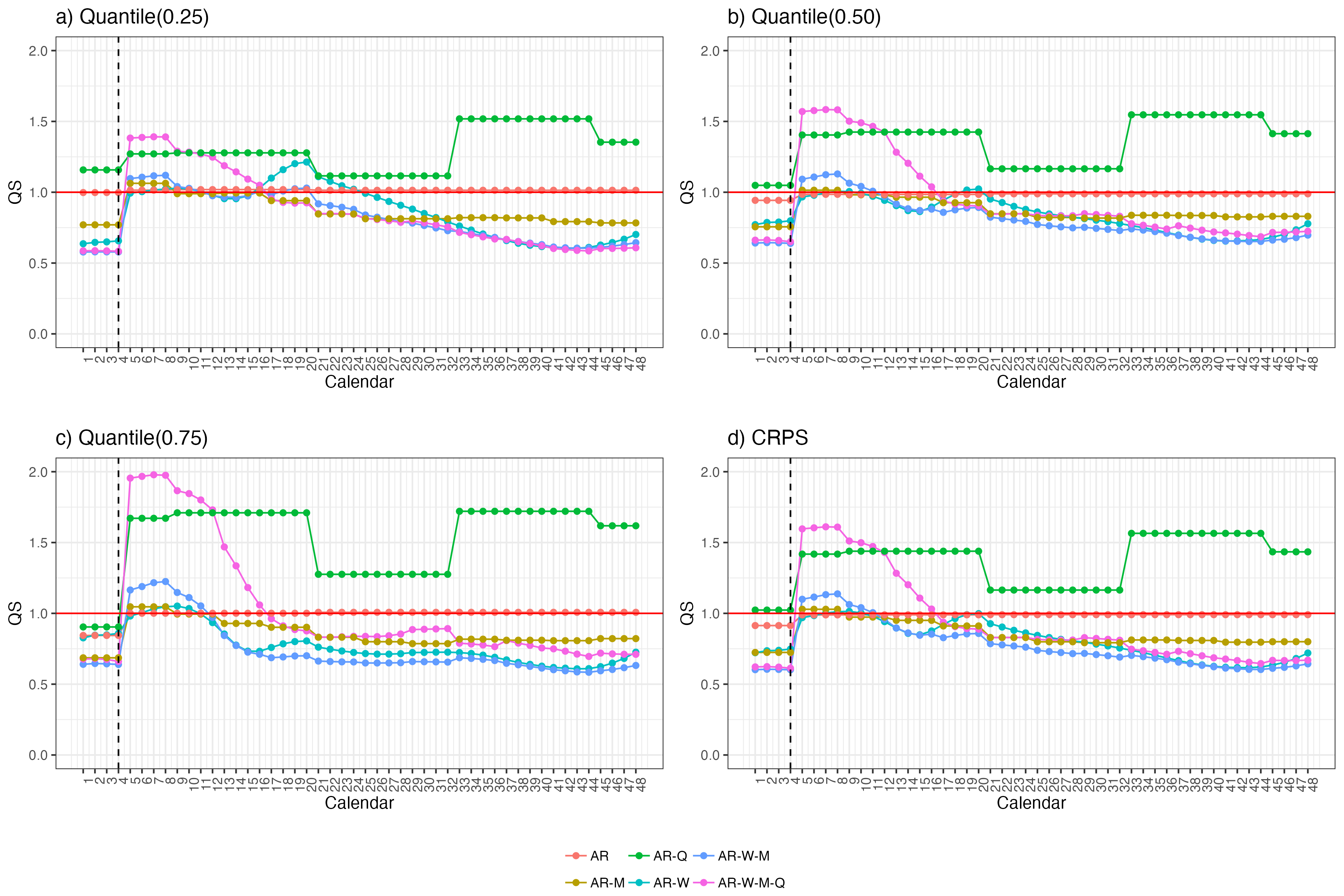}
      \textit{Notes: CO2 per capita growth nowcasts based on the AR-W-M-Q model for EC growth, which incorporates an autoregressive (AR) component, the weekly economic condition index (W), the monthly electricity sales (M), and the quarterly PI data (Q). The benchmark model is a historic mean for CO2 per capita growth. The benchmark normalizes the figures at the first release date. Consequently, any points below 1 indicate that the respective accuracy measures is lower than that of the benchmark.} 
    \label{fig_QuantEval_monthly_covid_19}
\end{figure}

\begin{figure}[H]
\caption{COVID-19: Density nowcast for per-capita CO2 emissions for California, Texas, and New York}
\centering
\begin{subfigure}{1\textwidth}
  \centering
  \includegraphics[width=.90\linewidth]{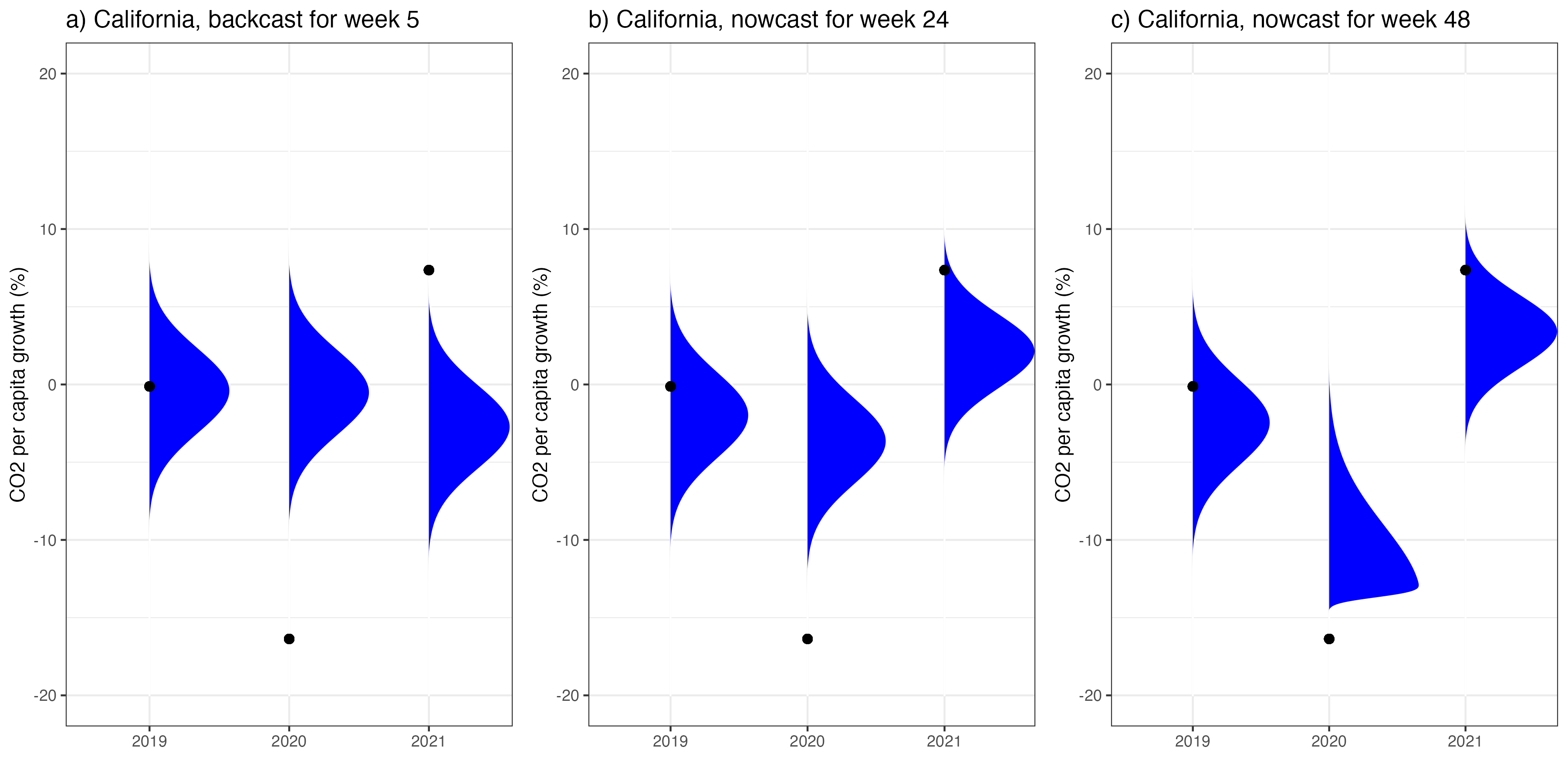}
\end{subfigure}%
  \hfill
\begin{subfigure}{1\textwidth}
  \centering
  \includegraphics[width=.90\linewidth]{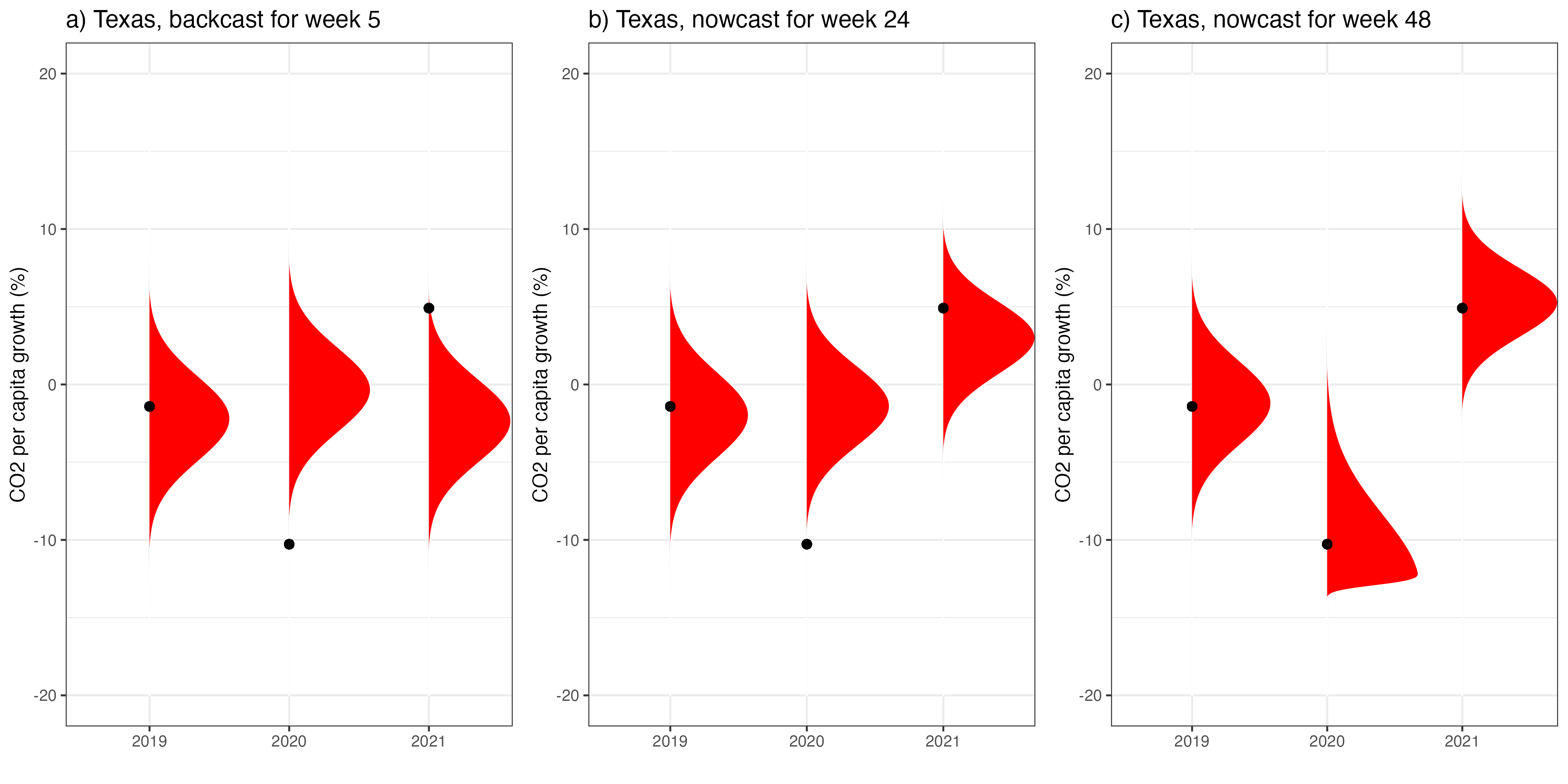}
\end{subfigure}
\begin{subfigure}{1\textwidth}
  \centering
  \includegraphics[width=.90\linewidth]{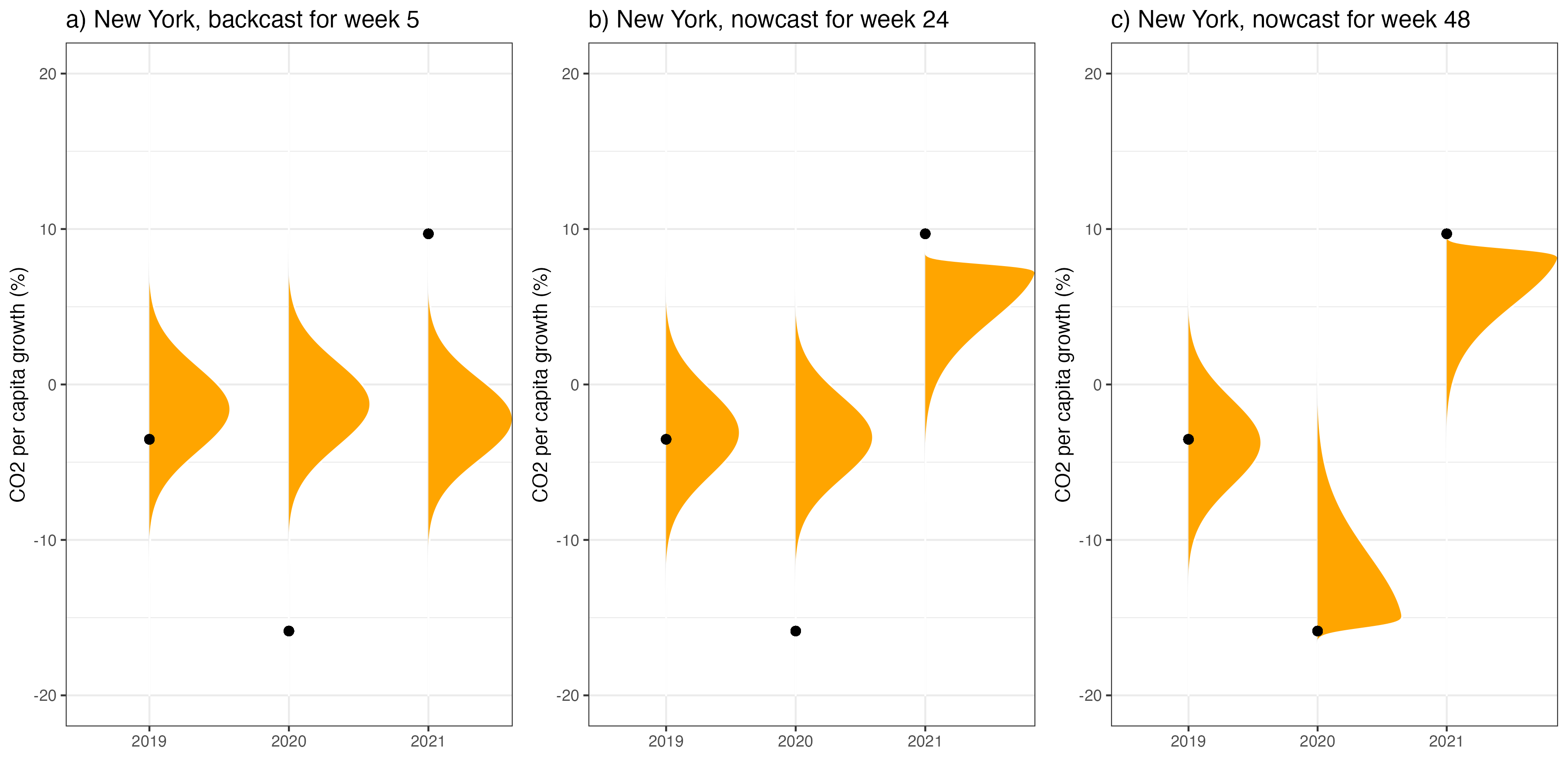}
\end{subfigure}
\label{Density_CO2_Emissions_covid}
      \textit{Notes: Black dots are the realized value of CO2 per capita growth. The densities are obtained from the quantile predictions obtained through the AR-W-M model using a bridge equation.} 
    \label{fig_nowcasts_states_COVID}
\end{figure}

We evaluate how effectively the various nowcasting approaches capture the abrupt decline in CO2 emissions during the COVID-19 period and the subsequent recovery. To this end, we extend our out-of-sample analysis to include the years 2019 through 2021 and compute the accuracy measures reported in Figures \ref{fig_RMSE_monthly_covid_19} and \ref{fig_QuantEval_monthly_covid_19}. For both target variables, incorporating quarterly PI data significantly worsens the predictions compared to the historical benchmark. In contrast, including monthly and weekly predictors yields substantial improvements, specially after week 15. The models that include the WECI achieve the lowest RMSFE and CRPS, with gains surpassing those observed in non-COVID periods. Figure \ref{fig_nowcasts_states_COVID} presents the densities constructed from the quantile predictions generated by the AR-W-M model via a bridge equation, illustrating how effectively the model nowcasts CO2 emissions across selected states as more information becomes available throughout the year. By week 48, the predicted densities are closer to the observed values.

\section{Conclusions}

In this paper, we introduced a panel nowcasting methodology to obtain high-frequency nowcasts of state-level per-capita energy consumption and CO2 emissions growth in the U.S. Our methodology extends the approach of \cite{Fosten_Nandi} by incorporating the state-level weekly economic conditions index of \cite{Baumeister}, and by utilizing panel quantile regressions to produce density nowcasts of per-capita CO2 emissions growth. This enhancement enables us to report not only the expected trajectory of per-capita CO2 emissions growth but also the uncertainty surrounding this central path. In our out-of-sample exercise, covering the period from 2009 to 2018, we found that the model incorporating economic predictors at mixed frequencies consistently outperformed other models. Specifically, the most effective model includes data at weekly, monthly, and quarterly intervals, and applies a restricted Almon lag polynomial to approximate the high-frequency weekly component. The inclusion of weekly data is particularly useful to better nowcast during the COVID-19 period. 

The application of our nowcasting methodology to the domains of energy consumption and CO2 emissions is highly pertinent, particularly due to the significant delays in the publication of official data. The publication delay for CO2 emissions data extends to approximately two years and three months after the end of the reference year, while the delay for energy consumption data is around 18 months. Our methodology leverages the more prompt availability of economic data to provide early insights to policymakers on the evolution of critical environmental variables. As the year progresses and more data becomes available, the accuracy of our predictions improves. This approach enables a timely and precise tracking of anthropogenic CO2 emissions at both national and sub-national levels, which is crucial for the development of effective climate policies and for meeting long-term international commitments to combat climate change.

\clearpage
\begin{spacing}{0.8}
\bibliographystyle{chicago}

\end{spacing}

\newpage

\appendix
\setcounter{table}{0}
\setcounter{figure}{0}
\section{Appendix}
\renewcommand{\thetable}{A\arabic{table}} 
\renewcommand{\thefigure}{A\arabic{figure}} 

\subsection{Results of the nowcasting exercise of the growth rate of energy consumption}\label{appendix_3}

The baseline analysis of this document addresses the nowcasting of the \textbf{per-capita} growth rate of energy consumption and CO2 emissions. In this Appendix, we present the results using the original quantities of these variables without normalization by population. The findings for the growth rate of energy consumption align qualitatively with those of the baseline analysis. Specifically, models that incorporate both weekly and monthly indicators simultaneously exhibit the best predictive performance. Quantitatively, these models demonstrate even stronger gains in predictive accuracy compared to the per-capita analysis, achieving an average improvement of around 40\% with respect to the historical mean benchmark by the end of the calendar year with the AR-W-M-Q model. These gains are translated to the nowcast of the growth rate of CO2 emissions. These results are omitted from this appendix due to space limitations but can be shared on request.

\begin{figure}[ht!]
    \caption{RMSFE across states for EC growth}
    \includegraphics[width=15 cm,keepaspectratio]{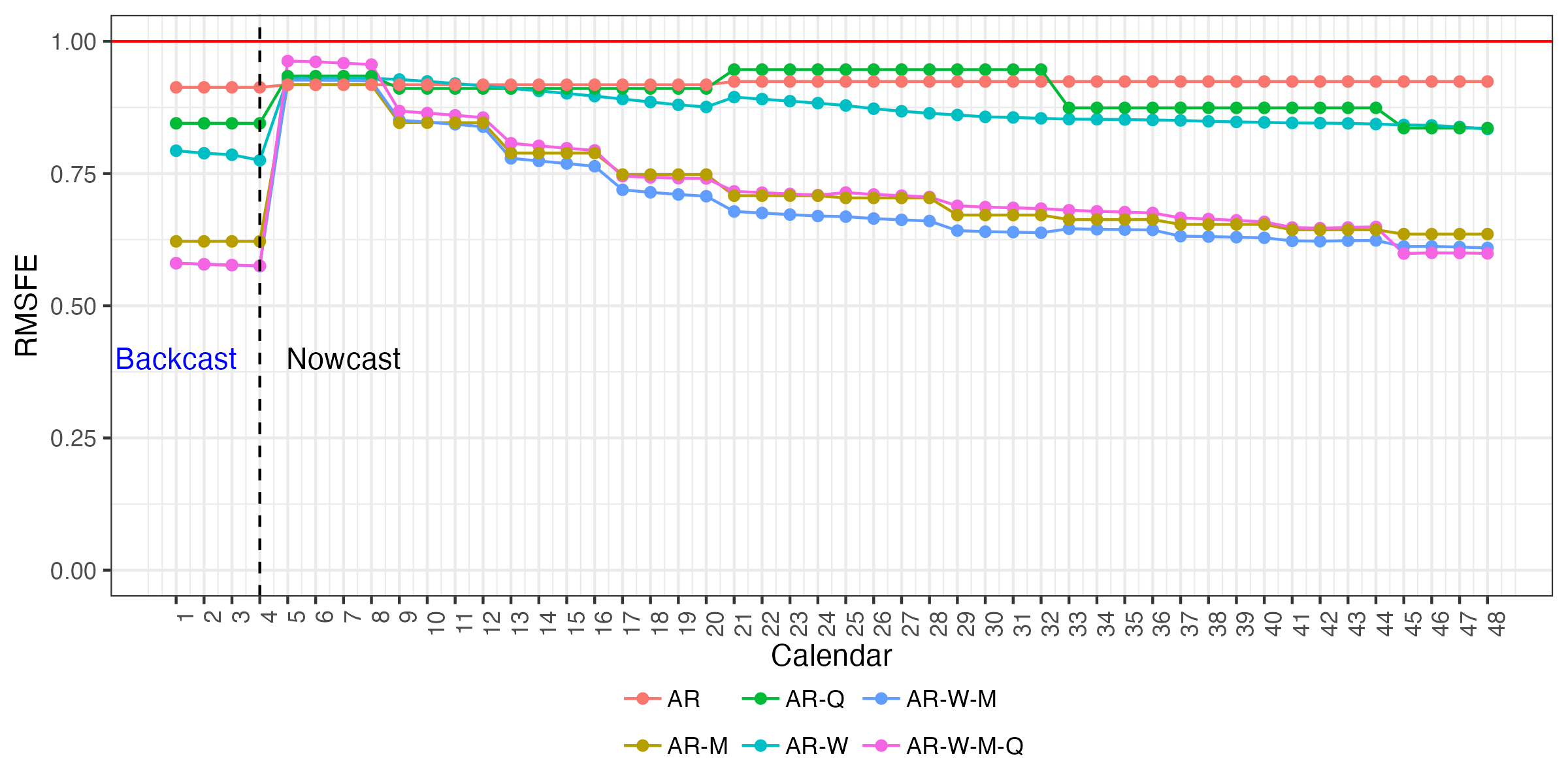}
    
    \textit{Notes: The AR-W-M-Q model incorporates an autoregressive (AR) component, the weekly economic condition index (W), the monthly electricity sales (M), and the quarterly PI data (Q). The benchmark model is a historic mean for electricity consumption growth. The benchmark normalizes the RMSFE figures at the first release date. Consequently, any points below 1 indicate that the RMSFE is lower than that of the benchmark.} 
\end{figure}

\begin{table}[!htbp]
\caption{Distribution of relative RMSFE across states for AR-W model}
\centering
\begin{adjustbox}{width=0.5\textwidth}
\begin{threeparttable}
\begin{tabular}{llccccccc}
\hline \hline
\multicolumn{2}{c}{\textbf{Calendar ($v$})} & \textbf{10\%} & \textbf{25\%} & \textbf{50\%} & \textbf{75\%} & \textbf{90\%} & \textbf{RMSE} \\ \hline
Backcast          & 2021:W1           & 0.624 & 0.705  & 0.790 & 0.874  & 0.955 & 0.793 \\
                  & 2021:W2           & 0.613 & 0.691  & 0.785 & 0.868  & 0.961 & 0.788 \\
                  & 2021:W3           & 0.612 & 0.682  & 0.771 & 0.866  & 0.955 & 0.786 \\
                  & 2021:W4           & 0.610 & 0.670  & 0.761 & 0.847  & 0.971 & 0.775 \\ \hline
Nowcast           & 2021:W5           & 0.803 & 0.870  & 0.949 & 1.011  & 1.040 & 0.930 \\
                  & 2021:W6           & 0.804 & 0.874  & 0.951 & 1.009  & 1.042 & 0.930 \\
                  & 2021:W7           & 0.804 & 0.876  & 0.951 & 1.005  & 1.045 & 0.931 \\
                  & 2021:W8           & 0.802 & 0.877  & 0.955 & 1.006  & 1.047 & 0.930 \\
                  & 2021:W9           & 0.798 & 0.878  & 0.957 & 0.999  & 1.049 & 0.928 \\
                  & 2021:W10          & 0.793 & 0.876  & 0.959 & 0.991  & 1.044 & 0.924 \\
                  & 2021:W11          & 0.787 & 0.873  & 0.951 & 0.990  & 1.039 & 0.920 \\
                  & 2021:W12          & 0.781 & 0.870  & 0.943 & 0.990  & 1.034 & 0.916 \\
                  & 2021:W13          & 0.776 & 0.865  & 0.938 & 0.982  & 1.028 & 0.911 \\
                  & 2021:W14          & 0.770 & 0.857  & 0.934 & 0.975  & 1.021 & 0.906 \\
                  & 2021:W15          & 0.756 & 0.854  & 0.932 & 0.968  & 1.021 & 0.902 \\
                  & 2021:W16          & 0.740 & 0.845  & 0.932 & 0.964  & 1.018 & 0.896 \\
                  & 2021:W17          & 0.732 & 0.838  & 0.923 & 0.963  & 1.016 & 0.891 \\
                  & 2021:W18          & 0.735 & 0.830  & 0.912 & 0.957  & 1.008 & 0.885 \\
                  & 2021:W19          & 0.740 & 0.822  & 0.903 & 0.952  & 1.003 & 0.880 \\
                  & 2021:W20          & 0.739 & 0.816  & 0.901 & 0.946  & 0.998 & 0.876 \\
                  & 2021:W21          & 0.748 & 0.828  & 0.942 & 0.975  & 1.010 & 0.895 \\
                  & 2021:W22          & 0.737 & 0.825  & 0.939 & 0.977  & 1.009 & 0.891 \\
                  & 2021:W23          & 0.727 & 0.814  & 0.930 & 0.977  & 1.009 & 0.887 \\
                  & 2021:W24          & 0.718 & 0.811  & 0.918 & 0.974  & 1.007 & 0.883 \\
                  & 2021:W25          & 0.709 & 0.812  & 0.912 & 0.967  & 1.006 & 0.879 \\
                  & 2021:W26          & 0.701 & 0.810  & 0.898 & 0.960  & 1.009 & 0.873 \\
                  & 2021:W27          & 0.695 & 0.806  & 0.887 & 0.948  & 1.012 & 0.868 \\
                  & 2021:W28          & 0.690 & 0.806  & 0.878 & 0.941  & 1.016 & 0.864 \\
                  & 2021:W29          & 0.686 & 0.801  & 0.873 & 0.938  & 1.019 & 0.861 \\
                  & 2021:W30          & 0.681 & 0.797  & 0.867 & 0.936  & 1.018 & 0.857 \\
                  & 2021:W31          & 0.678 & 0.797  & 0.870 & 0.938  & 1.017 & 0.856 \\
                  & 2021:W32          & 0.677 & 0.794  & 0.869 & 0.939  & 1.015 & 0.854 \\
                  & 2021:W33          & 0.675 & 0.787  & 0.863 & 0.942  & 1.013 & 0.853 \\
                  & 2021:W34          & 0.676 & 0.783  & 0.862 & 0.943  & 1.011 & 0.853 \\
                  & 2021:W35          & 0.677 & 0.779  & 0.860 & 0.942  & 1.008 & 0.852 \\
                  & 2021:W36          & 0.679 & 0.775  & 0.858 & 0.939  & 1.007 & 0.851 \\
                  & 2021:W37          & 0.681 & 0.770  & 0.856 & 0.940  & 1.005 & 0.850 \\
                  & 2021:W38          & 0.682 & 0.766  & 0.854 & 0.944  & 1.003 & 0.849 \\
                  & 2021:W39          & 0.684 & 0.760  & 0.851 & 0.946  & 1.001 & 0.848 \\
                  & 2021:W40          & 0.686 & 0.755  & 0.848 & 0.948  & 0.999 & 0.847 \\
                  & 2021:W41          & 0.687 & 0.754  & 0.849 & 0.950  & 0.997 & 0.846 \\
                  & 2021:W42          & 0.689 & 0.755  & 0.851 & 0.952  & 0.995 & 0.845 \\
                  & 2021:W43          & 0.689 & 0.753  & 0.857 & 0.947  & 0.992 & 0.845 \\
                  & 2021:W44          & 0.688 & 0.749  & 0.859 & 0.944  & 0.993 & 0.844 \\
                  & 2021:W45          & 0.686 & 0.745  & 0.852 & 0.944  & 0.995 & 0.842 \\
                  & 2021:W46          & 0.684 & 0.743  & 0.845 & 0.947  & 0.996 & 0.841 \\
                  & 2021:W47          & 0.682 & 0.741  & 0.840 & 0.941  & 0.994 & 0.838 \\
                  & 2021:W48          & 0.678 & 0.740  & 0.836 & 0.933  & 0.990 & 0.834 \\ \hline \hline
\end{tabular}
\begin{tablenotes}
\item \textit{Note: The dependent variable is the state-level EC growth.}
\end{tablenotes} 
\end{threeparttable}
\end{adjustbox}
\end{table}

\begin{table}[!htbp]
\caption{Distribution of relative RMSFE across states for AR-W-M model}
\centering
\begin{adjustbox}{width=0.6\textwidth}
\begin{threeparttable}
\begin{tabular}{llccccccc}
\hline \hline
\multicolumn{2}{c}{\textbf{Calendar ($v$})} & \textbf{10\%} & \textbf{25\%} & \textbf{50\%} & \textbf{75\%} & \textbf{90\%} & \textbf{RMSE} \\ \hline
Backcast          & 2021:W1           & 0.340 & 0.423  & 0.567 & 0.703  & 0.862 & 0.580 \\
                  & 2021:W2           & 0.340 & 0.416  & 0.565 & 0.698  & 0.868 & 0.578 \\
                  & 2021:W3           & 0.337 & 0.414  & 0.562 & 0.697  & 0.877 & 0.577 \\
                  & 2021:W4           & 0.337 & 0.414  & 0.558 & 0.695  & 0.885 & 0.575 \\ \hline
Nowcast           & 2021:W5           & 0.784 & 0.899  & 0.938 & 1.005  & 1.032 & 0.927 \\
                  & 2021:W6           & 0.782 & 0.892  & 0.940 & 1.003  & 1.029 & 0.927 \\
                  & 2021:W7           & 0.779 & 0.887  & 0.941 & 1.004  & 1.042 & 0.926 \\
                  & 2021:W8           & 0.776 & 0.887  & 0.940 & 1.001  & 1.043 & 0.925 \\
                  & 2021:W9           & 0.703 & 0.779  & 0.855 & 0.954  & 0.988 & 0.851 \\
                  & 2021:W10          & 0.695 & 0.778  & 0.854 & 0.949  & 0.980 & 0.847 \\
                  & 2021:W11          & 0.688 & 0.771  & 0.853 & 0.942  & 0.979 & 0.843 \\
                  & 2021:W12          & 0.679 & 0.767  & 0.853 & 0.940  & 0.975 & 0.839 \\
                  & 2021:W13          & 0.621 & 0.651  & 0.796 & 0.888  & 0.937 & 0.779 \\
                  & 2021:W14          & 0.614 & 0.648  & 0.791 & 0.883  & 0.944 & 0.774 \\
                  & 2021:W15          & 0.608 & 0.644  & 0.787 & 0.877  & 0.943 & 0.769 \\
                  & 2021:W16          & 0.603 & 0.632  & 0.783 & 0.877  & 0.943 & 0.764 \\
                  & 2021:W17          & 0.550 & 0.583  & 0.728 & 0.834  & 0.925 & 0.719 \\
                  & 2021:W18          & 0.546 & 0.587  & 0.725 & 0.824  & 0.937 & 0.715 \\
                  & 2021:W19          & 0.543 & 0.589  & 0.724 & 0.824  & 0.946 & 0.710 \\
                  & 2021:W20          & 0.539 & 0.579  & 0.724 & 0.823  & 0.950 & 0.707 \\
                  & 2021:W21          & 0.472 & 0.571  & 0.682 & 0.803  & 0.908 & 0.678 \\
                  & 2021:W22          & 0.460 & 0.573  & 0.674 & 0.805  & 0.900 & 0.675 \\
                  & 2021:W23          & 0.454 & 0.575  & 0.660 & 0.805  & 0.901 & 0.672 \\
                  & 2021:W24          & 0.450 & 0.578  & 0.654 & 0.805  & 0.902 & 0.670 \\
                  & 2021:W25          & 0.499 & 0.578  & 0.659 & 0.787  & 0.897 & 0.668 \\
                  & 2021:W26          & 0.490 & 0.573  & 0.646 & 0.789  & 0.894 & 0.665 \\
                  & 2021:W27          & 0.482 & 0.573  & 0.644 & 0.789  & 0.890 & 0.662 \\
                  & 2021:W28          & 0.474 & 0.570  & 0.635 & 0.783  & 0.886 & 0.660 \\
                  & 2021:W29          & 0.434 & 0.547  & 0.630 & 0.777  & 0.868 & 0.642 \\
                  & 2021:W30          & 0.431 & 0.543  & 0.628 & 0.774  & 0.867 & 0.640 \\
                  & 2021:W31          & 0.429 & 0.538  & 0.627 & 0.770  & 0.869 & 0.639 \\
                  & 2021:W32          & 0.426 & 0.537  & 0.627 & 0.767  & 0.870 & 0.638 \\
                  & 2021:W33          & 0.409 & 0.514  & 0.653 & 0.804  & 0.871 & 0.646 \\
                  & 2021:W34          & 0.406 & 0.513  & 0.650 & 0.800  & 0.871 & 0.645 \\
                  & 2021:W35          & 0.402 & 0.512  & 0.649 & 0.797  & 0.872 & 0.644 \\
                  & 2021:W36          & 0.398 & 0.514  & 0.647 & 0.795  & 0.873 & 0.643 \\
                  & 2021:W37          & 0.375 & 0.502  & 0.638 & 0.771  & 0.904 & 0.632 \\
                  & 2021:W38          & 0.377 & 0.499  & 0.633 & 0.768  & 0.903 & 0.631 \\
                  & 2021:W39          & 0.379 & 0.495  & 0.630 & 0.766  & 0.901 & 0.630 \\
                  & 2021:W40          & 0.377 & 0.493  & 0.632 & 0.764  & 0.898 & 0.629 \\
                  & 2021:W41          & 0.377 & 0.479  & 0.639 & 0.738  & 0.928 & 0.623 \\
                  & 2021:W42          & 0.379 & 0.478  & 0.639 & 0.739  & 0.927 & 0.622 \\
                  & 2021:W43          & 0.380 & 0.481  & 0.643 & 0.750  & 0.926 & 0.623 \\
                  & 2021:W44          & 0.382 & 0.483  & 0.641 & 0.760  & 0.925 & 0.624 \\
                  & 2021:W45          & 0.381 & 0.465  & 0.628 & 0.754  & 0.895 & 0.612 \\
                  & 2021:W46          & 0.380 & 0.467  & 0.629 & 0.754  & 0.895 & 0.612 \\
                  & 2021:W47          & 0.378 & 0.468  & 0.625 & 0.757  & 0.893 & 0.611 \\
                  & 2021:W48          & 0.377 & 0.465  & 0.621 & 0.757  & 0.889 & 0.609  \\ \hline \hline
\end{tabular}
\begin{tablenotes}
\item \textit{Note: The dependent variable is the state-level EC growth.}
\end{tablenotes} 
\end{threeparttable}
\end{adjustbox}
\end{table}

\begin{table}[!htbp]
\caption{Distribution of relative RMSFE across states for AR-W-M-Q model}
\centering
\begin{adjustbox}{width=0.6\textwidth}
\begin{threeparttable}
\begin{tabular}{llccccccc}
\hline \hline
\multicolumn{2}{c}{\textbf{Calendar ($v$})} & \textbf{10\%} & \textbf{25\%} & \textbf{50\%} & \textbf{75\%} & \textbf{90\%} & \textbf{RMSE} \\ \hline
Backcast          & 2021:W1           & 0.345 & 0.445  & 0.560 & 0.678  & 0.856 & 0.581 \\
                  & 2021:W2           & 0.355 & 0.445  & 0.557 & 0.679  & 0.863 & 0.579 \\
                  & 2021:W3           & 0.351 & 0.446  & 0.553 & 0.679  & 0.872 & 0.577 \\
                  & 2021:W4           & 0.354 & 0.439  & 0.548 & 0.676  & 0.886 & 0.576 \\ \hline
Nowcast           & 2021:W5           & 0.824 & 0.906  & 0.968 & 1.045  & 1.173 & 0.963 \\
                  & 2021:W6           & 0.819 & 0.901  & 0.966 & 1.044  & 1.173 & 0.961 \\
                  & 2021:W7           & 0.811 & 0.896  & 0.965 & 1.041  & 1.172 & 0.959 \\
                  & 2021:W8           & 0.805 & 0.890  & 0.965 & 1.041  & 1.167 & 0.956 \\
                  & 2021:W9           & 0.698 & 0.770  & 0.878 & 0.965  & 1.061 & 0.868 \\
                  & 2021:W10          & 0.694 & 0.763  & 0.867 & 0.959  & 1.049 & 0.864 \\
                  & 2021:W11          & 0.689 & 0.760  & 0.857 & 0.953  & 1.037 & 0.860 \\
                  & 2021:W12          & 0.686 & 0.756  & 0.846 & 0.955  & 1.035 & 0.856 \\
                  & 2021:W13          & 0.640 & 0.694  & 0.815 & 0.909  & 1.031 & 0.807 \\
                  & 2021:W14          & 0.637 & 0.687  & 0.806 & 0.905  & 1.032 & 0.802 \\
                  & 2021:W15          & 0.628 & 0.680  & 0.798 & 0.901  & 1.032 & 0.798 \\
                  & 2021:W16          & 0.611 & 0.675  & 0.784 & 0.899  & 1.032 & 0.794 \\
                  & 2021:W17          & 0.566 & 0.636  & 0.739 & 0.841  & 1.004 & 0.745 \\
                  & 2021:W18          & 0.561 & 0.633  & 0.736 & 0.834  & 1.004 & 0.743 \\
                  & 2021:W19          & 0.556 & 0.631  & 0.733 & 0.829  & 1.002 & 0.741 \\
                  & 2021:W20          & 0.552 & 0.632  & 0.733 & 0.830  & 1.000 & 0.741 \\
                  & 2021:W21          & 0.519 & 0.622  & 0.732 & 0.818  & 0.950 & 0.716 \\
                  & 2021:W22          & 0.515 & 0.621  & 0.726 & 0.814  & 0.951 & 0.714 \\
                  & 2021:W23          & 0.508 & 0.616  & 0.715 & 0.812  & 0.953 & 0.711 \\
                  & 2021:W24          & 0.501 & 0.608  & 0.702 & 0.813  & 0.953 & 0.709 \\
                  & 2021:W25          & 0.517 & 0.597  & 0.718 & 0.814  & 0.963 & 0.714 \\
                  & 2021:W26          & 0.511 & 0.597  & 0.710 & 0.815  & 0.969 & 0.710 \\
                  & 2021:W27          & 0.506 & 0.597  & 0.703 & 0.816  & 0.971 & 0.708 \\
                  & 2021:W28          & 0.498 & 0.598  & 0.699 & 0.815  & 0.965 & 0.706 \\
                  & 2021:W29          & 0.439 & 0.570  & 0.692 & 0.808  & 0.928 & 0.689 \\
                  & 2021:W30          & 0.434 & 0.571  & 0.694 & 0.809  & 0.921 & 0.687 \\
                  & 2021:W31          & 0.429 & 0.567  & 0.699 & 0.808  & 0.917 & 0.685 \\
                  & 2021:W32          & 0.426 & 0.565  & 0.703 & 0.808  & 0.918 & 0.684 \\
                  & 2021:W33          & 0.422 & 0.552  & 0.683 & 0.845  & 0.907 & 0.681 \\
                  & 2021:W34          & 0.417 & 0.549  & 0.679 & 0.839  & 0.907 & 0.679 \\
                  & 2021:W35          & 0.411 & 0.549  & 0.676 & 0.834  & 0.906 & 0.677 \\
                  & 2021:W36          & 0.406 & 0.548  & 0.671 & 0.829  & 0.905 & 0.675 \\
                  & 2021:W37          & 0.405 & 0.530  & 0.683 & 0.803  & 0.938 & 0.666 \\
                  & 2021:W38          & 0.408 & 0.525  & 0.676 & 0.799  & 0.935 & 0.664 \\
                  & 2021:W39          & 0.407 & 0.520  & 0.669 & 0.796  & 0.929 & 0.661 \\
                  & 2021:W40          & 0.406 & 0.517  & 0.660 & 0.793  & 0.923 & 0.659 \\
                  & 2021:W41          & 0.407 & 0.508  & 0.658 & 0.773  & 0.954 & 0.648 \\
                  & 2021:W42          & 0.403 & 0.506  & 0.655 & 0.776  & 0.950 & 0.647 \\
                  & 2021:W43          & 0.402 & 0.507  & 0.655 & 0.779  & 0.948 & 0.648 \\
                  & 2021:W44          & 0.400 & 0.507  & 0.657 & 0.784  & 0.947 & 0.649 \\
                  & 2021:W45          & 0.374 & 0.468  & 0.594 & 0.737  & 0.922 & 0.599 \\
                  & 2021:W46          & 0.374 & 0.470  & 0.597 & 0.744  & 0.923 & 0.600 \\
                  & 2021:W47          & 0.373 & 0.469  & 0.599 & 0.743  & 0.922 & 0.600 \\
                  & 2021:W48          & 0.372 & 0.467  & 0.599 & 0.741  & 0.918 & 0.599 \\ \hline \hline
\end{tabular}
\begin{tablenotes}
\item \textit{Note: The dependent variable is the state-level EC growth.}
\end{tablenotes} 
\end{threeparttable}
\end{adjustbox}
\end{table}

\begin{table}[!htbp]
\caption{Distribution of relative RMSFE across states for AR-M model}
\centering
\begin{adjustbox}{width=0.6\textwidth}
\begin{threeparttable}
\begin{tabular}{llccccccc}
\hline \hline
\multicolumn{2}{c}{\textbf{Calendar ($v$})} & \textbf{10\%} & \textbf{25\%} & \textbf{50\%} & \textbf{75\%} & \textbf{90\%} & \textbf{RMSE} \\ \hline
Backcast          & 2021:W1           & 0.410 & 0.482  & 0.629 & 0.751  & 0.872 & 0.622 \\
                  & 2021:W2           & 0.410 & 0.482  & 0.629 & 0.751  & 0.872 & 0.622 \\
                  & 2021:W3           & 0.410 & 0.482  & 0.629 & 0.751  & 0.872 & 0.622 \\
                  & 2021:W4           & 0.410 & 0.482  & 0.629 & 0.751  & 0.872 & 0.622 \\ \hline
Nowcast           & 2021:W5           & 0.773 & 0.884  & 0.937 & 0.986  & 1.030 & 0.918 \\
                  & 2021:W6           & 0.773 & 0.884  & 0.937 & 0.986  & 1.030 & 0.918 \\
                  & 2021:W7           & 0.773 & 0.884  & 0.937 & 0.986  & 1.030 & 0.918 \\
                  & 2021:W8           & 0.773 & 0.884  & 0.937 & 0.986  & 1.030 & 0.918 \\
                  & 2021:W9           & 0.692 & 0.771  & 0.856 & 0.947  & 0.991 & 0.846 \\
                  & 2021:W10          & 0.692 & 0.771  & 0.856 & 0.947  & 0.991 & 0.846 \\
                  & 2021:W11          & 0.692 & 0.771  & 0.856 & 0.947  & 0.991 & 0.846 \\
                  & 2021:W12          & 0.692 & 0.771  & 0.856 & 0.947  & 0.991 & 0.846 \\
                  & 2021:W13          & 0.616 & 0.677  & 0.825 & 0.908  & 0.957 & 0.789 \\
                  & 2021:W14          & 0.616 & 0.677  & 0.825 & 0.908  & 0.957 & 0.789 \\
                  & 2021:W15          & 0.616 & 0.677  & 0.825 & 0.908  & 0.957 & 0.789 \\
                  & 2021:W16          & 0.616 & 0.677  & 0.825 & 0.908  & 0.957 & 0.789 \\
                  & 2021:W17          & 0.556 & 0.616  & 0.763 & 0.864  & 0.932 & 0.748 \\
                  & 2021:W18          & 0.556 & 0.616  & 0.763 & 0.864  & 0.932 & 0.748 \\
                  & 2021:W19          & 0.556 & 0.616  & 0.763 & 0.864  & 0.932 & 0.748 \\
                  & 2021:W20          & 0.556 & 0.616  & 0.763 & 0.864  & 0.932 & 0.748 \\
                  & 2021:W21          & 0.511 & 0.588  & 0.722 & 0.820  & 0.904 & 0.708 \\
                  & 2021:W22          & 0.511 & 0.588  & 0.722 & 0.820  & 0.904 & 0.708 \\
                  & 2021:W23          & 0.511 & 0.588  & 0.722 & 0.820  & 0.904 & 0.708 \\
                  & 2021:W24          & 0.511 & 0.588  & 0.722 & 0.820  & 0.904 & 0.708 \\
                  & 2021:W25          & 0.530 & 0.592  & 0.718 & 0.829  & 0.892 & 0.704 \\
                  & 2021:W26          & 0.530 & 0.592  & 0.718 & 0.829  & 0.892 & 0.704 \\
                  & 2021:W27          & 0.530 & 0.592  & 0.718 & 0.829  & 0.892 & 0.704 \\
                  & 2021:W28          & 0.530 & 0.592  & 0.718 & 0.829  & 0.892 & 0.704 \\
                  & 2021:W29          & 0.502 & 0.531  & 0.675 & 0.800  & 0.874 & 0.672 \\
                  & 2021:W30          & 0.502 & 0.531  & 0.675 & 0.800  & 0.874 & 0.672 \\
                  & 2021:W31          & 0.502 & 0.531  & 0.675 & 0.800  & 0.874 & 0.672 \\
                  & 2021:W32          & 0.502 & 0.531  & 0.675 & 0.800  & 0.874 & 0.672 \\
                  & 2021:W33          & 0.461 & 0.542  & 0.681 & 0.787  & 0.885 & 0.663 \\
                  & 2021:W34          & 0.461 & 0.542  & 0.681 & 0.787  & 0.885 & 0.663 \\
                  & 2021:W35          & 0.461 & 0.542  & 0.681 & 0.787  & 0.885 & 0.663 \\
                  & 2021:W36          & 0.461 & 0.542  & 0.681 & 0.787  & 0.885 & 0.663 \\
                  & 2021:W37          & 0.461 & 0.521  & 0.668 & 0.765  & 0.869 & 0.654 \\
                  & 2021:W38          & 0.461 & 0.521  & 0.668 & 0.765  & 0.869 & 0.654 \\
                  & 2021:W39          & 0.461 & 0.521  & 0.668 & 0.765  & 0.869 & 0.654 \\
                  & 2021:W40          & 0.461 & 0.521  & 0.668 & 0.765  & 0.869 & 0.654 \\
                  & 2021:W41          & 0.441 & 0.477  & 0.678 & 0.767  & 0.906 & 0.644 \\
                  & 2021:W42          & 0.441 & 0.477  & 0.678 & 0.767  & 0.906 & 0.644 \\
                  & 2021:W43          & 0.441 & 0.477  & 0.678 & 0.767  & 0.906 & 0.644 \\
                  & 2021:W44          & 0.441 & 0.477  & 0.678 & 0.767  & 0.906 & 0.644 \\
                  & 2021:W45          & 0.424 & 0.471  & 0.685 & 0.746  & 0.908 & 0.635 \\
                  & 2021:W46          & 0.424 & 0.471  & 0.685 & 0.746  & 0.908 & 0.635 \\
                  & 2021:W47          & 0.424 & 0.471  & 0.685 & 0.746  & 0.908 & 0.635 \\
                  & 2021:W48          & 0.424 & 0.471  & 0.685 & 0.746  & 0.908 & 0.635 \\ \hline \hline
\end{tabular}
\begin{tablenotes}
\item \textit{Note: The dependent variable is the state-level EC growth.}
\end{tablenotes} 
\end{threeparttable}
\end{adjustbox}
\end{table}

\begin{table}[!htbp]
\caption{Distribution of relative RMSFE across states for AR-Q model}
\centering
\begin{adjustbox}{width=0.6\textwidth}
\begin{threeparttable}
\begin{tabular}{llccccccc}
\hline \hline
\multicolumn{2}{c}{\textbf{Calendar ($v$})} & \textbf{10\%} & \textbf{25\%} & \textbf{50\%} & \textbf{75\%} & \textbf{90\%} & \textbf{RMSE} \\ \hline
Backcast          & 2021:W1           & 0.657 & 0.761  & 0.887 & 0.950  & 0.986 & 0.845 \\
                  & 2021:W2           & 0.657 & 0.761  & 0.887 & 0.950  & 0.986 & 0.845 \\
                  & 2021:W3           & 0.657 & 0.761  & 0.887 & 0.950  & 0.986 & 0.845 \\
                  & 2021:W4           & 0.657 & 0.761  & 0.887 & 0.950  & 0.986 & 0.845 \\ \hline
Nowcast           & 2021:W5           & 0.751 & 0.861  & 0.946 & 1.017  & 1.090 & 0.934 \\
                  & 2021:W6           & 0.751 & 0.861  & 0.946 & 1.017  & 1.090 & 0.934 \\
                  & 2021:W7           & 0.751 & 0.861  & 0.946 & 1.017  & 1.090 & 0.934 \\
                  & 2021:W8           & 0.751 & 0.861  & 0.946 & 1.017  & 1.090 & 0.934 \\
                  & 2021:W9           & 0.700 & 0.825  & 0.921 & 1.002  & 1.047 & 0.911 \\
                  & 2021:W10          & 0.700 & 0.825  & 0.921 & 1.002  & 1.047 & 0.911 \\
                  & 2021:W11          & 0.700 & 0.825  & 0.921 & 1.002  & 1.047 & 0.911 \\
                  & 2021:W12          & 0.700 & 0.825  & 0.921 & 1.002  & 1.047 & 0.911 \\
                  & 2021:W13          & 0.700 & 0.825  & 0.921 & 1.002  & 1.047 & 0.911 \\
                  & 2021:W14          & 0.700 & 0.825  & 0.921 & 1.002  & 1.047 & 0.911 \\
                  & 2021:W15          & 0.700 & 0.825  & 0.921 & 1.002  & 1.047 & 0.911 \\
                  & 2021:W16          & 0.700 & 0.825  & 0.921 & 1.002  & 1.047 & 0.911 \\
                  & 2021:W17          & 0.700 & 0.825  & 0.921 & 1.002  & 1.047 & 0.911 \\
                  & 2021:W18          & 0.700 & 0.825  & 0.921 & 1.002  & 1.047 & 0.911 \\
                  & 2021:W19          & 0.700 & 0.825  & 0.921 & 1.002  & 1.047 & 0.911 \\
                  & 2021:W20          & 0.700 & 0.825  & 0.921 & 1.002  & 1.047 & 0.911 \\
                  & 2021:W21          & 0.831 & 0.877  & 0.961 & 1.009  & 1.097 & 0.947 \\
                  & 2021:W22          & 0.831 & 0.877  & 0.961 & 1.009  & 1.097 & 0.947 \\
                  & 2021:W23          & 0.831 & 0.877  & 0.961 & 1.009  & 1.097 & 0.947 \\
                  & 2021:W24          & 0.831 & 0.877  & 0.961 & 1.009  & 1.097 & 0.947 \\
                  & 2021:W25          & 0.831 & 0.877  & 0.961 & 1.009  & 1.097 & 0.947 \\
                  & 2021:W26          & 0.831 & 0.877  & 0.961 & 1.009  & 1.097 & 0.947 \\
                  & 2021:W27          & 0.831 & 0.877  & 0.961 & 1.009  & 1.097 & 0.947 \\
                  & 2021:W28          & 0.831 & 0.877  & 0.961 & 1.009  & 1.097 & 0.947 \\
                  & 2021:W29          & 0.831 & 0.877  & 0.961 & 1.009  & 1.097 & 0.947 \\
                  & 2021:W30          & 0.831 & 0.877  & 0.961 & 1.009  & 1.097 & 0.947 \\
                  & 2021:W31          & 0.831 & 0.877  & 0.961 & 1.009  & 1.097 & 0.947 \\
                  & 2021:W32          & 0.831 & 0.877  & 0.961 & 1.009  & 1.097 & 0.947 \\
                  & 2021:W33          & 0.705 & 0.792  & 0.911 & 0.949  & 1.028 & 0.874 \\
                  & 2021:W34          & 0.705 & 0.792  & 0.911 & 0.949  & 1.028 & 0.874 \\
                  & 2021:W35          & 0.705 & 0.792  & 0.911 & 0.949  & 1.028 & 0.874 \\
                  & 2021:W36          & 0.705 & 0.792  & 0.911 & 0.949  & 1.028 & 0.874 \\
                  & 2021:W37          & 0.705 & 0.792  & 0.911 & 0.949  & 1.028 & 0.874 \\
                  & 2021:W38          & 0.705 & 0.792  & 0.911 & 0.949  & 1.028 & 0.874 \\
                  & 2021:W39          & 0.705 & 0.792  & 0.911 & 0.949  & 1.028 & 0.874 \\
                  & 2021:W40          & 0.705 & 0.792  & 0.911 & 0.949  & 1.028 & 0.874 \\
                  & 2021:W41          & 0.705 & 0.792  & 0.911 & 0.949  & 1.028 & 0.874 \\
                  & 2021:W42          & 0.705 & 0.792  & 0.911 & 0.949  & 1.028 & 0.874 \\
                  & 2021:W43          & 0.705 & 0.792  & 0.911 & 0.949  & 1.028 & 0.874 \\
                  & 2021:W44          & 0.705 & 0.792  & 0.911 & 0.949  & 1.028 & 0.874 \\
                  & 2021:W45          & 0.666 & 0.762  & 0.878 & 0.927  & 0.978 & 0.836 \\
                  & 2021:W46          & 0.666 & 0.762  & 0.878 & 0.927  & 0.978 & 0.836 \\
                  & 2021:W47          & 0.666 & 0.762  & 0.878 & 0.927  & 0.978 & 0.836 \\
                  & 2021:W48          & 0.666 & 0.762  & 0.878 & 0.927  & 0.978 & 0.836 \\ \hline \hline
\end{tabular}
\begin{tablenotes}
\item \textit{Note: The dependent variable is the state-level EC growth.}
\end{tablenotes} 
\end{threeparttable}
\end{adjustbox}
\end{table}

\begin{table}[!htbp]
\caption{Distribution of relative RMSFE across states for AR model}
\centering
\begin{adjustbox}{width=0.6\textwidth}
\begin{threeparttable}
\begin{tabular}{llccccccc}
\hline \hline
\multicolumn{2}{c}{\textbf{Calendar ($v$})} & \textbf{10\%} & \textbf{25\%} & \textbf{50\%} & \textbf{75\%} & \textbf{90\%} & \textbf{RMSE} \\ \hline
Backcast          & 2021:W1           & 0.779 & 0.853  & 0.940 & 0.990  & 1.039 & 0.913 \\
                  & 2021:W2           & 0.779 & 0.853  & 0.940 & 0.990  & 1.039 & 0.913 \\
                  & 2021:W3           & 0.779 & 0.853  & 0.940 & 0.990  & 1.039 & 0.913 \\
                  & 2021:W4           & 0.779 & 0.853  & 0.940 & 0.990  & 1.039 & 0.913 \\ \hline
Nowcast           & 2021:W5           & 0.806 & 0.864  & 0.949 & 0.979  & 1.015 & 0.918 \\
                  & 2021:W6           & 0.806 & 0.864  & 0.949 & 0.979  & 1.015 & 0.918 \\
                  & 2021:W7           & 0.806 & 0.864  & 0.949 & 0.979  & 1.015 & 0.918 \\
                  & 2021:W8           & 0.806 & 0.864  & 0.949 & 0.979  & 1.015 & 0.918 \\
                  & 2021:W9           & 0.806 & 0.864  & 0.949 & 0.979  & 1.015 & 0.918 \\
                  & 2021:W10          & 0.806 & 0.864  & 0.949 & 0.979  & 1.015 & 0.918 \\
                  & 2021:W11          & 0.806 & 0.864  & 0.949 & 0.979  & 1.015 & 0.918 \\
                  & 2021:W12          & 0.806 & 0.864  & 0.949 & 0.979  & 1.015 & 0.918 \\
                  & 2021:W13          & 0.806 & 0.864  & 0.949 & 0.979  & 1.015 & 0.918 \\
                  & 2021:W14          & 0.806 & 0.864  & 0.949 & 0.979  & 1.015 & 0.918 \\
                  & 2021:W15          & 0.806 & 0.864  & 0.949 & 0.979  & 1.015 & 0.918 \\
                  & 2021:W16          & 0.806 & 0.864  & 0.949 & 0.979  & 1.015 & 0.918 \\
                  & 2021:W17          & 0.806 & 0.864  & 0.949 & 0.979  & 1.015 & 0.918 \\
                  & 2021:W18          & 0.806 & 0.864  & 0.949 & 0.979  & 1.015 & 0.918 \\
                  & 2021:W19          & 0.806 & 0.864  & 0.949 & 0.979  & 1.015 & 0.918 \\
                  & 2021:W20          & 0.806 & 0.864  & 0.949 & 0.979  & 1.015 & 0.918 \\
                  & 2021:W21          & 0.812 & 0.881  & 0.955 & 0.986  & 1.013 & 0.924 \\
                  & 2021:W22          & 0.812 & 0.881  & 0.955 & 0.986  & 1.013 & 0.924 \\
                  & 2021:W23          & 0.812 & 0.881  & 0.955 & 0.986  & 1.013 & 0.924 \\
                  & 2021:W24          & 0.812 & 0.881  & 0.955 & 0.986  & 1.013 & 0.924 \\
                  & 2021:W25          & 0.812 & 0.881  & 0.955 & 0.986  & 1.013 & 0.924 \\
                  & 2021:W26          & 0.812 & 0.881  & 0.955 & 0.986  & 1.013 & 0.924 \\
                  & 2021:W27          & 0.812 & 0.881  & 0.955 & 0.986  & 1.013 & 0.924 \\
                  & 2021:W28          & 0.812 & 0.881  & 0.955 & 0.986  & 1.013 & 0.924 \\
                  & 2021:W29          & 0.812 & 0.881  & 0.955 & 0.986  & 1.013 & 0.924 \\
                  & 2021:W30          & 0.812 & 0.881  & 0.955 & 0.986  & 1.013 & 0.924 \\
                  & 2021:W31          & 0.812 & 0.881  & 0.955 & 0.986  & 1.013 & 0.924 \\
                  & 2021:W32          & 0.812 & 0.881  & 0.955 & 0.986  & 1.013 & 0.924 \\
                  & 2021:W33          & 0.812 & 0.881  & 0.955 & 0.986  & 1.013 & 0.924 \\
                  & 2021:W34          & 0.812 & 0.881  & 0.955 & 0.986  & 1.013 & 0.924 \\
                  & 2021:W35          & 0.812 & 0.881  & 0.955 & 0.986  & 1.013 & 0.924 \\
                  & 2021:W36          & 0.812 & 0.881  & 0.955 & 0.986  & 1.013 & 0.924 \\
                  & 2021:W37          & 0.812 & 0.881  & 0.955 & 0.986  & 1.013 & 0.924 \\
                  & 2021:W38          & 0.812 & 0.881  & 0.955 & 0.986  & 1.013 & 0.924 \\
                  & 2021:W39          & 0.812 & 0.881  & 0.955 & 0.986  & 1.013 & 0.924 \\
                  & 2021:W40          & 0.812 & 0.881  & 0.955 & 0.986  & 1.013 & 0.924 \\
                  & 2021:W41          & 0.812 & 0.881  & 0.955 & 0.986  & 1.013 & 0.924 \\
                  & 2021:W42          & 0.812 & 0.881  & 0.955 & 0.986  & 1.013 & 0.924 \\
                  & 2021:W43          & 0.812 & 0.881  & 0.955 & 0.986  & 1.013 & 0.924 \\
                  & 2021:W44          & 0.812 & 0.881  & 0.955 & 0.986  & 1.013 & 0.924 \\
                  & 2021:W45          & 0.812 & 0.881  & 0.955 & 0.986  & 1.013 & 0.924 \\
                  & 2021:W46          & 0.812 & 0.881  & 0.955 & 0.986  & 1.013 & 0.924 \\
                  & 2021:W47          & 0.812 & 0.881  & 0.955 & 0.986  & 1.013 & 0.924 \\
                  & 2021:W48          & 0.812 & 0.881  & 0.955 & 0.986  & 1.013 & 0.924  \\  \hline \hline
\end{tabular}
\begin{tablenotes}
\item \textit{Note: The dependent variable is the state-level EC growth.}
\end{tablenotes} 
\end{threeparttable}
\end{adjustbox}
\end{table}

\newpage

\subsection{Supplementary tables for the out-of-sample nowcast of the growth rate of per-capita energy consumption}\label{appendix_1}

This Appendix provides supplementary tables detailing the quantiles of the distribution of the relative RMSFE across states for the additional models proposed to nowcast the growth rate of per-capita energy consumption. 

\begin{table}[!htbp]
\caption{Distribution of relative RMSFE across states for AR-W model}
\centering
\begin{adjustbox}{width=0.6\textwidth}
\begin{threeparttable}
\begin{tabular}{llccccccc}
\hline \hline
\multicolumn{2}{c}{\textbf{Calendar ($v$})} & \textbf{10\%} & \textbf{25\%} & \textbf{50\%} & \textbf{75\%} & \textbf{90\%} & \textbf{RMSE} \\ \hline
Backcast          & 2021:W1           & 0.679 & 0.703  & 0.776 & 0.914  & 0.977 & 0.810 \\
                  & 2021:W2           & 0.677 & 0.701  & 0.773 & 0.909  & 0.972 & 0.805 \\
                  & 2021:W3           & 0.672 & 0.697  & 0.771 & 0.901  & 0.971 & 0.802 \\
                  & 2021:W4           & 0.651 & 0.678  & 0.756 & 0.886  & 0.984 & 0.791 \\ \hline
Nowcast           & 2021:W5           & 0.871 & 0.934  & 0.982 & 1.016  & 1.051 & 0.966 \\
                  & 2021:W6           & 0.871 & 0.934  & 0.986 & 1.015  & 1.052 & 0.967 \\
                  & 2021:W7           & 0.869 & 0.939  & 0.989 & 1.019  & 1.053 & 0.968 \\
                  & 2021:W8           & 0.868 & 0.941  & 0.987 & 1.016  & 1.051 & 0.968 \\
                  & 2021:W9           & 0.867 & 0.941  & 0.987 & 1.014  & 1.052 & 0.967 \\
                  & 2021:W10          & 0.866 & 0.939  & 0.984 & 1.014  & 1.047 & 0.964 \\
                  & 2021:W11          & 0.865 & 0.934  & 0.977 & 1.013  & 1.040 & 0.961 \\
                  & 2021:W12          & 0.858 & 0.928  & 0.972 & 1.011  & 1.033 & 0.957 \\
                  & 2021:W13          & 0.857 & 0.924  & 0.967 & 1.007  & 1.026 & 0.954 \\
                  & 2021:W14          & 0.856 & 0.916  & 0.958 & 1.001  & 1.024 & 0.949 \\
                  & 2021:W15          & 0.854 & 0.912  & 0.952 & 0.990  & 1.024 & 0.945 \\
                  & 2021:W16          & 0.848 & 0.909  & 0.949 & 0.982  & 1.025 & 0.941 \\
                  & 2021:W17          & 0.841 & 0.906  & 0.944 & 0.976  & 1.024 & 0.936 \\
                  & 2021:W18          & 0.837 & 0.901  & 0.940 & 0.973  & 1.021 & 0.932 \\
                  & 2021:W19          & 0.832 & 0.893  & 0.937 & 0.971  & 1.020 & 0.927 \\
                  & 2021:W20          & 0.827 & 0.885  & 0.935 & 0.972  & 1.020 & 0.923 \\
                  & 2021:W21          & 0.810 & 0.879  & 0.937 & 0.972  & 1.028 & 0.923 \\
                  & 2021:W22          & 0.805 & 0.873  & 0.935 & 0.967  & 1.027 & 0.919 \\
                  & 2021:W23          & 0.799 & 0.872  & 0.926 & 0.965  & 1.026 & 0.916 \\
                  & 2021:W24          & 0.794 & 0.863  & 0.918 & 0.971  & 1.024 & 0.912 \\
                  & 2021:W25          & 0.789 & 0.858  & 0.914 & 0.963  & 1.028 & 0.907 \\
                  & 2021:W26          & 0.785 & 0.844  & 0.906 & 0.957  & 1.034 & 0.900 \\
                  & 2021:W27          & 0.791 & 0.838  & 0.892 & 0.950  & 1.033 & 0.895 \\
                  & 2021:W28          & 0.791 & 0.833  & 0.880 & 0.949  & 1.032 & 0.891 \\
                  & 2021:W29          & 0.786 & 0.830  & 0.874 & 0.955  & 1.031 & 0.888 \\
                  & 2021:W30          & 0.779 & 0.822  & 0.866 & 0.954  & 1.029 & 0.883 \\
                  & 2021:W31          & 0.770 & 0.820  & 0.861 & 0.958  & 1.027 & 0.882 \\
                  & 2021:W32          & 0.767 & 0.812  & 0.857 & 0.952  & 1.025 & 0.880 \\
                  & 2021:W33          & 0.764 & 0.810  & 0.859 & 0.949  & 1.022 & 0.878 \\
                  & 2021:W34          & 0.758 & 0.808  & 0.868 & 0.952  & 1.029 & 0.877 \\
                  & 2021:W35          & 0.752 & 0.802  & 0.878 & 0.955  & 1.045 & 0.877 \\
                  & 2021:W36          & 0.748 & 0.794  & 0.885 & 0.959  & 1.061 & 0.876 \\
                  & 2021:W37          & 0.745 & 0.785  & 0.887 & 0.960  & 1.062 & 0.875 \\
                  & 2021:W38          & 0.744 & 0.775  & 0.883 & 0.955  & 1.060 & 0.873 \\
                  & 2021:W39          & 0.746 & 0.767  & 0.883 & 0.952  & 1.056 & 0.871 \\
                  & 2021:W40          & 0.746 & 0.762  & 0.887 & 0.948  & 1.052 & 0.870 \\
                  & 2021:W41          & 0.745 & 0.760  & 0.892 & 0.947  & 1.048 & 0.869 \\
                  & 2021:W42          & 0.749 & 0.761  & 0.889 & 0.945  & 1.045 & 0.868 \\
                  & 2021:W43          & 0.744 & 0.762  & 0.883 & 0.947  & 1.046 & 0.868 \\
                  & 2021:W44          & 0.744 & 0.762  & 0.871 & 0.947  & 1.049 & 0.867 \\
                  & 2021:W45          & 0.739 & 0.764  & 0.861 & 0.948  & 1.049 & 0.866 \\
                  & 2021:W46          & 0.737 & 0.766  & 0.858 & 0.944  & 1.046 & 0.865 \\
                  & 2021:W47          & 0.733 & 0.766  & 0.856 & 0.938  & 1.043 & 0.863 \\
                  & 2021:W48          & 0.729 & 0.765  & 0.857 & 0.940  & 1.042 & 0.859 \\ \hline \hline
\end{tabular}
\begin{tablenotes}
\item \textit{Note: The dependent variables is the state-level EC per capita growth.}
\end{tablenotes} 
\end{threeparttable}
\end{adjustbox}
\end{table}

\begin{table}[!htbp]
\caption{Distribution of relative RMSFE across states for AR-W-M model}
\centering
\begin{adjustbox}{width=0.6\textwidth}
\begin{threeparttable}
\begin{tabular}{llccccccc}
\hline \hline
\multicolumn{2}{c}{\textbf{Calendar ($v$})} & \textbf{10\%} & \textbf{25\%} & \textbf{50\%} & \textbf{75\%} & \textbf{90\%} & \textbf{RMSE} \\ \hline
Backcast          & 2021:W1           & 0.407 & 0.453  & 0.620 & 0.752  & 0.926 & 0.627 \\
                  & 2021:W2           & 0.404 & 0.448  & 0.610 & 0.752  & 0.922 & 0.625 \\
                  & 2021:W3           & 0.399 & 0.449  & 0.605 & 0.752  & 0.930 & 0.623 \\
                  & 2021:W4           & 0.393 & 0.453  & 0.604 & 0.750  & 0.932 & 0.622 \\ \hline
Nowcast           & 2021:W5           & 0.892 & 0.919  & 0.966 & 1.016  & 1.050 & 0.961 \\
                  & 2021:W6           & 0.897 & 0.920  & 0.964 & 1.014  & 1.046 & 0.962 \\
                  & 2021:W7           & 0.898 & 0.926  & 0.963 & 1.013  & 1.044 & 0.962 \\
                  & 2021:W8           & 0.895 & 0.926  & 0.966 & 1.011  & 1.039 & 0.961 \\
                  & 2021:W9           & 0.772 & 0.833  & 0.897 & 0.961  & 1.002 & 0.888 \\
                  & 2021:W10          & 0.770 & 0.832  & 0.894 & 0.960  & 0.999 & 0.885 \\
                  & 2021:W11          & 0.767 & 0.829  & 0.899 & 0.956  & 0.999 & 0.882 \\
                  & 2021:W12          & 0.764 & 0.822  & 0.891 & 0.949  & 0.994 & 0.878 \\
                  & 2021:W13          & 0.652 & 0.740  & 0.843 & 0.909  & 0.979 & 0.825 \\
                  & 2021:W14          & 0.646 & 0.730  & 0.839 & 0.902  & 0.976 & 0.821 \\
                  & 2021:W15          & 0.641 & 0.720  & 0.836 & 0.902  & 0.974 & 0.817 \\
                  & 2021:W16          & 0.634 & 0.709  & 0.830 & 0.899  & 0.974 & 0.812 \\
                  & 2021:W17          & 0.592 & 0.662  & 0.797 & 0.850  & 0.954 & 0.773 \\
                  & 2021:W18          & 0.587 & 0.657  & 0.791 & 0.845  & 0.950 & 0.769 \\
                  & 2021:W19          & 0.582 & 0.655  & 0.786 & 0.844  & 0.955 & 0.766 \\
                  & 2021:W20          & 0.577 & 0.655  & 0.782 & 0.845  & 0.955 & 0.763 \\
                  & 2021:W21          & 0.532 & 0.615  & 0.731 & 0.824  & 0.929 & 0.724 \\
                  & 2021:W22          & 0.522 & 0.616  & 0.722 & 0.819  & 0.935 & 0.720 \\
                  & 2021:W23          & 0.512 & 0.614  & 0.712 & 0.822  & 0.939 & 0.717 \\
                  & 2021:W24          & 0.504 & 0.612  & 0.702 & 0.825  & 0.945 & 0.714 \\
                  & 2021:W25          & 0.513 & 0.644  & 0.717 & 0.821  & 0.915 & 0.716 \\
                  & 2021:W26          & 0.512 & 0.637  & 0.706 & 0.821  & 0.921 & 0.711 \\
                  & 2021:W27          & 0.514 & 0.632  & 0.699 & 0.821  & 0.929 & 0.708 \\
                  & 2021:W28          & 0.517 & 0.627  & 0.691 & 0.821  & 0.937 & 0.705 \\
                  & 2021:W29          & 0.492 & 0.605  & 0.694 & 0.813  & 0.953 & 0.693 \\
                  & 2021:W30          & 0.488 & 0.605  & 0.685 & 0.808  & 0.955 & 0.691 \\
                  & 2021:W31          & 0.485 & 0.601  & 0.683 & 0.806  & 0.948 & 0.689 \\
                  & 2021:W32          & 0.482 & 0.593  & 0.679 & 0.803  & 0.944 & 0.687 \\
                  & 2021:W33          & 0.482 & 0.569  & 0.703 & 0.817  & 0.928 & 0.690 \\
                  & 2021:W34          & 0.483 & 0.568  & 0.702 & 0.818  & 0.928 & 0.689 \\
                  & 2021:W35          & 0.484 & 0.567  & 0.699 & 0.818  & 0.929 & 0.688 \\
                  & 2021:W36          & 0.484 & 0.565  & 0.695 & 0.813  & 0.930 & 0.687 \\
                  & 2021:W37          & 0.467 & 0.553  & 0.679 & 0.794  & 0.925 & 0.682 \\
                  & 2021:W38          & 0.468 & 0.552  & 0.675 & 0.791  & 0.943 & 0.680 \\
                  & 2021:W39          & 0.469 & 0.547  & 0.672 & 0.787  & 0.947 & 0.679 \\
                  & 2021:W40          & 0.468 & 0.544  & 0.669 & 0.783  & 0.950 & 0.677 \\
                  & 2021:W41          & 0.425 & 0.532  & 0.669 & 0.785  & 0.999 & 0.672 \\
                  & 2021:W42          & 0.424 & 0.532  & 0.671 & 0.788  & 1.001 & 0.671 \\
                  & 2021:W43          & 0.424 & 0.532  & 0.669 & 0.793  & 0.998 & 0.671 \\
                  & 2021:W44          & 0.424 & 0.532  & 0.667 & 0.798  & 0.998 & 0.671 \\
                  & 2021:W45          & 0.402 & 0.515  & 0.680 & 0.793  & 0.981 & 0.661 \\
                  & 2021:W46          & 0.403 & 0.514  & 0.681 & 0.804  & 0.983 & 0.661 \\
                  & 2021:W47          & 0.403 & 0.511  & 0.677 & 0.809  & 0.986 & 0.659 \\
                  & 2021:W48          & 0.404 & 0.507  & 0.676 & 0.807  & 0.985 & 0.657  \\ \hline \hline
\end{tabular}
\begin{tablenotes}
\item \textit{Note: The dependent variable is the state-level EC per capita growth.}
\end{tablenotes} 
\end{threeparttable}
\end{adjustbox}
\end{table}

\begin{table}[!htbp]
\caption{Distribution of relative RMSFE across states for AR-M model}
\centering
\begin{adjustbox}{width=0.6\textwidth}
\begin{threeparttable}
\begin{tabular}{llccccccc}
\hline \hline
\multicolumn{2}{c}{\textbf{Calendar ($v$})} & \textbf{10\%} & \textbf{25\%} & \textbf{50\%} & \textbf{75\%} & \textbf{90\%} & \textbf{RMSE} \\ \hline
Backcast          & 2021:W1           & 0.469 & 0.528  & 0.688 & 0.832  & 0.944 & 0.684 \\
                  & 2021:W2           & 0.469 & 0.528  & 0.688 & 0.832  & 0.944 & 0.684 \\
                  & 2021:W3           & 0.469 & 0.528  & 0.688 & 0.832  & 0.944 & 0.684 \\
                  & 2021:W4           & 0.469 & 0.528  & 0.688 & 0.832  & 0.944 & 0.684 \\ \hline
Nowcast           & 2021:W5           & 0.877 & 0.913  & 0.954 & 0.991  & 1.031 & 0.950 \\
                  & 2021:W6           & 0.877 & 0.913  & 0.954 & 0.991  & 1.031 & 0.950 \\
                  & 2021:W7           & 0.877 & 0.913  & 0.954 & 0.991  & 1.031 & 0.950 \\
                  & 2021:W8           & 0.877 & 0.913  & 0.954 & 0.991  & 1.031 & 0.950 \\
                  & 2021:W9           & 0.773 & 0.817  & 0.874 & 0.957  & 0.988 & 0.877 \\
                  & 2021:W10          & 0.773 & 0.817  & 0.874 & 0.957  & 0.988 & 0.877 \\
                  & 2021:W11          & 0.773 & 0.817  & 0.874 & 0.957  & 0.988 & 0.877 \\
                  & 2021:W12          & 0.773 & 0.817  & 0.874 & 0.957  & 0.988 & 0.877 \\
                  & 2021:W13          & 0.652 & 0.735  & 0.861 & 0.905  & 0.985 & 0.826 \\
                  & 2021:W14          & 0.652 & 0.735  & 0.861 & 0.905  & 0.985 & 0.826 \\
                  & 2021:W15          & 0.652 & 0.735  & 0.861 & 0.905  & 0.985 & 0.826 \\
                  & 2021:W16          & 0.652 & 0.735  & 0.861 & 0.905  & 0.985 & 0.826 \\
                  & 2021:W17          & 0.609 & 0.680  & 0.807 & 0.884  & 0.956 & 0.789 \\
                  & 2021:W18          & 0.609 & 0.680  & 0.807 & 0.884  & 0.956 & 0.789 \\
                  & 2021:W19          & 0.609 & 0.680  & 0.807 & 0.884  & 0.956 & 0.789 \\
                  & 2021:W20          & 0.609 & 0.680  & 0.807 & 0.884  & 0.956 & 0.789 \\
                  & 2021:W21          & 0.542 & 0.664  & 0.750 & 0.853  & 0.913 & 0.747 \\
                  & 2021:W22          & 0.542 & 0.664  & 0.750 & 0.853  & 0.913 & 0.747 \\
                  & 2021:W23          & 0.542 & 0.664  & 0.750 & 0.853  & 0.913 & 0.747 \\
                  & 2021:W24          & 0.542 & 0.664  & 0.750 & 0.853  & 0.913 & 0.747 \\
                  & 2021:W25          & 0.541 & 0.656  & 0.753 & 0.850  & 0.899 & 0.743 \\
                  & 2021:W26          & 0.541 & 0.656  & 0.753 & 0.850  & 0.899 & 0.743 \\
                  & 2021:W27          & 0.541 & 0.656  & 0.753 & 0.850  & 0.899 & 0.743 \\
                  & 2021:W28          & 0.541 & 0.656  & 0.753 & 0.850  & 0.899 & 0.743 \\
                  & 2021:W29          & 0.521 & 0.644  & 0.733 & 0.823  & 0.904 & 0.718 \\
                  & 2021:W30          & 0.521 & 0.644  & 0.733 & 0.823  & 0.904 & 0.718 \\
                  & 2021:W31          & 0.521 & 0.644  & 0.733 & 0.823  & 0.904 & 0.718 \\
                  & 2021:W32          & 0.521 & 0.644  & 0.733 & 0.823  & 0.904 & 0.718 \\
                  & 2021:W33          & 0.519 & 0.598  & 0.756 & 0.811  & 0.946 & 0.713 \\
                  & 2021:W34          & 0.519 & 0.598  & 0.756 & 0.811  & 0.946 & 0.713 \\
                  & 2021:W35          & 0.519 & 0.598  & 0.756 & 0.811  & 0.946 & 0.713 \\
                  & 2021:W36          & 0.519 & 0.598  & 0.756 & 0.811  & 0.946 & 0.713 \\
                  & 2021:W37          & 0.509 & 0.614  & 0.730 & 0.800  & 0.933 & 0.706 \\
                  & 2021:W38          & 0.509 & 0.614  & 0.730 & 0.800  & 0.933 & 0.706 \\
                  & 2021:W39          & 0.509 & 0.614  & 0.730 & 0.800  & 0.933 & 0.706 \\
                  & 2021:W40          & 0.509 & 0.614  & 0.730 & 0.800  & 0.933 & 0.706 \\
                  & 2021:W41          & 0.476 & 0.567  & 0.720 & 0.824  & 0.935 & 0.701 \\
                  & 2021:W42          & 0.476 & 0.567  & 0.720 & 0.824  & 0.935 & 0.701 \\
                  & 2021:W43          & 0.476 & 0.567  & 0.720 & 0.824  & 0.935 & 0.701 \\
                  & 2021:W44          & 0.476 & 0.567  & 0.720 & 0.824  & 0.935 & 0.701 \\
                  & 2021:W45          & 0.468 & 0.558  & 0.710 & 0.819  & 0.933 & 0.696 \\
                  & 2021:W46          & 0.468 & 0.558  & 0.710 & 0.819  & 0.933 & 0.696 \\
                  & 2021:W47          & 0.468 & 0.558  & 0.710 & 0.819  & 0.933 & 0.696 \\
                  & 2021:W48          & 0.468 & 0.558  & 0.710 & 0.819  & 0.933 & 0.696 \\ \hline \hline
\end{tabular}
\begin{tablenotes}
\item \textit{Note: The dependent variable is the state-level EC per capita growth.}
\end{tablenotes} 
\end{threeparttable}
\end{adjustbox}
\end{table}

\begin{table}[!htbp]
\caption{Distribution of relative RMSFE across states for AR-Q model}
\centering
\begin{adjustbox}{width=0.6\textwidth}
\begin{threeparttable}
\begin{tabular}{llccccccc}
\hline \hline
\multicolumn{2}{c}{\textbf{Calendar ($v$})} & \textbf{10\%} & \textbf{25\%} & \textbf{50\%} & \textbf{75\%} & \textbf{90\%} & \textbf{RMSE} \\ \hline
Backcast          & 2021:W1           & 0.759 & 0.836  & 0.908 & 0.965  & 0.991 & 0.892 \\
                  & 2021:W2           & 0.759 & 0.836  & 0.908 & 0.965  & 0.991 & 0.892 \\
                  & 2021:W3           & 0.759 & 0.836  & 0.908 & 0.965  & 0.991 & 0.892 \\
                  & 2021:W4           & 0.759 & 0.836  & 0.908 & 0.965  & 0.991 & 0.892 \\ \hline
Nowcast           & 2021:W5           & 0.782 & 0.888  & 0.944 & 1.016  & 1.131 & 0.950 \\
                  & 2021:W6           & 0.782 & 0.888  & 0.944 & 1.016  & 1.131 & 0.950 \\
                  & 2021:W7           & 0.782 & 0.888  & 0.944 & 1.016  & 1.131 & 0.950 \\
                  & 2021:W8           & 0.782 & 0.888  & 0.944 & 1.016  & 1.131 & 0.950 \\
                  & 2021:W9           & 0.768 & 0.865  & 0.920 & 1.001  & 1.081 & 0.930 \\
                  & 2021:W10          & 0.768 & 0.865  & 0.920 & 1.001  & 1.081 & 0.930 \\
                  & 2021:W11          & 0.768 & 0.865  & 0.920 & 1.001  & 1.081 & 0.930 \\
                  & 2021:W12          & 0.768 & 0.865  & 0.920 & 1.001  & 1.081 & 0.930 \\
                  & 2021:W13          & 0.768 & 0.865  & 0.920 & 1.001  & 1.081 & 0.930 \\
                  & 2021:W14          & 0.768 & 0.865  & 0.920 & 1.001  & 1.081 & 0.930 \\
                  & 2021:W15          & 0.768 & 0.865  & 0.920 & 1.001  & 1.081 & 0.930 \\
                  & 2021:W16          & 0.768 & 0.865  & 0.920 & 1.001  & 1.081 & 0.930 \\
                  & 2021:W17          & 0.768 & 0.865  & 0.920 & 1.001  & 1.081 & 0.930 \\
                  & 2021:W18          & 0.768 & 0.865  & 0.920 & 1.001  & 1.081 & 0.930 \\
                  & 2021:W19          & 0.768 & 0.865  & 0.920 & 1.001  & 1.081 & 0.930 \\
                  & 2021:W20          & 0.768 & 0.865  & 0.920 & 1.001  & 1.081 & 0.930 \\
                  & 2021:W21          & 0.854 & 0.914  & 0.971 & 1.001  & 1.064 & 0.954 \\
                  & 2021:W22          & 0.854 & 0.914  & 0.971 & 1.001  & 1.064 & 0.954 \\
                  & 2021:W23          & 0.854 & 0.914  & 0.971 & 1.001  & 1.064 & 0.954 \\
                  & 2021:W24          & 0.854 & 0.914  & 0.971 & 1.001  & 1.064 & 0.954 \\
                  & 2021:W25          & 0.854 & 0.914  & 0.971 & 1.001  & 1.064 & 0.954 \\
                  & 2021:W26          & 0.854 & 0.914  & 0.971 & 1.001  & 1.064 & 0.954 \\
                  & 2021:W27          & 0.854 & 0.914  & 0.971 & 1.001  & 1.064 & 0.954 \\
                  & 2021:W28          & 0.854 & 0.914  & 0.971 & 1.001  & 1.064 & 0.954 \\
                  & 2021:W29          & 0.854 & 0.914  & 0.971 & 1.001  & 1.064 & 0.954 \\
                  & 2021:W30          & 0.854 & 0.914  & 0.971 & 1.001  & 1.064 & 0.954 \\
                  & 2021:W31          & 0.854 & 0.914  & 0.971 & 1.001  & 1.064 & 0.954 \\
                  & 2021:W32          & 0.854 & 0.914  & 0.971 & 1.001  & 1.064 & 0.954 \\
                  & 2021:W33          & 0.792 & 0.847  & 0.917 & 0.958  & 1.002 & 0.906 \\
                  & 2021:W34          & 0.792 & 0.847  & 0.917 & 0.958  & 1.002 & 0.906 \\
                  & 2021:W35          & 0.792 & 0.847  & 0.917 & 0.958  & 1.002 & 0.906 \\
                  & 2021:W36          & 0.792 & 0.847  & 0.917 & 0.958  & 1.002 & 0.906 \\
                  & 2021:W37          & 0.792 & 0.847  & 0.917 & 0.958  & 1.002 & 0.906 \\
                  & 2021:W38          & 0.792 & 0.847  & 0.917 & 0.958  & 1.002 & 0.906 \\
                  & 2021:W39          & 0.792 & 0.847  & 0.917 & 0.958  & 1.002 & 0.906 \\
                  & 2021:W40          & 0.792 & 0.847  & 0.917 & 0.958  & 1.002 & 0.906 \\
                  & 2021:W41          & 0.792 & 0.847  & 0.917 & 0.958  & 1.002 & 0.906 \\
                  & 2021:W42          & 0.792 & 0.847  & 0.917 & 0.958  & 1.002 & 0.906 \\
                  & 2021:W43          & 0.792 & 0.847  & 0.917 & 0.958  & 1.002 & 0.906 \\
                  & 2021:W44          & 0.792 & 0.847  & 0.917 & 0.958  & 1.002 & 0.906 \\
                  & 2021:W45          & 0.756 & 0.819  & 0.894 & 0.962  & 0.996 & 0.887 \\
                  & 2021:W46          & 0.756 & 0.819  & 0.894 & 0.962  & 0.996 & 0.887 \\
                  & 2021:W47          & 0.756 & 0.819  & 0.894 & 0.962  & 0.996 & 0.887 \\
                  & 2021:W48          & 0.756 & 0.819  & 0.894 & 0.962  & 0.996 & 0.887 \\ \hline \hline
\end{tabular}
\begin{tablenotes}
\item \textit{Note: The dependent variable is the state-level EC per capita growth.}
\end{tablenotes} 
\end{threeparttable}
\end{adjustbox}
\end{table}

\begin{table}[!htbp]
\caption{Distribution of relative RMSFE across states for AR model}
\centering
\begin{adjustbox}{width=0.6\textwidth}
\begin{threeparttable}
\begin{tabular}{llccccccc}
\hline \hline
\multicolumn{2}{c}{\textbf{Calendar ($v$})} & \textbf{10\%} & \textbf{25\%} & \textbf{50\%} & \textbf{75\%} & \textbf{90\%} & \textbf{RMSE} \\ \hline
Backcast          & 2021:W1           & 0.846 & 0.892  & 0.957 & 0.983  & 1.002 & 0.929 \\
                  & 2021:W2           & 0.846 & 0.892  & 0.957 & 0.983  & 1.002 & 0.929 \\
                  & 2021:W3           & 0.846 & 0.892  & 0.957 & 0.983  & 1.002 & 0.929 \\
                  & 2021:W4           & 0.846 & 0.892  & 0.957 & 0.983  & 1.002 & 0.929 \\ \hline
Nowcast           & 2021:W5           & 0.872 & 0.926  & 0.966 & 0.994  & 1.007 & 0.952 \\
                  & 2021:W6           & 0.872 & 0.926  & 0.966 & 0.994  & 1.007 & 0.952 \\
                  & 2021:W7           & 0.872 & 0.926  & 0.966 & 0.994  & 1.007 & 0.952 \\
                  & 2021:W8           & 0.872 & 0.926  & 0.966 & 0.994  & 1.007 & 0.952 \\
                  & 2021:W9           & 0.872 & 0.926  & 0.966 & 0.994  & 1.007 & 0.952 \\
                  & 2021:W10          & 0.872 & 0.926  & 0.966 & 0.994  & 1.007 & 0.952 \\
                  & 2021:W11          & 0.872 & 0.926  & 0.966 & 0.994  & 1.007 & 0.952 \\
                  & 2021:W12          & 0.872 & 0.926  & 0.966 & 0.994  & 1.007 & 0.952 \\
                  & 2021:W13          & 0.872 & 0.926  & 0.966 & 0.994  & 1.007 & 0.952 \\
                  & 2021:W14          & 0.872 & 0.926  & 0.966 & 0.994  & 1.007 & 0.952 \\
                  & 2021:W15          & 0.872 & 0.926  & 0.966 & 0.994  & 1.007 & 0.952 \\
                  & 2021:W16          & 0.872 & 0.926  & 0.966 & 0.994  & 1.007 & 0.952 \\
                  & 2021:W17          & 0.872 & 0.926  & 0.966 & 0.994  & 1.007 & 0.952 \\
                  & 2021:W18          & 0.872 & 0.926  & 0.966 & 0.994  & 1.007 & 0.952 \\
                  & 2021:W19          & 0.872 & 0.926  & 0.966 & 0.994  & 1.007 & 0.952 \\
                  & 2021:W20          & 0.872 & 0.926  & 0.966 & 0.994  & 1.007 & 0.952 \\
                  & 2021:W21          & 0.867 & 0.921  & 0.950 & 0.990  & 1.016 & 0.944 \\
                  & 2021:W22          & 0.867 & 0.921  & 0.950 & 0.990  & 1.016 & 0.944 \\
                  & 2021:W23          & 0.867 & 0.921  & 0.950 & 0.990  & 1.016 & 0.944 \\
                  & 2021:W24          & 0.867 & 0.921  & 0.950 & 0.990  & 1.016 & 0.944 \\
                  & 2021:W25          & 0.867 & 0.921  & 0.950 & 0.990  & 1.016 & 0.944 \\
                  & 2021:W26          & 0.867 & 0.921  & 0.950 & 0.990  & 1.016 & 0.944 \\
                  & 2021:W27          & 0.867 & 0.921  & 0.950 & 0.990  & 1.016 & 0.944 \\
                  & 2021:W28          & 0.867 & 0.921  & 0.950 & 0.990  & 1.016 & 0.944 \\
                  & 2021:W29          & 0.867 & 0.921  & 0.950 & 0.990  & 1.016 & 0.944 \\
                  & 2021:W30          & 0.867 & 0.921  & 0.950 & 0.990  & 1.016 & 0.944 \\
                  & 2021:W31          & 0.867 & 0.921  & 0.950 & 0.990  & 1.016 & 0.944 \\
                  & 2021:W32          & 0.867 & 0.921  & 0.950 & 0.990  & 1.016 & 0.944 \\
                  & 2021:W33          & 0.867 & 0.921  & 0.950 & 0.990  & 1.016 & 0.944 \\
                  & 2021:W34          & 0.867 & 0.921  & 0.950 & 0.990  & 1.016 & 0.944 \\
                  & 2021:W35          & 0.867 & 0.921  & 0.950 & 0.990  & 1.016 & 0.944 \\
                  & 2021:W36          & 0.867 & 0.921  & 0.950 & 0.990  & 1.016 & 0.944 \\
                  & 2021:W37          & 0.867 & 0.921  & 0.950 & 0.990  & 1.016 & 0.944 \\
                  & 2021:W38          & 0.867 & 0.921  & 0.950 & 0.990  & 1.016 & 0.944 \\
                  & 2021:W39          & 0.867 & 0.921  & 0.950 & 0.990  & 1.016 & 0.944 \\
                  & 2021:W40          & 0.867 & 0.921  & 0.950 & 0.990  & 1.016 & 0.944 \\
                  & 2021:W41          & 0.867 & 0.921  & 0.950 & 0.990  & 1.016 & 0.944 \\
                  & 2021:W42          & 0.867 & 0.921  & 0.950 & 0.990  & 1.016 & 0.944 \\
                  & 2021:W43          & 0.867 & 0.921  & 0.950 & 0.990  & 1.016 & 0.944 \\
                  & 2021:W44          & 0.867 & 0.921  & 0.950 & 0.990  & 1.016 & 0.944 \\
                  & 2021:W45          & 0.867 & 0.921  & 0.950 & 0.990  & 1.016 & 0.944 \\
                  & 2021:W46          & 0.867 & 0.921  & 0.950 & 0.990  & 1.016 & 0.944 \\
                  & 2021:W47          & 0.867 & 0.921  & 0.950 & 0.990  & 1.016 & 0.944 \\
                  & 2021:W48          & 0.867 & 0.921  & 0.950 & 0.990  & 1.016 & 0.944  \\  \hline \hline
\end{tabular}
\begin{tablenotes}
\item \textit{Note: The dependent variable is the state-level EC per capita growth.}
\end{tablenotes} 
\end{threeparttable}
\end{adjustbox}
\end{table}

\newpage

\subsection{Supplementary tables for the out-of-sample nowcast of the growth rate of per-capita CO2 emissions using the AR-W-M-Q model}\label{appendix_2}

This Appendix provides supplementary tables detailing the quantiles of the distribution of the relative QS across states for the AR-W-M-Q model, the model with the best performance in nowcasting the growth rate of per-capita CO2 emissions. 

\begin{table}[!htbp]
\caption{Distribution of relative QS across states for AR-W-M-Q model, $\tau=0.25$}
\centering
\begin{adjustbox}{width=0.6\textwidth}
\begin{threeparttable}
\begin{tabular}{llccccccc}
\hline \hline
\multicolumn{2}{c}{\textbf{Calendar ($v$})} & \textbf{10\%} & \textbf{25\%} & \textbf{50\%} & \textbf{75\%} & \textbf{90\%} & \textbf{QS}\\ \hline
Backcast          & 2021:W1           & 0.568 & 0.635  & 0.796 & 0.867  & 1.117 & 0.798 \\
                  & 2021:W2           & 0.568 & 0.636  & 0.797 & 0.862  & 1.110 & 0.796 \\
                  & 2021:W3           & 0.569 & 0.635  & 0.798 & 0.858  & 1.102 & 0.795 \\
                  & 2021:W4           & 0.572 & 0.634  & 0.793 & 0.857  & 1.091 & 0.794 \\ \hline
Nowcast           & 2021:W5           & 0.778 & 0.901  & 1.036 & 1.165  & 1.395 & 1.061 \\
                  & 2021:W6           & 0.775 & 0.901  & 1.038 & 1.164  & 1.395 & 1.061 \\
                  & 2021:W7           & 0.771 & 0.902  & 1.042 & 1.167  & 1.394 & 1.061 \\
                  & 2021:W8           & 0.769 & 0.903  & 1.046 & 1.170  & 1.391 & 1.060 \\
                  & 2021:W9           & 0.771 & 0.841  & 1.023 & 1.122  & 1.315 & 1.030 \\
                  & 2021:W10          & 0.768 & 0.840  & 1.023 & 1.124  & 1.311 & 1.027 \\
                  & 2021:W11          & 0.768 & 0.839  & 1.022 & 1.125  & 1.307 & 1.024 \\
                  & 2021:W12          & 0.768 & 0.837  & 1.021 & 1.130  & 1.302 & 1.021 \\
                  & 2021:W13          & 0.735 & 0.822  & 0.959 & 1.046  & 1.251 & 0.965 \\
                  & 2021:W14          & 0.732 & 0.820  & 0.954 & 1.041  & 1.246 & 0.961 \\
                  & 2021:W15          & 0.730 & 0.818  & 0.944 & 1.036  & 1.241 & 0.956 \\
                  & 2021:W16          & 0.728 & 0.814  & 0.933 & 1.030  & 1.237 & 0.952 \\
                  & 2021:W17          & 0.720 & 0.773  & 0.874 & 0.998  & 1.180 & 0.907 \\
                  & 2021:W18          & 0.718 & 0.767  & 0.874 & 0.999  & 1.176 & 0.904 \\
                  & 2021:W19          & 0.716 & 0.763  & 0.874 & 1.001  & 1.173 & 0.903 \\
                  & 2021:W20          & 0.714 & 0.759  & 0.873 & 1.004  & 1.174 & 0.902 \\
                  & 2021:W21          & 0.684 & 0.760  & 0.853 & 0.961  & 1.060 & 0.879 \\
                  & 2021:W22          & 0.681 & 0.757  & 0.849 & 0.961  & 1.059 & 0.876 \\
                  & 2021:W23          & 0.678 & 0.755  & 0.842 & 0.960  & 1.057 & 0.873 \\
                  & 2021:W24          & 0.676 & 0.752  & 0.836 & 0.959  & 1.056 & 0.870 \\
                  & 2021:W25          & 0.694 & 0.765  & 0.841 & 0.980  & 1.092 & 0.879 \\
                  & 2021:W26          & 0.688 & 0.758  & 0.831 & 0.978  & 1.087 & 0.875 \\
                  & 2021:W27          & 0.680 & 0.754  & 0.829 & 0.974  & 1.083 & 0.872 \\
                  & 2021:W28          & 0.681 & 0.741  & 0.827 & 0.971  & 1.078 & 0.868 \\
                  & 2021:W29          & 0.674 & 0.745  & 0.810 & 0.956  & 1.078 & 0.860 \\
                  & 2021:W30          & 0.667 & 0.735  & 0.805 & 0.957  & 1.088 & 0.858 \\
                  & 2021:W31          & 0.663 & 0.726  & 0.805 & 0.959  & 1.092 & 0.856 \\
                  & 2021:W32          & 0.659 & 0.716  & 0.805 & 0.959  & 1.091 & 0.854 \\
                  & 2021:W33          & 0.625 & 0.683  & 0.774 & 0.918  & 1.167 & 0.827 \\
                  & 2021:W34          & 0.629 & 0.678  & 0.769 & 0.915  & 1.161 & 0.826 \\
                  & 2021:W35          & 0.628 & 0.680  & 0.768 & 0.910  & 1.156 & 0.825 \\
                  & 2021:W36          & 0.625 & 0.688  & 0.769 & 0.902  & 1.150 & 0.823 \\
                  & 2021:W37          & 0.608 & 0.694  & 0.771 & 0.911  & 1.056 & 0.819 \\
                  & 2021:W38          & 0.606 & 0.687  & 0.768 & 0.906  & 1.052 & 0.816 \\
                  & 2021:W39          & 0.603 & 0.691  & 0.765 & 0.905  & 1.048 & 0.813 \\
                  & 2021:W40          & 0.600 & 0.691  & 0.761 & 0.905  & 1.044 & 0.810 \\
                  & 2021:W41          & 0.596 & 0.669  & 0.744 & 0.877  & 1.070 & 0.795 \\
                  & 2021:W42          & 0.592 & 0.669  & 0.743 & 0.874  & 1.067 & 0.793 \\
                  & 2021:W43          & 0.589 & 0.669  & 0.743 & 0.874  & 1.069 & 0.793 \\
                  & 2021:W44          & 0.586 & 0.668  & 0.743 & 0.874  & 1.069 & 0.793 \\
                  & 2021:W45          & 0.561 & 0.646  & 0.727 & 0.841  & 1.088 & 0.780 \\
                  & 2021:W46          & 0.567 & 0.644  & 0.723 & 0.842  & 1.086 & 0.780 \\
                  & 2021:W47          & 0.568 & 0.641  & 0.719 & 0.842  & 1.083 & 0.779 \\
                  & 2021:W48          & 0.567 & 0.638  & 0.716 & 0.844  & 1.075 & 0.778  \\ \hline \hline
\end{tabular}
\begin{tablenotes}
\item \textit{Note: The dependent variable is the state-level CO2 per capita growth.}
\end{tablenotes} 
\end{threeparttable}
\end{adjustbox}
\end{table}

\begin{table}[!htbp]
\caption{Distribution of relative QS across states for AR-W-M-Q model, $\tau=0.50$}
\centering
\begin{adjustbox}{width=0.7\textwidth}
\begin{threeparttable}
\begin{tabular}{llccccccc}
\hline \hline
\multicolumn{2}{c}{\textbf{Calendar ($v$})} & \textbf{10\%} & \textbf{25\%} & \textbf{50\%} & \textbf{75\%} & \textbf{90\%} & \textbf{QS}\\ \hline
Backcast          & 2021:W1           & 0.628 & 0.662  & 0.754 & 0.870  & 0.966 & 0.779 \\
                  & 2021:W2           & 0.625 & 0.659  & 0.756 & 0.868  & 0.963 & 0.778 \\
                  & 2021:W3           & 0.621 & 0.660  & 0.756 & 0.868  & 0.962 & 0.777 \\
                  & 2021:W4           & 0.620 & 0.663  & 0.756 & 0.868  & 0.960 & 0.776 \\ \hline
Nowcast           & 2021:W5           & 0.785 & 0.874  & 0.988 & 1.051  & 1.189 & 0.982 \\
                  & 2021:W6           & 0.783 & 0.876  & 0.986 & 1.052  & 1.189 & 0.982 \\
                  & 2021:W7           & 0.781 & 0.876  & 0.983 & 1.053  & 1.188 & 0.982 \\
                  & 2021:W8           & 0.778 & 0.878  & 0.982 & 1.054  & 1.186 & 0.981 \\
                  & 2021:W9           & 0.727 & 0.855  & 0.939 & 1.008  & 1.143 & 0.939 \\
                  & 2021:W10          & 0.723 & 0.856  & 0.938 & 1.001  & 1.144 & 0.937 \\
                  & 2021:W11          & 0.717 & 0.856  & 0.936 & 0.996  & 1.145 & 0.935 \\
                  & 2021:W12          & 0.714 & 0.852  & 0.933 & 0.994  & 1.144 & 0.933 \\
                  & 2021:W13          & 0.705 & 0.833  & 0.893 & 0.967  & 1.071 & 0.896 \\
                  & 2021:W14          & 0.702 & 0.831  & 0.887 & 0.964  & 1.071 & 0.893 \\
                  & 2021:W15          & 0.698 & 0.831  & 0.883 & 0.963  & 1.070 & 0.890 \\
                  & 2021:W16          & 0.693 & 0.832  & 0.878 & 0.962  & 1.069 & 0.888 \\
                  & 2021:W17          & 0.684 & 0.770  & 0.840 & 0.958  & 1.043 & 0.862 \\
                  & 2021:W18          & 0.681 & 0.772  & 0.841 & 0.957  & 1.045 & 0.861 \\
                  & 2021:W19          & 0.680 & 0.776  & 0.840 & 0.955  & 1.047 & 0.861 \\
                  & 2021:W20          & 0.680 & 0.777  & 0.839 & 0.958  & 1.050 & 0.861 \\
                  & 2021:W21          & 0.676 & 0.731  & 0.834 & 0.930  & 1.015 & 0.840 \\
                  & 2021:W22          & 0.669 & 0.730  & 0.836 & 0.927  & 1.016 & 0.838 \\
                  & 2021:W23          & 0.664 & 0.731  & 0.834 & 0.924  & 1.009 & 0.837 \\
                  & 2021:W24          & 0.661 & 0.727  & 0.831 & 0.924  & 1.016 & 0.836 \\
                  & 2021:W25          & 0.667 & 0.728  & 0.832 & 0.928  & 1.018 & 0.840 \\
                  & 2021:W26          & 0.663 & 0.725  & 0.829 & 0.928  & 1.018 & 0.838 \\
                  & 2021:W27          & 0.659 & 0.717  & 0.830 & 0.925  & 1.019 & 0.836 \\
                  & 2021:W28          & 0.656 & 0.710  & 0.829 & 0.919  & 1.018 & 0.835 \\
                  & 2021:W29          & 0.652 & 0.723  & 0.790 & 0.923  & 1.020 & 0.828 \\
                  & 2021:W30          & 0.650 & 0.718  & 0.786 & 0.920  & 1.026 & 0.826 \\
                  & 2021:W31          & 0.648 & 0.716  & 0.780 & 0.918  & 1.034 & 0.824 \\
                  & 2021:W32          & 0.645 & 0.714  & 0.776 & 0.913  & 1.039 & 0.823 \\
                  & 2021:W33          & 0.606 & 0.675  & 0.773 & 0.908  & 1.023 & 0.804 \\
                  & 2021:W34          & 0.612 & 0.678  & 0.770 & 0.905  & 1.024 & 0.803 \\
                  & 2021:W35          & 0.619 & 0.677  & 0.766 & 0.901  & 1.024 & 0.802 \\
                  & 2021:W36          & 0.621 & 0.674  & 0.765 & 0.896  & 1.024 & 0.801 \\
                  & 2021:W37          & 0.613 & 0.671  & 0.776 & 0.859  & 1.011 & 0.795 \\
                  & 2021:W38          & 0.610 & 0.664  & 0.776 & 0.856  & 1.013 & 0.793 \\
                  & 2021:W39          & 0.607 & 0.661  & 0.777 & 0.856  & 1.014 & 0.792 \\
                  & 2021:W40          & 0.605 & 0.662  & 0.776 & 0.857  & 1.016 & 0.790 \\
                  & 2021:W41          & 0.597 & 0.670  & 0.769 & 0.851  & 1.005 & 0.777 \\
                  & 2021:W42          & 0.595 & 0.672  & 0.767 & 0.849  & 1.002 & 0.776 \\
                  & 2021:W43          & 0.595 & 0.668  & 0.767 & 0.846  & 1.003 & 0.776 \\
                  & 2021:W44          & 0.595 & 0.667  & 0.766 & 0.843  & 1.002 & 0.776 \\
                  & 2021:W45          & 0.567 & 0.663  & 0.751 & 0.812  & 0.972 & 0.763 \\
                  & 2021:W46          & 0.568 & 0.660  & 0.749 & 0.812  & 0.973 & 0.763 \\
                  & 2021:W47          & 0.567 & 0.657  & 0.747 & 0.812  & 0.973 & 0.763 \\
                  & 2021:W48          & 0.566 & 0.653  & 0.746 & 0.811  & 0.972 & 0.762  \\ \hline \hline
\end{tabular}
\begin{tablenotes}
\item \textit{Note: The dependent variable is the state-level CO2 per capita growth.}
\end{tablenotes} 
\end{threeparttable}
\end{adjustbox}
\end{table}

\begin{table}[!htbp]
\caption{Distribution of relative QS across states for AR-W-M-Q model, $\tau=0.75$}
\centering
\begin{adjustbox}{width=0.7\textwidth}
\begin{threeparttable}
\begin{tabular}{llccccccc}
\hline \hline
\multicolumn{2}{c}{\textbf{Calendar ($v$})} & \textbf{10\%} & \textbf{25\%} & \textbf{50\%} & \textbf{75\%} & \textbf{90\%} & \textbf{QS}\\ \hline
Backcast          & 2021:W1           & 0.544 & 0.694  & 0.771 & 0.896  & 1.036 & 0.792 \\
                  & 2021:W2           & 0.546 & 0.692  & 0.768 & 0.889  & 1.044 & 0.792 \\
                  & 2021:W3           & 0.549 & 0.688  & 0.769 & 0.885  & 1.047 & 0.792 \\
                  & 2021:W4           & 0.553 & 0.689  & 0.778 & 0.882  & 1.059 & 0.792 \\ \hline
Nowcast           & 2021:W5           & 0.772 & 0.821  & 0.918 & 1.047  & 1.172 & 0.945 \\
                  & 2021:W6           & 0.773 & 0.823  & 0.920 & 1.048  & 1.170 & 0.945 \\
                  & 2021:W7           & 0.773 & 0.825  & 0.921 & 1.049  & 1.166 & 0.945 \\
                  & 2021:W8           & 0.770 & 0.826  & 0.922 & 1.051  & 1.165 & 0.945 \\
                  & 2021:W9           & 0.727 & 0.770  & 0.910 & 1.013  & 1.138 & 0.900 \\
                  & 2021:W10          & 0.725 & 0.769  & 0.911 & 1.011  & 1.140 & 0.899 \\
                  & 2021:W11          & 0.722 & 0.770  & 0.909 & 1.010  & 1.141 & 0.898 \\
                  & 2021:W12          & 0.719 & 0.769  & 0.901 & 1.010  & 1.144 & 0.896 \\
                  & 2021:W13          & 0.693 & 0.753  & 0.866 & 0.973  & 1.130 & 0.881 \\
                  & 2021:W14          & 0.690 & 0.750  & 0.862 & 0.971  & 1.127 & 0.879 \\
                  & 2021:W15          & 0.688 & 0.748  & 0.862 & 0.969  & 1.126 & 0.877 \\
                  & 2021:W16          & 0.685 & 0.746  & 0.863 & 0.966  & 1.126 & 0.876 \\
                  & 2021:W17          & 0.657 & 0.742  & 0.853 & 0.960  & 1.095 & 0.859 \\
                  & 2021:W18          & 0.654 & 0.742  & 0.856 & 0.960  & 1.090 & 0.858 \\
                  & 2021:W19          & 0.650 & 0.739  & 0.859 & 0.960  & 1.086 & 0.858 \\
                  & 2021:W20          & 0.650 & 0.737  & 0.861 & 0.957  & 1.083 & 0.858 \\
                  & 2021:W21          & 0.649 & 0.720  & 0.825 & 0.903  & 1.035 & 0.832 \\
                  & 2021:W22          & 0.648 & 0.723  & 0.820 & 0.904  & 1.036 & 0.831 \\
                  & 2021:W23          & 0.646 & 0.723  & 0.815 & 0.898  & 1.038 & 0.831 \\
                  & 2021:W24          & 0.644 & 0.722  & 0.813 & 0.895  & 1.039 & 0.830 \\
                  & 2021:W25          & 0.648 & 0.732  & 0.834 & 0.904  & 1.035 & 0.833 \\
                  & 2021:W26          & 0.645 & 0.730  & 0.829 & 0.912  & 1.039 & 0.832 \\
                  & 2021:W27          & 0.643 & 0.728  & 0.826 & 0.914  & 1.046 & 0.831 \\
                  & 2021:W28          & 0.640 & 0.726  & 0.821 & 0.912  & 1.054 & 0.830 \\
                  & 2021:W29          & 0.617 & 0.708  & 0.823 & 0.910  & 1.068 & 0.827 \\
                  & 2021:W30          & 0.614 & 0.703  & 0.821 & 0.910  & 1.067 & 0.824 \\
                  & 2021:W31          & 0.611 & 0.700  & 0.819 & 0.906  & 1.068 & 0.822 \\
                  & 2021:W32          & 0.611 & 0.697  & 0.816 & 0.904  & 1.065 & 0.821 \\
                  & 2021:W33          & 0.593 & 0.680  & 0.809 & 0.882  & 1.030 & 0.809 \\
                  & 2021:W34          & 0.591 & 0.679  & 0.813 & 0.884  & 1.029 & 0.808 \\
                  & 2021:W35          & 0.588 & 0.678  & 0.813 & 0.882  & 1.030 & 0.807 \\
                  & 2021:W36          & 0.585 & 0.677  & 0.813 & 0.880  & 1.030 & 0.806 \\
                  & 2021:W37          & 0.604 & 0.687  & 0.813 & 0.889  & 1.047 & 0.807 \\
                  & 2021:W38          & 0.602 & 0.688  & 0.810 & 0.887  & 1.044 & 0.806 \\
                  & 2021:W39          & 0.599 & 0.686  & 0.806 & 0.885  & 1.040 & 0.805 \\
                  & 2021:W40          & 0.596 & 0.686  & 0.802 & 0.883  & 1.036 & 0.804 \\
                  & 2021:W41          & 0.580 & 0.661  & 0.799 & 0.891  & 1.008 & 0.797 \\
                  & 2021:W42          & 0.581 & 0.661  & 0.797 & 0.889  & 1.005 & 0.796 \\
                  & 2021:W43          & 0.584 & 0.663  & 0.796 & 0.886  & 1.007 & 0.796 \\
                  & 2021:W44          & 0.588 & 0.665  & 0.795 & 0.883  & 1.007 & 0.796 \\
                  & 2021:W45          & 0.583 & 0.649  & 0.780 & 0.864  & 0.979 & 0.785 \\
                  & 2021:W46          & 0.584 & 0.648  & 0.783 & 0.864  & 0.977 & 0.785 \\
                  & 2021:W47          & 0.585 & 0.646  & 0.784 & 0.865  & 0.978 & 0.785 \\
                  & 2021:W48          & 0.586 & 0.644  & 0.785 & 0.864  & 0.978 & 0.785  \\ \hline \hline
\end{tabular}
\begin{tablenotes}
\item \textit{Note: The dependent variable is the state-level CO2 per capita growth.}
\end{tablenotes} 
\end{threeparttable}
\end{adjustbox}
\end{table}

\newpage

\end{document}